\documentclass[12pt]{article}

\usepackage{amsmath,amssymb,amsfonts,graphics,graphicx,amscd,amsfonts,epsf,epsfig,color,url}
\allowdisplaybreaks[4]

\date{}
\begin{document}

\begin{flushright} 

\end{flushright} 

\vspace{0.1cm}

\begin{center}
  {\LARGE
  
 Markov Chain Monte Carlo 
for Dummies
}
\vspace{5mm}

\end{center}
\vspace{0.1cm}
\vspace{0.1cm}
\begin{center}

Masanori H{\sc anada}

\vspace{0.3cm}

%
%
%
%
hanada@yukawa.kyoto-u.ac.jp

\end{center}


\vspace{1.5cm}

\begin{center}
  {\bf abstract}
\end{center}

This is an introductory article about Markov Chain Monte Carlo (MCMC) simulation for pedestrians.
Actual simulation codes are provided, and necessary practical details, 
which are skipped in most textbooks, are shown.  
The second half is written for hep-th and hep-lat audience. 
It explains specific methods needed for simulations with dynamical fermions, especially supersymmetric Yang-Mills. 
The examples include QCD and matrix integral, in addition to SYM. 

\newpage
\tableofcontents
\section{Introduction}
\hspace{0.51cm}
Markov Chain Monte Carlo (MCMC) simulation is a very powerful tool for studying the dynamics of quantum field theory (QFT). 
But in hep-th community people tend to think it is a very complicated thing which is beyond their imagination \cite{Joao}. 
They tend to think that a simulation code requires a very complicated and long computer program, 
they need to hire special postdocs with mysterious skill sets, 
very expensive supercomputers which they will never have access are needed, etc. 
It is a pity, because MCMC is actually (at least conceptually) very simple,\footnote{
Because I don't like black boxes, I usually code everything by myself from scratch. 
Still it is extremely rare to use anything more than $+,-,\times,\div,\sin,\cos,\exp,\log,\sqrt{\ \ }$, ``if" and loop. 
Sometimes a few linear algebra routines from LAPACK \cite{LAPACK} are needed, but you can copy and paste them. 
For the Matrix Model of M-theory \cite{Banks:1996vh,deWit:1988wri}
you don't even need LAPACK. 
In short: {\it nothing more than high school math is needed}.
We just have to remove bugs patiently.  
} and 
a lot of nontrivial simulations can be done by using a laptop. 
Indeed I have several papers for which the coding took at most an hour and 
crucial parts of simulations were done on a laptop, e.g. \cite{Azeyanagi:2007su,Azeyanagi:2010ne,Hanada:2012si}. 
You can quickly write a simple code, say a simple integral with the Metropolis algorithm, 
and it teaches you all important concepts. 

Of course for certain theories we have to invest a lot of computational resources.  
If you wanted to compete with lattice QCD experts, a lot of sophisticated optimizations, 
sometimes at the level of hardware, would be needed. However there are many other subjects 
--- including many problems in hep-th field --- which are not yet at that stage. 

There are many sophisticated techniques which enables us to perform large scale simulations 
with realistic computational resources. They are sometimes technically very complicated
but almost always conceptually very simple. 
It is not easy to invent new techniques by ourselves, but it is not hard to learn and use 
something experts have already invented.
In case you have to do a serious simulation, 
you may not be able to code everything by yourself. 
But you can use open-source simulation codes,\footnote{See e.g.~\cite{MCSMC-code,Schaich:2014pda,David-Github,Catterall:2011cea} for supersymmetric theories.}
or you can work with somebody who can write a code. 
And running the code and getting results are rather straightforward, once you understand 
the very basics. 

In this introductory article, I will present basic knowledge needed for the Monte Carlo study of SYM. 
I have two kinds of audience in mind: string theorists who have no idea what is lattice Monte Carlo,
and lattice QCD practitioners who know QCD but not SYM. 
For the former, I provide plenty of examples, including sample codes, which are sufficient 
for running actual simulation codes for SYM.
(Sec.~\ref{sec:MCMC}, Sec.~\ref{sec:Gaussian} and a part of Sec.\ref{sec:multi-variables}
would be useful for much broader audience including non-physicists.)
These materials can also be useful for students and postdocs already working with MCMC; 
the materials presented here are something all senior people working in MCMC expect
their students/postdocs to know, but many students/postdocs do not have chance to learn. 
I also explain the technical differences between lattice QCD and SYM simulations,
which are useful for both string and lattice people. 
\begin{center}
\section*{Note}
\hspace{0.51cm}
\end{center}
This version is (probably) not final; more examples and sample codes will be added. 
I have decided to post it to arXiv because lately I am too busy and do not have much time to work on this. 

Sample codes can be downloaded from GitHub,  \url{https://github.com/MCSMC/MCMC_sample_codes}. 
The latest version of this review will be uploaded there as well. 

Comments, requests and bug/typo reports will be appreciated. 

\section{Markov Chain Monte Carlo (MCMC)}\label{sec:MCMC}
\hspace{0.51cm}

\begin{figure}[htbp]
  \begin{center}
  \rotatebox{0}{
   \includegraphics[width=50mm]{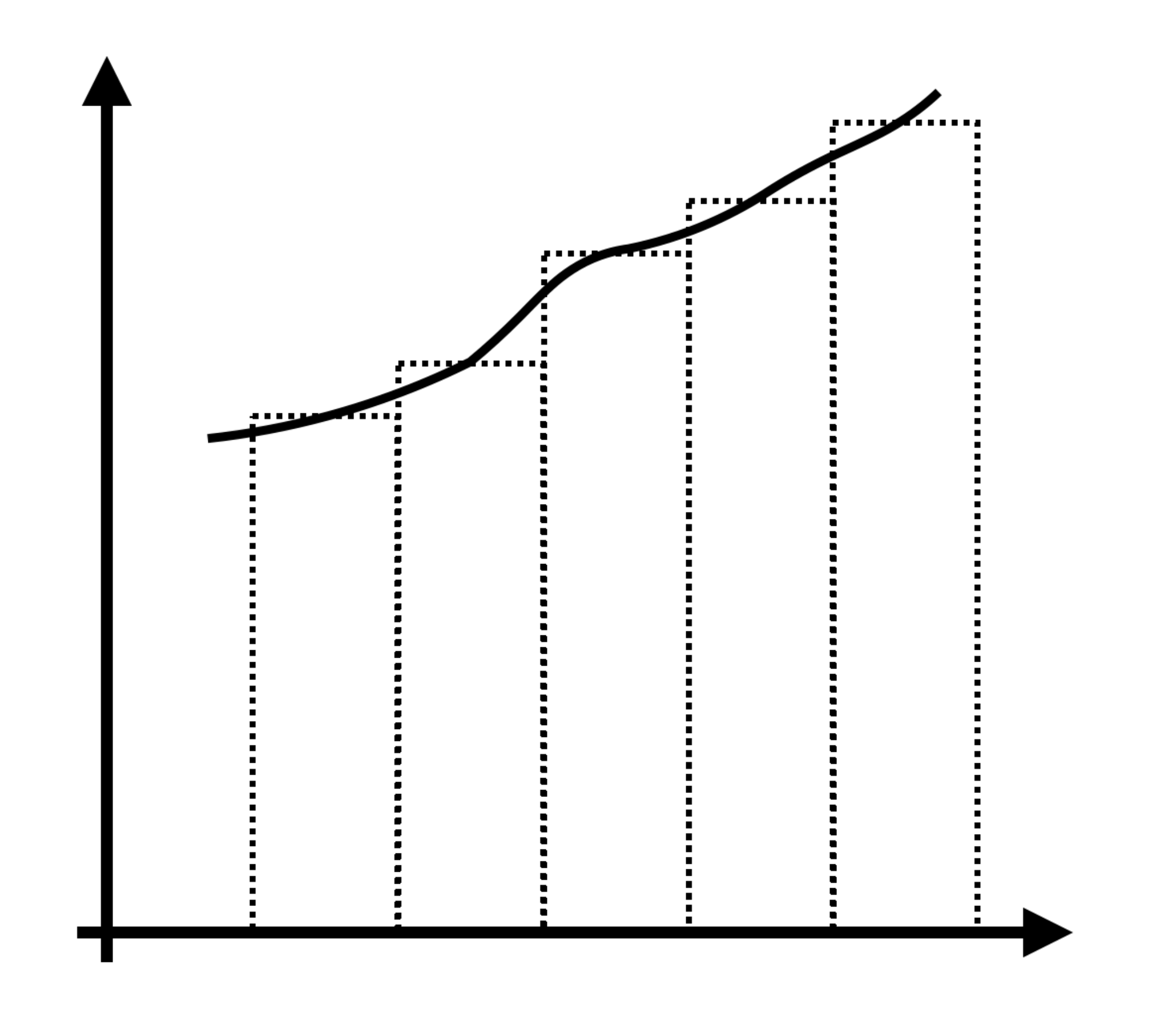}}
  \end{center}
  \caption{Approximate the integral by a sum of the area of rectangles. 
  }\label{fig:integral-sum}
\end{figure}

Suppose we want to perform a Euclidean path-integral with a partition function 
\begin{eqnarray}
Z
=
\int [d\phi] e^{-S[\phi]}, 
\end{eqnarray}
where the action $S[\phi]$ depends only on bosonic field(s) $\phi$. (In later sections I will explain how to include fermions.)
Usually we are interested in the expectation values of operators, 
\begin{eqnarray}
\langle{\cal O}\rangle
=
\frac{1}{Z}\int [d\phi] e^{-S[\phi]}{\cal O}(\phi). 
\end{eqnarray}
In order to make sense of this expression, typically we regularize the theory on a lattice, 
so that the path-integral reduces to a usual integral with respect to finitely many variables.
Let's call these variables $x_1,x_2,\cdots,x_p$. 
Typically, the lattice action is so complicated that it is impossible to estimate the integral analytically. 
Because we have to send $p$ to infinity in order to take the continuum limit or the large-volume limit, 
a naive numerical integral does not work either. If we approximate the integral by a sum by dividing the integral region of each $x_i$ to $n$ intervals like in Fig.~\ref{fig:integral-sum}, 
the calculation cost is proportional to $n^p$, simply because 
we have to take a sum of $n^p$ numbers. 
This is hopelessly hard for realistic values of $n$ and $p$. 
Suppose $n=100$ and $p=10$. Then we have to take a sum of $10^{20}$ numbers. 
Let us convince my collaborators in Livermore Laboratory that this integral is extremely important, 
and use their supercomputer {\it Sequoia}, which was the fastest in the world from 2012 to 2013. Its performance is 20PFLOPS, namely it can process $2\times 10^{16}$ double-precision floating-point arithmetics every second. 
Let's ignore the cost for calculating the value of the function at each point, 
and consider only a sum of given numbers. (This is an unrealistic assumption, of course.) 
But already it takes  $10^{20}/(2\cdot 10^{16})=5000$ seconds. 
Well, it may be acceptable... but if you take $p=15$, it takes
$10^{30}/(2\cdot 10^{16})=5\times 10^{13}$ seconds, which is about 634,000 years. 
For a 4d pure SU$(3)$ Yang-Mills on lattice with $10^4$ points, $p$ is $4\times (3^2-1)\times 10^4$. 
We cannot even take $n=2$. This is the notorious {\it curse of dimensionality}; when the dimension $p$ is large 
it is practically impossible to scan the phase space.     

Perhaps when you are reading this article you have access to much better machines, but it will not give you much gain. 

Markov Chain Monte Carlo (MCMC) circumvents the curse of dimensionality based on the idea of {\it importance sampling}. 
In most cases of our interest, the majority of the phase space is irrelevant because the action $S$ is large 
and the weight $e^{-S}$ is very small. If we can find {\it important} regions in the phase space and invest our resources there, 
we can avoid the curse of dimensionality. MCMC enables us to actually do it.

We assume $S[x_1,x_2,\cdots,x_p]$ is real and the partition function $Z=\int dx_1\cdots dx_p e^{-S[x_1,x_2,\cdots,x_p]}$ is finite.  
In MCMC simulations, 
we construct a chain of sets of variables $\{x^{(0)}\}\to\{x^{(1)}\}\to\{x^{(2)}\}\to\cdots
\{x^{(k)}\}\to\{x^{(k+1)}\}\to\cdots$ satisfying the following conditions: 

\begin{itemize}
\item
{\bf Markov Chain}. 
The probability of obtaining $\{x^{(k+1)}\}$ from $\{x^{(k)}\}$ does not depend on the previous configurations 
$\{x^{(0)}\},\{x^{(1)}\},\cdots, \{x^{(k-1)}\}$. We denote this transition probability by $T[\{x^{(k)}\}\to\{x^{(k+1)}\}]$. 

\item
{\bf Irreducibility}. 
Any two configurations are connected by finite steps. 

\item
{\bf Aperiodicity}.
The {\it period} of a configuration $\{x\}$ is given by the greatest common divisor of possible numbers of steps to come back to itself. 
When the period is 1 for all configurations, the Markov chain is called aperiodic.  

\item
{\bf Detailed balance condition}. 
The transition probability $T$ satisfies 
$e^{-S[\{x\}]}T[\{x\}\to\{x'\}]
=
e^{-S[\{x'\}]}T[\{x'\}\to\{x\}]$. 

\end{itemize}

Then, the probability distribution of $\{x^{(k)}\} (k=1,2,\cdots)$ converges to 
$P(x_1,x_2,\cdots,x_p)=e^{-S(x_1,x_2,\cdots,x_p)}/Z$ 
as the chain becomes longer. 
The expectation values are obtained by taking the average over the configurations, 
\begin{eqnarray}
\langle\hat{O}\rangle
=
\int dx_1\cdots dx_p O(x_1,\cdots,x_p)P(x_1,x_2,\cdots,x_p)
=
\lim_{n\to\infty}
\frac{1}{n}\sum_{k=1}^n O(x_1^{(k)},\cdots,x_p^{(k)}). 
\end{eqnarray}
Note that this is not an approximation; {\it this is exact}.  
Practically we can have only finitely many configurations, so we can only approximate 
the right hand side by a finite sum. However there is a systematic way to improve it to arbitrary precision: just make the chain longer. 
 
Although a proof is rather involved, the importance of each condition can easily be understood. 
Probably the most nontrivial condition for most readers is the detailed balance. Suppose the chain converged to a certain distribution $P[\{x\}]$. 
Then it has to be `stationary', or equivalently, it should be invariant when shifted one step:
\begin{eqnarray}
\sum_{\{x\}}P[\{x\}]T[\{x\}\to\{x'\}]
=
P[\{x'\}]
\end{eqnarray}
If $P[\{x\}]\propto e^{-S[\{x\}]}$, 
it follows from the detailed balance condition as 
\begin{eqnarray}
\sum_{\{x\}}P[\{x\}]T[\{x\}\to\{x'\}]
&=&
\sum_{\{x\}}P[\{x'\}]T[\{x'\}\to\{x\}]
\nonumber\\
&=&
P[\{x'\}]\sum_{\{x\}}T[\{x'\}\to\{x\}]
\nonumber\\
&=&
P[\{x'\}]. 
\end{eqnarray}
I recommend you to follow a complete proof once by looking at an appropriate textbook, 
but you don't have to keep it in your brain. You will need it only when you try to invent something better than MCMC.  
 
Note that, even if you use exactly the same simulation code, 
 if you take different initial condition or use different sequence of random numbers, you get different chain. 
 Still, the chain always converges to the same statistical distribution. 

\subsection{Off-topic: Bayesian analysis}\label{sec:Bayes}
\hspace{0.51cm}
MCMC is powerful outside physics as well. To see a little bit of flavor, 
let us consider the Bayes's theorem, 
\begin{eqnarray}
P(B_i|A)=\frac{P(A|B_i)P(B_i)}{\sum_jP(A|B_j)P(B_j)}. 
\end{eqnarray}
Here $P(A|B)$ is the conditional probability:
Probability that $A$ is true when the condition $B$ is satisfied.
For example $B_1,B_2,B_3\cdots$ are physicists, high tech engineers, MLB players etc, 
and $A$ is millionaires. 

Suppose $P(A|B_i)$ and $P(B_i)$ are given (e.~g. $P(A|B_1)=10^{-4}, P(A|B_2)=0.05, P(A|B_3)=0.8,\cdots$), and we want to derive $P(B_i|A)$. 
We can identify $B_j$ and $P(A|B_j)P(B_j)$ with the value of the field $\phi$, the path integral weight $e^{-S[\phi]}[d\phi]$. 
The denominator $\sum_jP(A|B_j)P(B_j)=P(A)$ is regarded as the partition function $Z$. 
Then we can use MCMC to obtain $P(B_i|A)\sim \frac{e^{-S[\phi]}[d\phi]}{Z}$ via the Bayes's theorem; 
namely we can collect many samples and see the distribution.

Also if we know $f(B_i)\equiv P(C|B_i)$ you can calculate $P(C|A)$ as 
\begin{eqnarray}
P(C|A)=\langle f\rangle. 
\end{eqnarray}
For example $C$ is nice muscle and $P(C|B_1)=P(C|B_2)=0.01, P(C|B_3)=0.99,\cdots$. 
\section{Integration of one-variable functions with Metropolis algorithm}\label{sec:Gaussian}
\hspace{0.51cm}
Let us start with the integration of a one-variable function with the Metropolis algorithm \cite{Metropolis}. 
In particular, we will consider the simplest example we can imagine: the Gaussian integral, $S(x)=x^2/2$. 
This very basic example contains essentially all important ingredients; 
all other cases are, ultimately, just technical improvements of this example. 

Of course we can handle the Gaussian integral analytically. 
Also there is a much better algorithm for generating Gaussian random numbers 
(see Appendix~\ref{sec:Box-Muller}). 
We use it just for an educational purpose.

\subsection{Metropolis Algorithm}\label{sec:Metropolis-Gaussian}
\hspace{0.51cm}
Let us consider the weight $e^{-S(x)}$, where $S(x)$ is a continuous function of $x\in{\mathbb R}$ bounded from below. 
We further assume that $\int e^{-S(x)}dx$ is finite. 
The Metropolis algorithm gives us a chain of configurations (or just `values' in the case of single variable)
$x^{(0)}\to x^{(1)}\to x^{(2)}\to\cdots$ which satisfies the conditions listed above: 
\begin{enumerate}
\item
Randomly choose $\Delta x\in {\mathbb R}$, and shift $x^{(k)}$ as $x^{(k)}\to x'\equiv x^{(k)}+\Delta x$. 
($\Delta x$ and $-\Delta x$ must appear with the same probability, so that the detailed balance condition is satisfied. 
Here we use the uniform random number between $\pm c$, 
where $c>0$ is the `step size'.) 

\item
Metropolis test: Generate a uniform random number $r$ between 0 and 1. 
If $r<e^{-\Delta S}$, where  $\Delta S = S(x')-S(x^{(k)})$, then 
$x^{(k+1)}=x'$, i.e. the new value is `accepted.' 
Otherwise 
$x^{(k+1)}=x^{(k)}$,  i.e. the new value is `rejected.' 

\item
Repeat the same for $k+1, k+2, \cdots$. 
\end{enumerate}
It is an easy exercise to see that all conditions explained above are satisfied:
\begin{itemize}
\item
It is a Markov Chain, because the past history is not referred
either for the selection of $\Delta x$ or the Metropolis test.   

\item
It is irreducible; for example, any $x$ and $x'$, by taking $n$ large we can make $\frac{x-x'}{n}$ to be in $[-c,c]$, 
and there is a nonzero probability that $\Delta x=\frac{x-x'}{n}$ appears $n$ times in a row and passes the Metropolis test every time. 

\item
For any $x$, there is a nonzero probability of $\Delta x=0$. Hence the period is one for any $x$. 

\item
If $|x-x'|>c$, $T[x\to x']=T[x'\to x]=0$. 
When $|x-x'|\le c$, both $\Delta x = x'-x$ and $\Delta x' = x-x'$ are chosen with probability $\frac{1}{2c}$. 
Let us assume $\Delta S=S[x']-S[x]>0$, without a loss of generality. Then the change $x\to x'$ passes the Metropolis test with probability $e^{-\Delta S}$, 
while $x'\to x$ is always accepted. Hence $T[x\to x']=\frac{e^{-\Delta S}}{2c}$ and $T[x'\to x]=\frac{1}{2c}$, 
and $e^{-S[x]}T[x\to x']=e^{-S[x']}T[x'\to x]=\frac{e^{-S[x']}}{2c}$. 

\end{itemize}

\subsubsection{How it works}
\hspace{0.51cm}

Let me show a sample code written in C:

\begin{verbatim}
#include <stdio.h>
#include <stdlib.h>
#include <math.h>
#include <time.h>

int main(void){
  int iter,niter=100;
  int naccept;
  double step_size=0.5e0;
  double x,backup_x,dx;
  double action_init, action_fin;
  double metropolis;

  srand((unsigned)time(NULL)); 

  /*********************************/
  /* Set the initial configuration */
  /*********************************/      
  x=0e0;
  naccept=0;
  /*************/
  /* Main loop */
  /*************/
  for(iter=1;iter<niter+1;iter++){
    backup_x=x;    
    action_init=0.5e0*x*x;
    
    dx = (double)rand()/RAND_MAX;
    dx=(dx-0.5e0)*step_size*2e0;
    x=x+dx;
    
    action_fin=0.5e0*x*x;
    /*******************/
    /* Metropolis test */
    /*******************/
    metropolis = (double)rand()/RAND_MAX;    
    if(exp(action_init-action_fin) > metropolis)
      /* accept */
      naccept=naccept+1;
    else 
      /* reject */
      x=backup_x;
    /***************/
    /* data output */
    /***************/	
    printf("%f\n",x);}
}
\end{verbatim}

Let me explain the code line by line. Firstly, by 
 \begin{verbatim}
  srand((unsigned)time(NULL)); 
 \end{verbatim}
the seed for the random number generator is set. 
A default random number generator is used, by using the system clock time to set the seed randomly. 
For more serious simulations, it is better to use a good generator, say the Mersenne twister.

Then we specify an initial configuration; here we took $x^{(0)}=0$. 
\textbf{naccept} counts how many times the new values are accepted. 
 \begin{verbatim}
  x=0e0;
  naccept=0;
 \end{verbatim}
Then we move on to the main part of the simulation,  
which is inside the following loop:
 \begin{verbatim}
  for(iter=1;iter<niter+1;iter++){ .... }
 \end{verbatim}
Here, \textbf{iter} corresponds to $k$, and \textbf{niter} is the number of configurations
we will collect during the simulation. 

Inside the loop, the first thing we have to do is to save the value of \textbf{x}$=x^{(k)}$, 
because it may or may not be updated: 
\begin{verbatim}
    backup_x=x; 
\end{verbatim}
Then \textbf{action$\underbar{\ }$init}$=S(x^{(k)})$ is calculated. 

Now we have to generate a random variation \textbf{dx}$=\Delta x$ with an appropriate step size, 
and shift $x$ to $x'=x^{(k)}+\Delta x$.
We can generate a uniform random number in $[0,1]$ by \textbf{rand()/RAND}$\underbar{\ }$\textbf{MAX}. 
From this we can easily get $-c<\Delta x<c$. 
\begin{verbatim}
    dx = (double)rand()/RAND_MAX;
    dx=(dx-0.5e0)*step_size*2e0;
    x=x+dx;
\end{verbatim}
By using $x'$, \textbf{action$\underbar{\ }$fin}$=S(x')$ is calculated.
Note that \textbf{x} and \textbf{backup$\underbar{\ }$x} in the code correspond to $x'$ and $x^{(k)}$. 

Finally we perform the Metropolis test: 
\begin{verbatim}
    metropolis = (double)rand()/RAND_MAX;    
    if(exp(action_init-action_fin) > metropolis)
      /* accept */
      naccept=naccept+1;
    else 
      /* reject */
      x=backup_x;
\end{verbatim}
\textbf{metropolis} is a uniform random number in $[0,1]$, which corresponds to $r$. 
Depending on the result of the test, we accept or reject $x'$. 

We emphasize again that {\it all MCMC simulations have exactly the same structure}; 
there are many fancy algorithms, but essentially, they are all about improving the step $x\to x+\Delta x$. 

\begin{figure}[htbp]
 \begin{minipage}{0.32\hsize}
 \begin{center}
   \rotatebox{0}{
  \includegraphics[width=40mm]{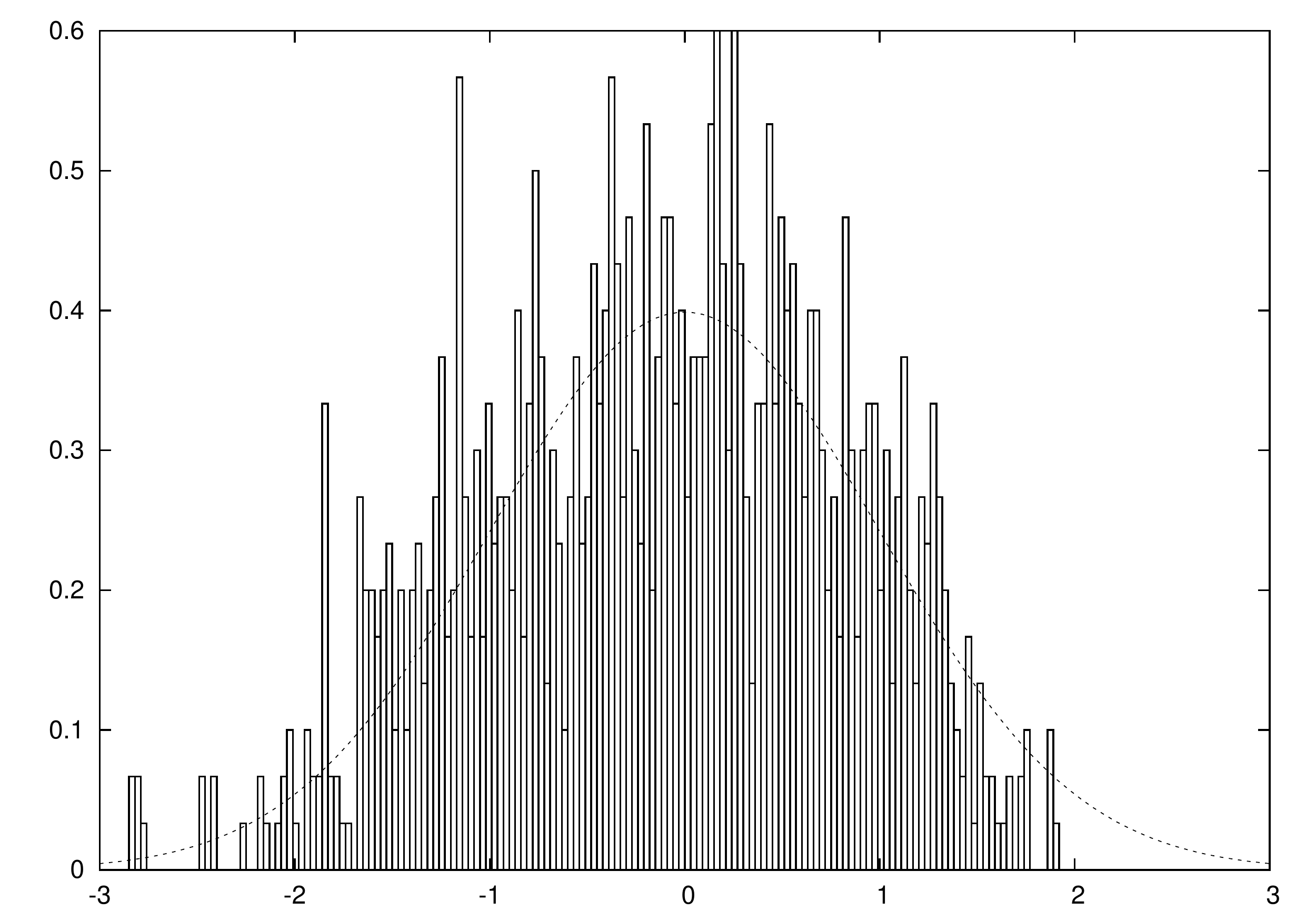}}
 \end{center}
 \end{minipage}
 \begin{minipage}{0.32\hsize}
  \begin{center}
  \rotatebox{0}{
   \includegraphics[width=40mm]{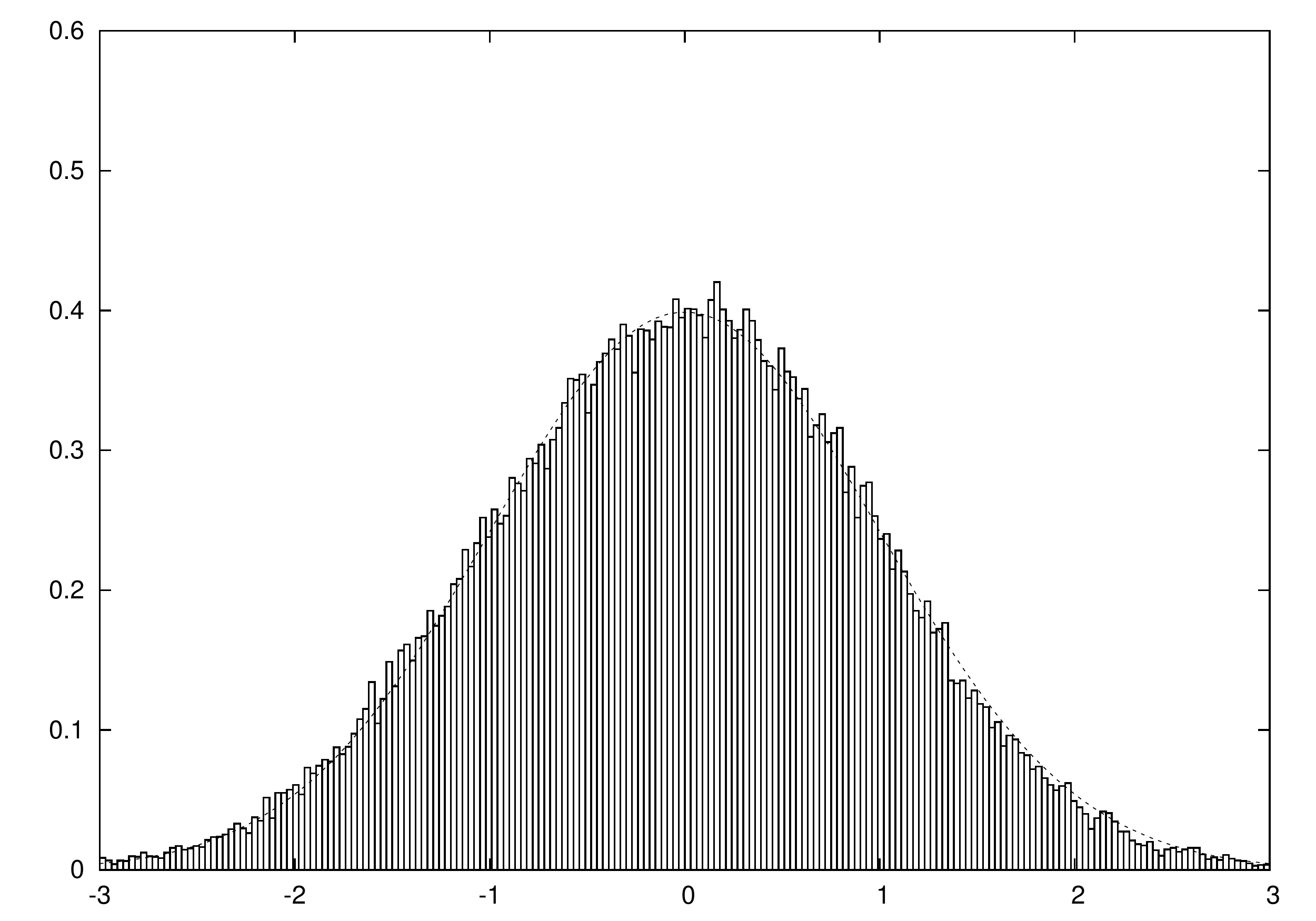}}
  \end{center}
 \end{minipage}
 \begin{minipage}{0.32\hsize}
 \begin{center}
   \rotatebox{0}{
  \includegraphics[width=40mm]{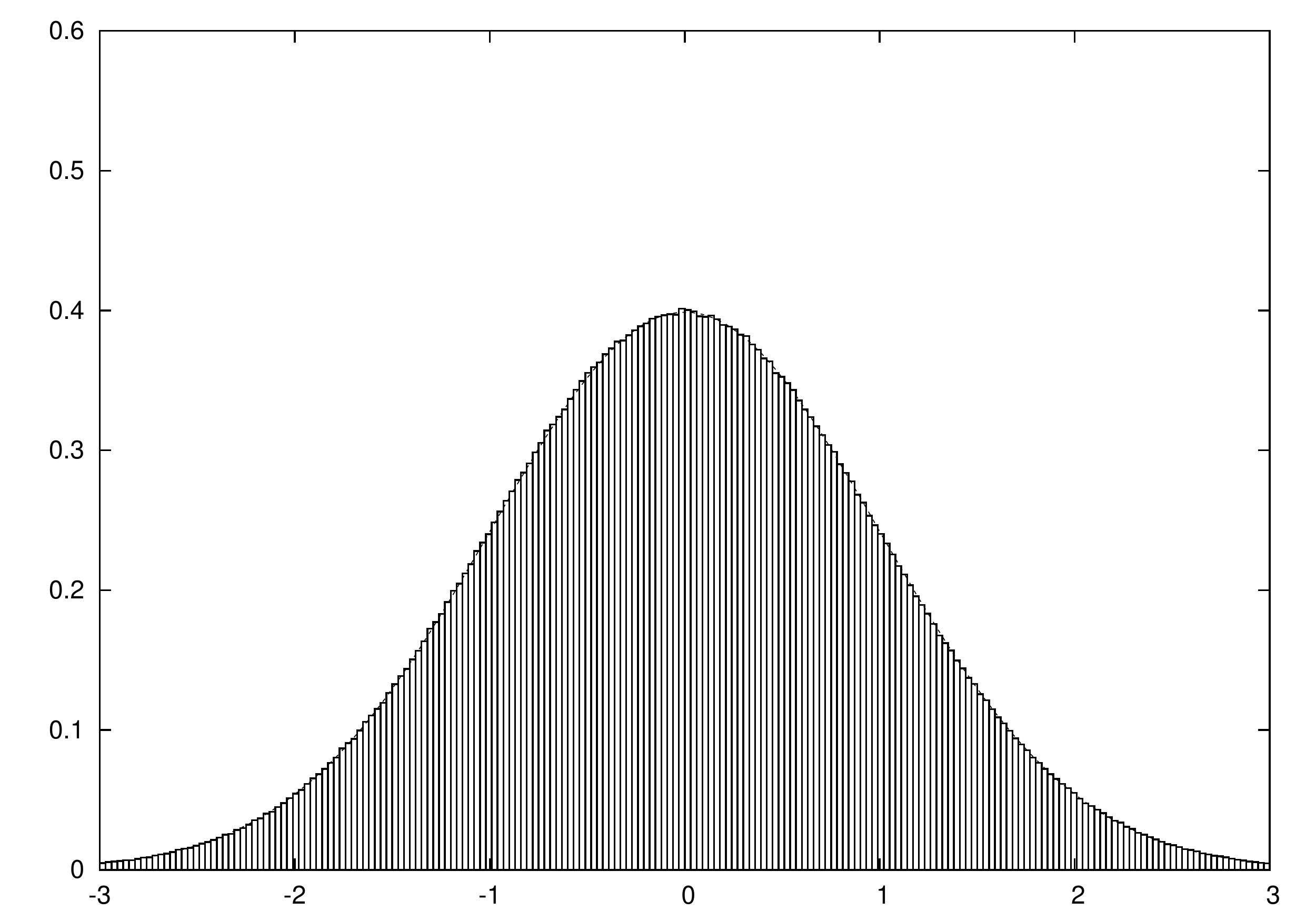}}
 \end{center}
 \end{minipage}
 \caption{The distribution of $x^{(1)},x^{(2)},\cdots,x^{(n)}$, for $n=10^3, 10^5$ and $10^7$, 
 and $\frac{e^{-x^2/2}}{\sqrt{2\pi}}$.
 }\label{Gaussian_plot}
\end{figure}
\begin{figure}[htbp]
  \begin{center}
  \rotatebox{0}{
   \includegraphics[width=70mm]{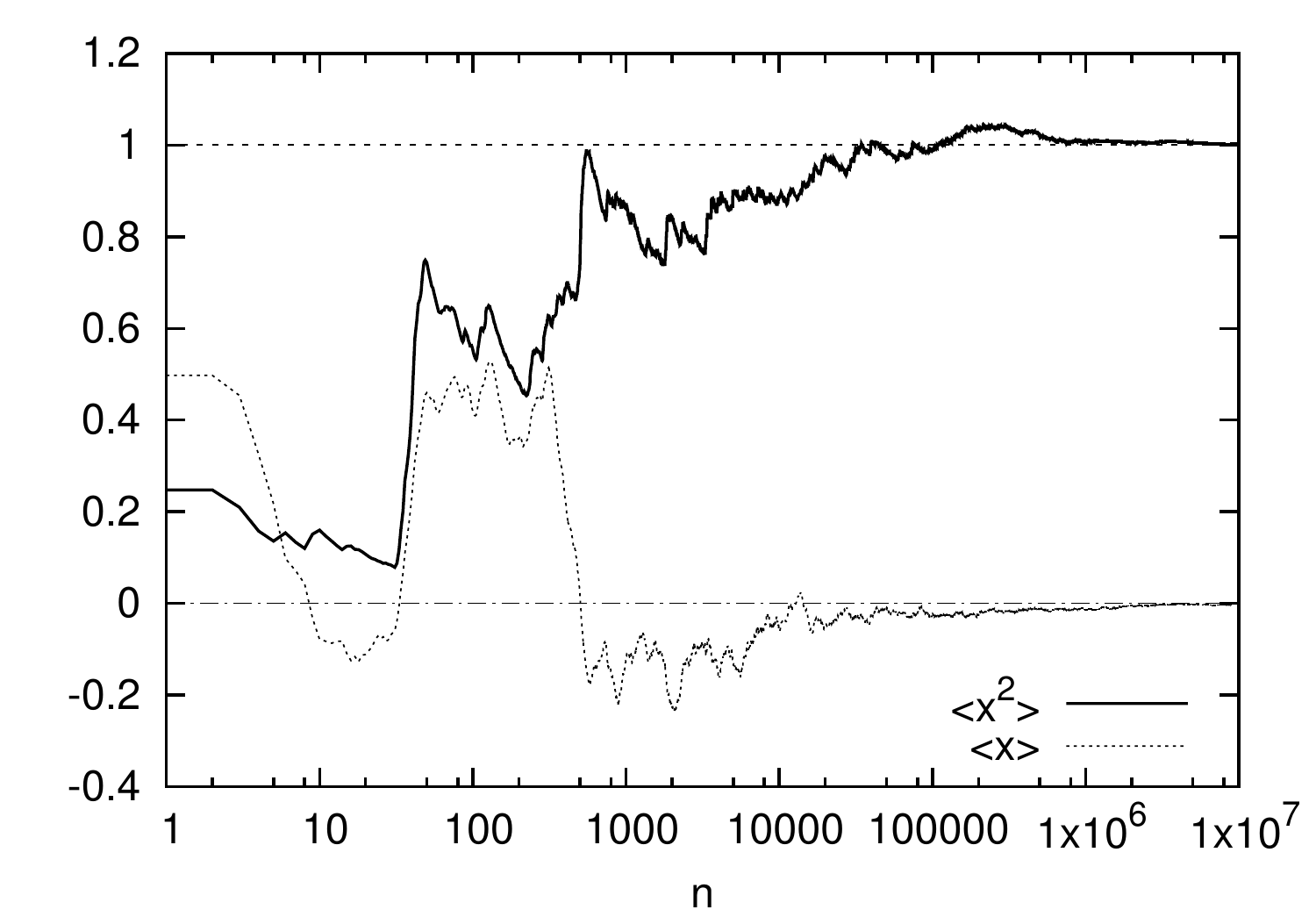}}
  \end{center}
  \caption{$\langle x\rangle = \frac{1}{n}\sum_{k=1}^n x^{(k)}$ and $\langle x^2\rangle = \frac{1}{n}\sum_{k=1}^n \left(x^{(k)}\right)^2$.
  As $n$ becomes large, they converge to the right values, $0$ and $1$. 
  }\label{fig:Gaussian-Metropolis-vev}
\end{figure}

We take $x^{(0)}=0$, and $\Delta x$ to be a uniform random number between $-0.5$ and $0.5$. 
(As we will see later, this parameter choice is not optimal.)
In Fig.~\ref{Gaussian_plot}, we show the distribution of $x^{(1)},x^{(2)},\cdots,x^{(n)}$, for $n=10^3, 10^5$ and $10^7$. 
We can see that the distribution converges to $e^{-x^2/2}/\sqrt{2\pi}$. 
The expectation values $\langle x\rangle = \frac{1}{n}\sum_{k=1}^n x^{(k)}$ and $\langle x^2\rangle = \frac{1}{n}\sum_{k=1}^n \left(x^{(k)}\right)^2$ are plotted in Fig.~\ref{fig:Gaussian-Metropolis-vev}. 
As $n$ becomes large, they converge to the right values, $0$ and $1$. 

Note that the step size $c$ should be chosen so that the acceptance rate is not too high, not too low. 
If $c$ is too large, the acceptance rate becomes extremely low, then the value is rarely updated. If $c$ is too small, the acceptance rate is almost 1, but the change of the value at each step is extremely small. In both cases, huge amount of configurations are needed in order to approximate 
the integration measure accurately. The readers can confirm it by changing the step size in the sample code. (We will demonstrate it in Sec.~\ref{sec:Autocorrelation}.) 

Typically the acceptance rate 30\% -- 80\% is good. But it can heavily depend on the detail of the system and algorithm;   
See Sec.~\ref{sec:choice-of-parameters} and Sec.~\ref{sec:HMC-how-it-works-matrix} for details. 

\subsubsection*{A bad example}
\begin{figure}[htbp]
  \begin{center}
  \rotatebox{0}{
   \includegraphics[width=60mm]{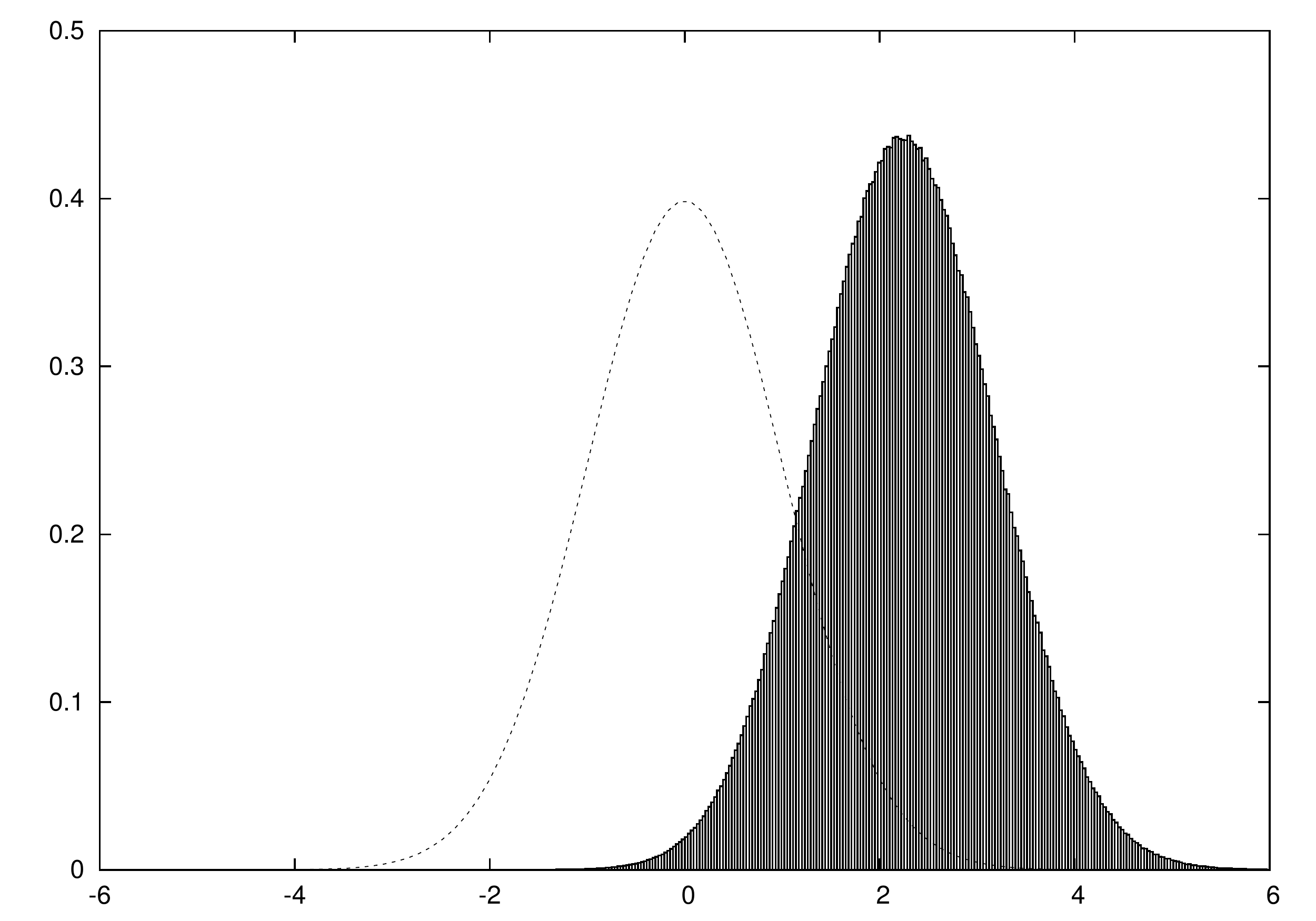}}
  \end{center}
  \caption{
  The distribution of $x^{(1)},x^{(2)},\cdots,x^{(n)}$ for $n=10^7$, with {\it wrong} algorithm with $\Delta x\in \left[-\frac{1}{2}, 1\right]$. 
  The dotted line is the right Gaussian distribution $\frac{e^{-x^2/2}}{\sqrt{2\pi}}$. 
    }\label{fig:Gaussian-Metropolis-wrong}
\end{figure}

\hspace{0.51cm}
It is instructive to see wrong examples. Let us take $\Delta x$ from $\left[-\frac{1}{2}, 1\right]$, so that  
the detailed balance condition is violated; for example $0\to 1$ has a finite probability while $1\to 0$ is impossible. 
Then, as we can see from Fig~\ref{fig:Gaussian-Metropolis-wrong}, the chain does not converge to the right probability distribution.

\subsection{Autocorrelation and Thermalization}\label{sec:Autocorrelation}
\hspace{0.51cm}
In MCMC, $x^{(k+1)}$ is obtained from $x^{(k)}$. In general, they are correlated. 
This correlation is called {\it autocorrelation}. The autocorrelation can exist over many steps. 
The autocorrelation length depends on the detail of the theory, algorithm and the parameter choice. 
Because of the autocorrelation, some cares are needed. 

\begin{figure}[htbp]
  \begin{center}
  \rotatebox{0}{
   \includegraphics[width=70mm]{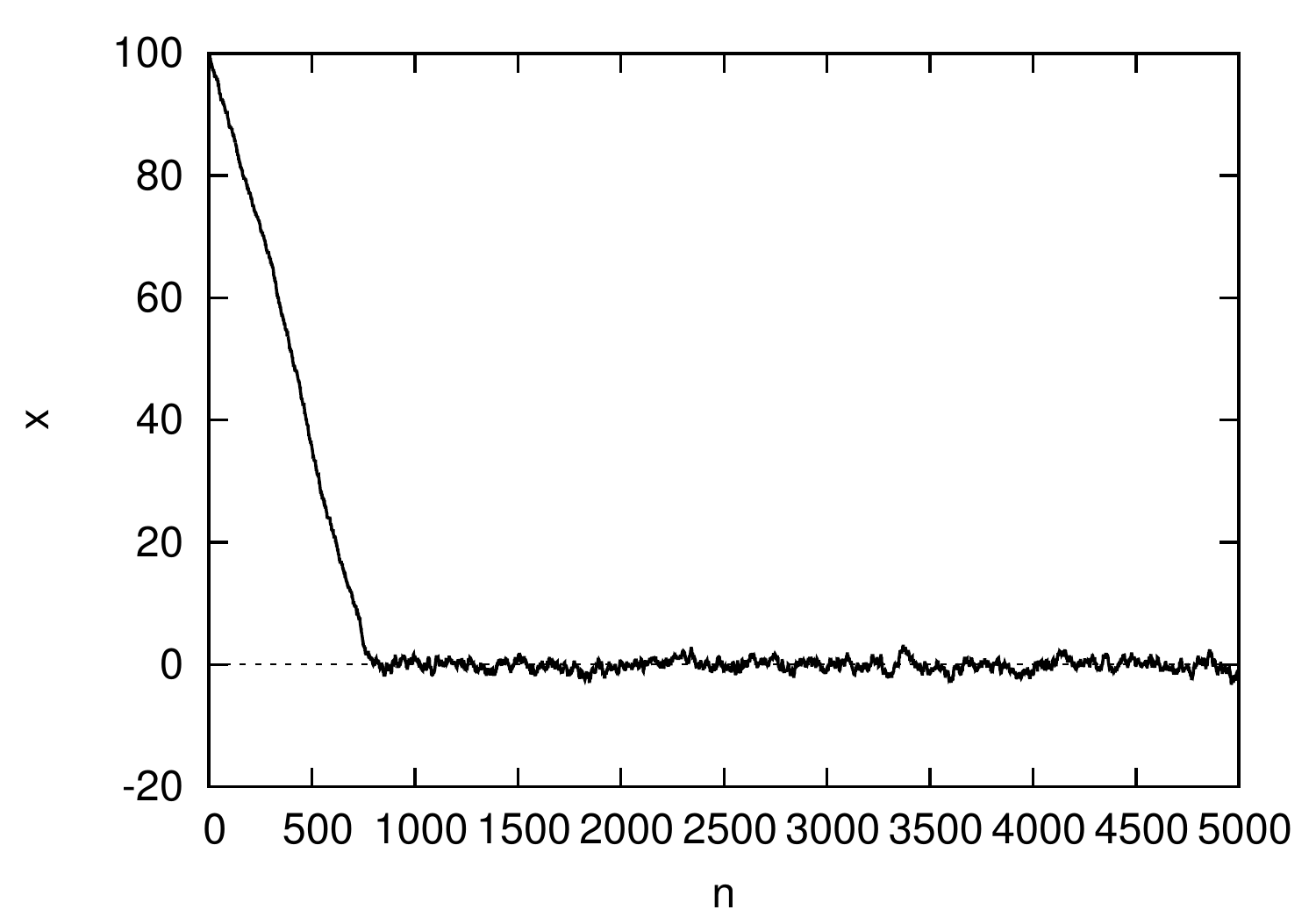}}
  \end{center}
  \caption{
  Monte Calro history of the Gaussian integral with the Metropolis algorithm, $\Delta x\in[-0.5,0.5]$. 
  We took the initial value to be a very atypical value, $x=100$. It takes a lot of steps to reach typical values $|x|\sim 1$. 
    }\label{fig:thermalization}
\end{figure}

In the above, we have set $x^{(0)}=0$, because we knew it is `the most important configuration'. 
What happens if we start with an atypical value, say $x^{(0)}=100$? 
It takes some time for typical values to appear, due to the autocorrelation. 
The history of the Monte Carlo simulation 
with this initial condition is shown in Fig.~\ref{fig:thermalization}. 
The value of $x$ eventually reaches to `typical values' $|x|\lesssim 1$  --- we often say 
`the configurations are thermalized' (note that the same term `thermalized' has another meaning as well, as we will see shortly) ---, but a lot of steps are needed. 
If we include `unthermalized' configurations when we estimate the expectation values, 
we will suffer from huge error unless the number of configurations are extremely large. 
We should discard unthermalized configurations, say $n\lesssim 1000$.\footnote{
Number of steps needed for the thermalization is sometimes called 
`burn-in time' or `mixing time'. 
} 

\begin{figure}[htbp]
  \begin{center}
  \rotatebox{0}{
   \includegraphics[width=70mm]{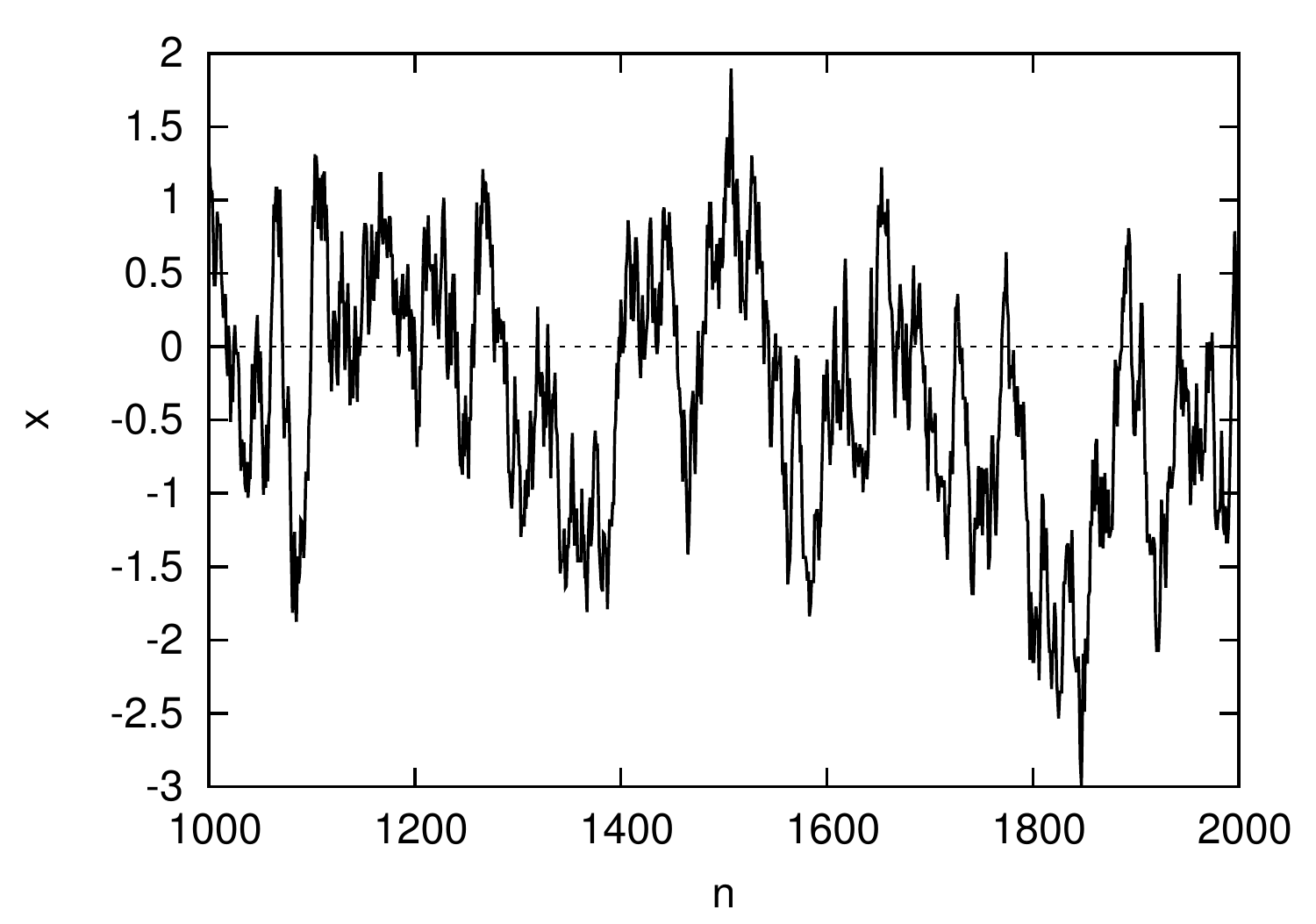}}
  \end{center}
  \caption{
A zoom-up of Fig.~\ref{fig:thermalization}, from $n=1000$ to $n=2000$. 
The values of $x$ can be strongly correlated unless they are at least $20$ or $30$ steps separated. 
  }\label{fig:autocorrelation}
\end{figure}

In generic, more complicated situation, we don't a priori know 
what the typical configurations look like. 
Still, whether the configurations are thermalized or not can be seen by looking at 
several observables. As long as they are changing monotonically, it is plausible that the configuration is moving toward a typical one. 
When they start to oscillate around certain values (see Fig.~\ref{fig:autocorrelation}, we can see a fluctuation around $x=0$),
it is reasonable to think the configuration has been thermalized.\footnote{
For more careful analysis, we can vary the number of configurations removed and take it large enough so that 
the average values do not change any more.  
}  

Fig.~\ref{fig:autocorrelation} is a zoom-up of Fig.~\ref{fig:thermalization}, from $n=1000$ to $n=2000$. 
We can see that the values of $x$ can be strongly correlated unless they are $20$ or $30$ step separated. 
(We will give a quantitative analysis in Sec.~\ref{sec:Jackknife}.)
When we estimate the statistical error, we should not treat all configuration as independent; rather we have only 1 independent configuration every 20 or 30 steps. 
Note also that we need sufficiently many independent configurations in order to estimate the expectation values reliably. 
We often say the simulation has been thermalized when we have sufficiently many independent configurations 
so that the expectation values are stabilized.

\subsubsection{Jackknife method}\label{sec:Jackknife}
\hspace{0.51cm}
The Jackknife method provides us with a simple way to estimate the autocorrelation length. 
Here we assume the quantity of interest can be calculated for each sample.\footnote{
The correlation function is in this class. 
The mass of particle excitation is not; we need to calculate two-point function 
by using many samples and then extract the mass from its exponential decay. 
} 
For more generic cases, see Appendix~\ref{sec:Jackknife-generic}. 

In the Jackknife method, we first divide the configurations to bins with width $w$; 
the first bin consists of $\{x^{(1)}\}, \{x^{(2)}\},\cdots,\{x^{(w)}\}$, the second bin is $\{x^{(w+1)}\}, \{x^{(w+2)}\},\cdots,\{x^{(2w)}\}$, and so on. 
Suppose we have $n$ bins. 
Then we define the average of an observable $f(x)$ with $k$-th bin removed, 
\begin{eqnarray}
\overline{f}^{(k,w)}
\equiv
\frac{1}{(n-1)w}\sum_{j\ \notin\ k\mathchar`-{\rm th\ bin}}f(x^{(j)}). 
\end{eqnarray}
The average value 
\begin{eqnarray}
\overline{f}
\equiv
\frac{1}{n}\sum_{k}\overline{f}^{(k,w)}
\end{eqnarray}
is the same as the average of all samples, $\frac{1}{nw}\sum_j f(x^{(j)})$, for the class of quantities we are discussing.  
The {\it Jackknife error} is defined by 
\begin{eqnarray}
\Delta_w
\equiv
\sqrt{\frac{n-1}{n}\sum_{k}\left(\overline{f}^{(k,w)}-\overline{f}\right)^2}. 
\end{eqnarray}
By using 
\begin{eqnarray}
\tilde{f}^{(k,w)}
\equiv
\frac{1}{w}\sum_{j\ \in\ k\mathchar`-{\rm th\ bin}}f(x^{(j)}),  
\end{eqnarray}
we can easily see 
\begin{eqnarray}
\overline{f}^{(k,w)}-\overline{f}
=
\frac{\overline{f}-\tilde{f}^{(k,w)}}{n-1}. 
\end{eqnarray}
Hence 
\begin{eqnarray}
\Delta_w
\equiv
\sqrt{\frac{1}{n(n-1)}\sum_{k}\left(\tilde{f}^{(k,w)}-\overline{f}\right)^2}. 
\end{eqnarray}
Namely $\Delta_w$ is the standard error obtained by treating $\tilde{f}^{(k,w)}$ to be independent samples.

Typically, as $w$ becomes large, $\Delta_w$ increases and then becomes almost constant at certain value of $w$, which we denote by $w_c$.
This $w_c$ and $\Delta_{w_c}$ give good estimates of the autocorrelation length and the error bar. 

It can be understood as follows. Let us consider two bin sizes $w$ and $2w$. Then
\begin{eqnarray}
\tilde{f}^{(k,2w)}
=
\frac{\tilde{f}^{(2k-1,w)}+\tilde{f}^{(2k,w)}}{2}, 
\end{eqnarray}
\begin{eqnarray}
\Delta_{2w}
&=&
\sqrt{\frac{1}{\frac{n}{2}\left(\frac{n}{2}-1\right)}\sum_{k=1}^{n/2}\left(\tilde{f}^{(k,2w)}-\overline{f}\right)^2}
\nonumber\\
&=&
\sqrt{\frac{4}{n(n-2)}\sum_{k=1}^{n/2}\left(
\frac{\left(\tilde{f}^{(2k-1,w)}-\overline{f}\right)}{2}
+
\frac{\left(\tilde{f}^{(2k,w)}-\overline{f}\right)}{2}
\right)^2}. 
\end{eqnarray}
If $w$ is sufficiently large, $\tilde{f}^{(2k-1,w)}-\overline{f}$ and $\tilde{f}^{(2k,w)}-\overline{f}$ 
should be independent, and the cross-term $\left(\tilde{f}^{(2k-1,w)}-\overline{f}\right)\cdot\left(\tilde{f}^{(2k,w)}-\overline{f}\right)$
should average to zero after summing up with respect to sufficiently many $k$. Then
\begin{eqnarray}
\Delta_{2w}
\sim
\sqrt{\frac{1}{n^2}\sum_{k=1}^{n}\left(\tilde{f}^{(k,w)}-\overline{f}\right)^2}
\sim
\Delta_{w}. 
\end{eqnarray}
In this way, $\Delta_{w}$ becomes approximately constant when $w$ is large enough
so that $\tilde{f}^{(k,w)}$ can be treated as independent samples. 
(Note that $n$ must also be large for the above estimate to hold.)
$\Delta_{w}$ is the standard error of these `independent samples'. 

\begin{figure}[htbp]
 \begin{center}
   \rotatebox{0}{
  \includegraphics[width=70mm]{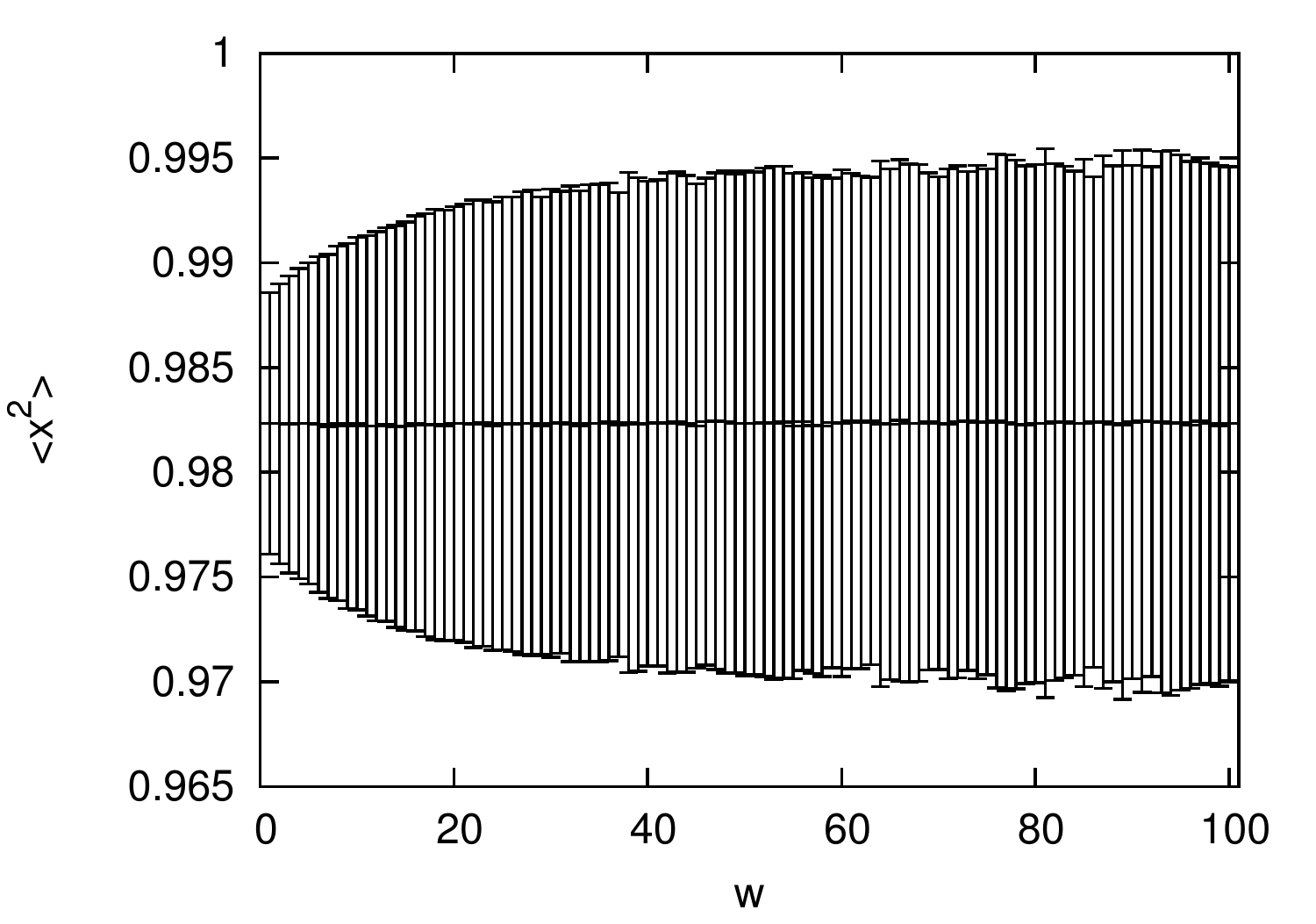}}
 \end{center}
 \caption{$\langle x^2\rangle$ and Jackknife error $\Delta_{w}$ with $50000$ samples. 
 }\label{fig:jackknife}
\end{figure}

\begin{figure}[htbp]
  \begin{center}
  \rotatebox{0}{
   \includegraphics[width=70mm]{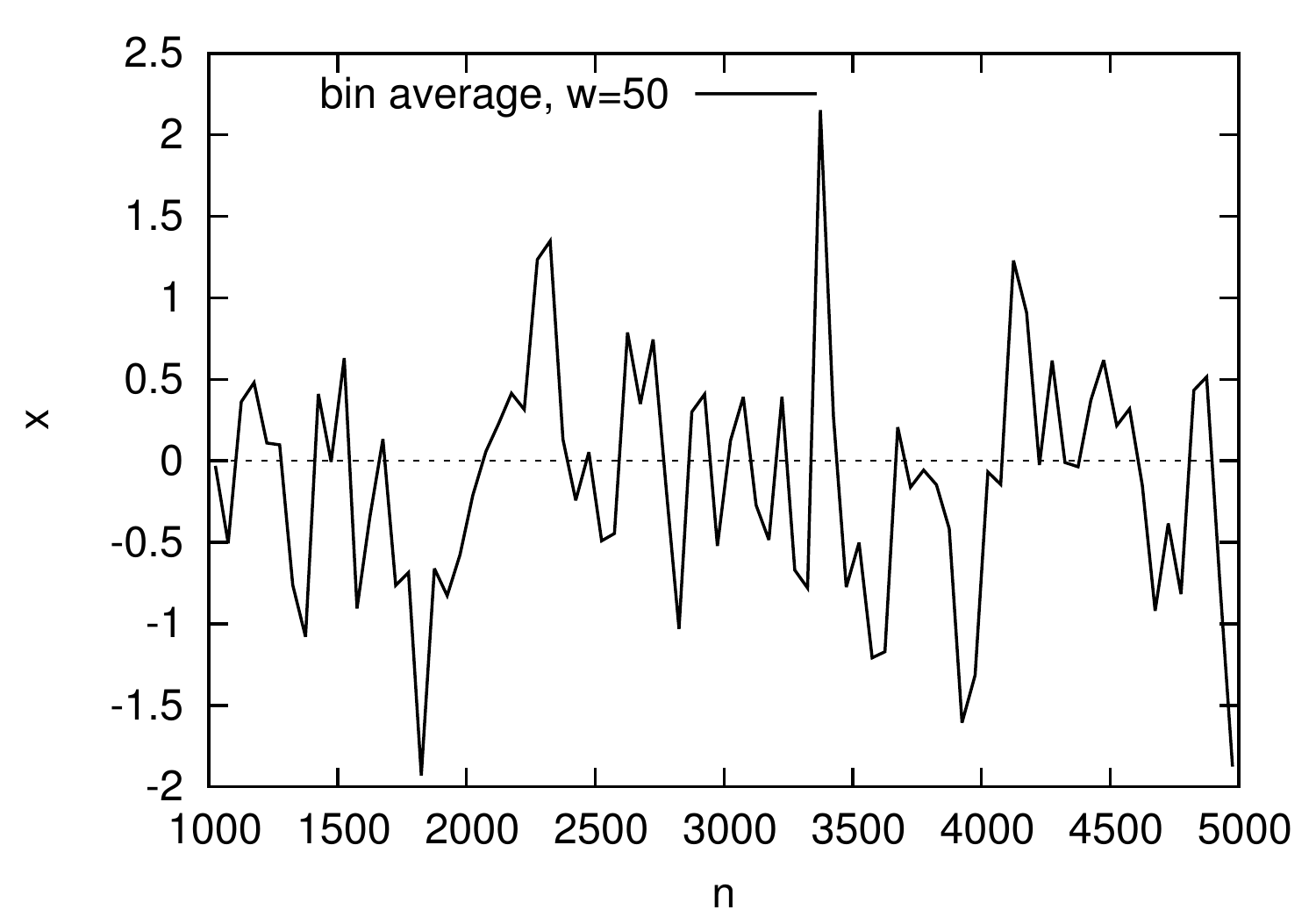}}
  \end{center}
  \caption{
  Bin-averaged version of Fig.~\ref{fig:autocorrelation}, with bigger window for $n$, with $w=50$. 
  }\label{fig:bin-average}
\end{figure}

In Fig.~\ref{fig:jackknife}, $\langle x^2\rangle$ and Jackknife error $\Delta_{w}$ are shown 
by using first $50000$ samples. 
We can see that $w_c=50$ is a reasonably safe choice; $w_c=20$ is already in the right ballpark. 
In Fig.~\ref{fig:bin-average}, bin-averaged values with $w=50$ are plotted. They do look independent. 
We obtained $\langle x^2\rangle=0.982\pm 0.012$, 
which agree reasonably well with the analytic answer, $\langle x^2\rangle=1$. 

\subsubsection{Tuning the simulation parameters}\label{sec:choice-of-parameters}
\hspace{0.51cm}
In order to run the simulation efficiently, we should tune parameters so that 
we can obtain {\it more independent samples with less cost}.\footnote{
In parallelized simulations, the notion of the {\it cost} is more nontrivial because time is money.  
Sometimes you may want to invest more electricity and machine resources to obtain the same result 
with shorter time.  
}

\begin{figure}[htbp]
  \begin{center}
  \rotatebox{0}{
   \includegraphics[width=70mm]{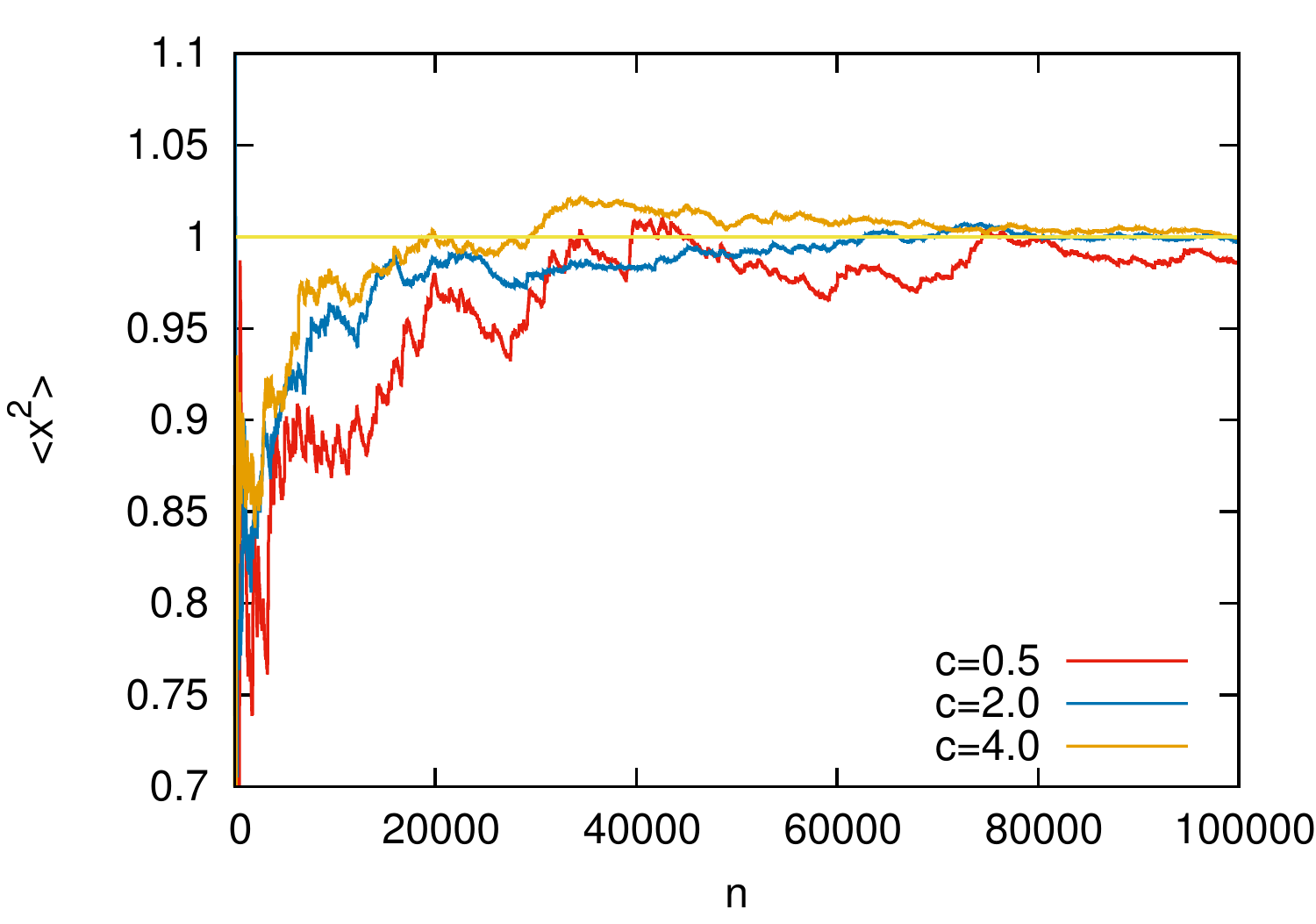}}
  \rotatebox{0}{
   \includegraphics[width=70mm]{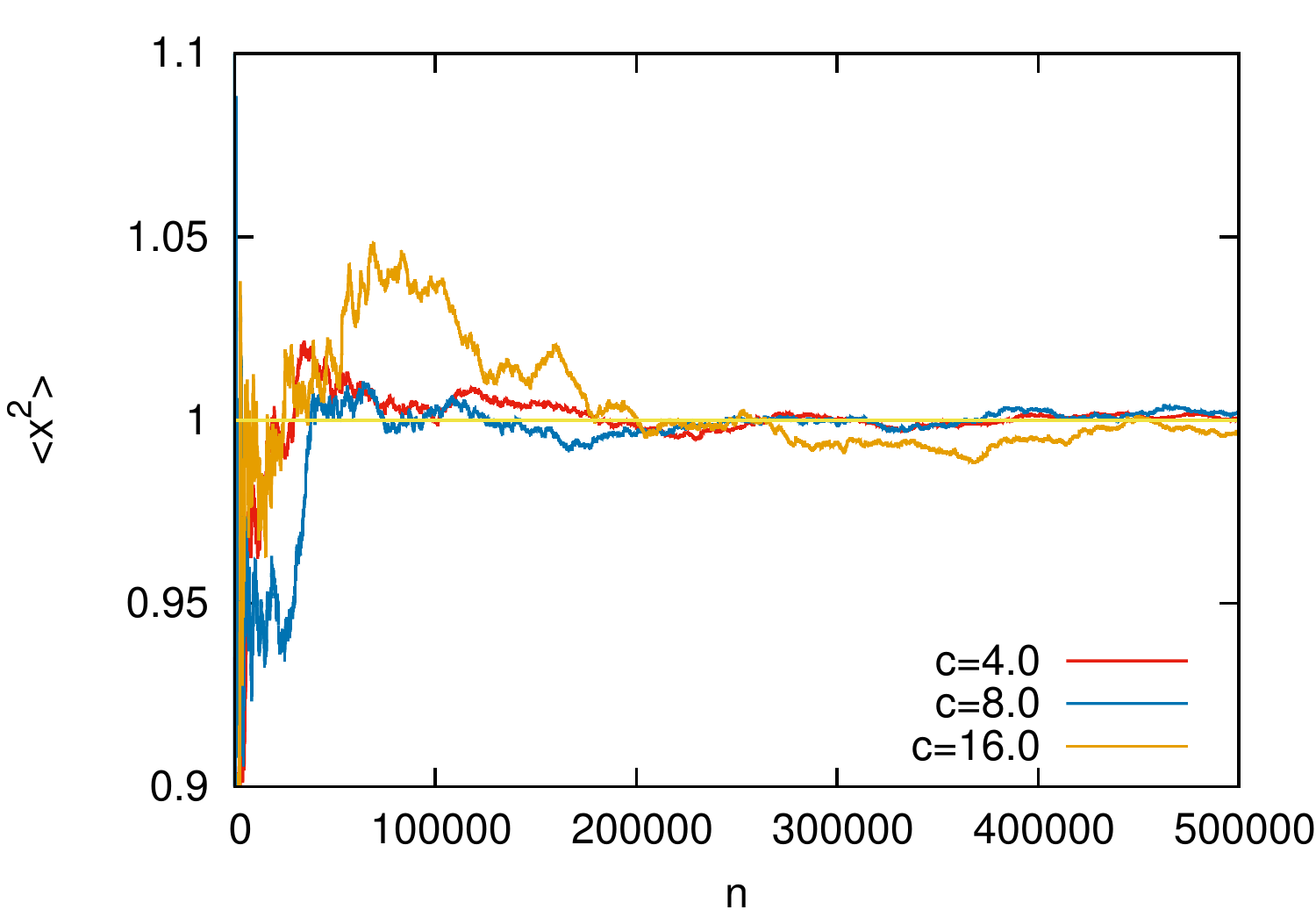}}
  \end{center}
  \caption{
  $\langle x^2\rangle = \frac{1}{n}\sum_{k=1}^n \left(x^{(k)}\right)^2$ for several different step sizes $c$. 
  }\label{fig:gaussian-metropolis-various-c}
\end{figure}

In the current example (Gaussian integral with uniform random number), when the step size $c$ is too large, 
unless $\Delta x\lesssim 1$ the configuration is rarely updated; the acceptance rate and the autocorrelation length
scale as $1/c$ and $c$, respectively. 
On the other hand, when $c$ is too small, the configurations are almost always updated, but only tiny amount.  
This is just a random walk with a step size $c$, and hence the average change after $n$ steps is $c\sqrt{n}$. 
Therefore the autocorrelation length should scale as $n\sim 1/c^2$. 
We expect the autocorrelation is minimized between these two regions.  
In Table~\ref{table-Metropolis-acceptance}, we have listed the acceptance rate for several values of $c$. 
We can see that the large-$c$ scaling ($c\times{\rm acceptance}\sim{\rm const}$) sets in at around $c=2$ $\sim$ $c=4$. 
In Fig.~\ref{fig:gaussian-metropolis-various-c} we have shown how $\langle x^2\rangle$ converges to 1 
as the number of configurations increases. We can actually see that $c=2$ and $c=4$ show faster convergence
compared to too small or too large $c$. 

\begin{table}
\begin{center}
\begin{tabular}{|c|c||c|}
\hline
step size $c$ & acceptance & $c$ $\times$ acceptance\\
\hline
\hline
0.5	& 0.9077& 0.454 \\
\hline
1.0 & 0.8098 & 0.810\\
\hline
2.0 & 0.6281 & 1.256\\
\hline
3.0 & 0.4864 & 1.459\\
\hline
4.0 & 0.3911 & 1.564\\
\hline
6.0 & 0.2643 & 1.586\\
\hline
8.0 & 0.1993 &  1.594\\
\hline
\end{tabular}
\caption{Step size vs acceptance rate, total 10000 samples.}\label{table-Metropolis-acceptance}
\end{center}
\end{table}

\subsection{How to calculate partition function}
\hspace{0.51cm}
In MCMC, we cannot directly calculate the partition function $Z$; we can only see the expectation values. 
Usually the partition function is merely a normalization factor of the path integral measure which does not affect the path integral, so we do not care. 
But sometimes it has interesting physical meanings; 
for example it can be used to test the conjectured dualities between supersymmetric theories. 

Suppose you want to calculate $Z=\int dx e^{-S(x)}$, where $S(x)$ is much more complicated than $S_0(x)=x^2/2$. 
By using MCMC, we can calculate the ratio between $Z$ and $Z_0=\int dx e^{-S_0(x)}=\sqrt{2\pi}$, 
\begin{eqnarray}
\frac{Z}{Z_0}
=
\frac{1}{Z_0}\int dx e^{-S_0}\cdot e^{S_0-S}
=
\left\langle
e^{S_0-S}
\right\rangle_0, 
\end{eqnarray}
where $\langle\ \cdot\ \rangle_0$ stands for the expectation value with respect to the action $S_0$. 
Because we know $Z_0$ analytically, we can determine $Z$.
\subsubsection{Overlapping problem and its cure}\label{sec:overlapping-problem}
\hspace{0.51cm}
The method described above can always work {\it in principle}. 
In practice, however, it fails when the probability distributions $\rho(x)=\frac{e^{-S(x)}}{Z}$ and $\rho_0(x)=\frac{e^{-S_0(x)}}{Z_0}$
do not have sufficiently large overlap. As a simple example, 
let us consider $S=(x-c)^2/2$ (though you can analytically handle it!). Then $\rho(x)$ and $\rho_0(x)$ have peaks around $x=c$ and $x=0$, 
respectively. When $c$ is very large, say $c=100$, the value of $e^{S_0-S}$ appearing in the simulation is almost always 
an extremely small number $\sim e^{-5000}$, and once every $e^{+5000}$ steps or so we get an extremely large number $\sim e^{+5000}$. 
And they average to $\frac{Z}{Z_0}=1$. 
Clearly, we cannot get an accurate number if we truncate the sum at a realistic number of configurations.  
It happens because of the absence of the overlap of $\rho(x)$ and $\rho_0(x)$, or equivalently, 
because important configurations in two different theories are different; 
hence the `operator' $e^{S_0-S}$ behaves badly at the tail of $\rho_0(x)$. 
This is so-called {\it overlapping problem}.\footnote{ 
In SYM, the overlapping problem can appear combined with the {\it sign problem}; we will revisit this point in Sec.~\ref{sec:sign-problem}.} 

In the current situation, the overlapping problem can easily be solved as follows. 
Let us introduce a series of actions $S_0$, $S_1$, $S_2$, ..., $S_k=S$. 
We choose them so that $S_i$ and $S_{i+1}$ are sufficiently close
and the ratio $\frac{Z_{i+1}}{Z_i}$, where $Z_i=\int dx e^{-S_i(x)}$, can be calculated without the overlapping problem. 
For example we can take $S_i=\frac{1}{2}\left(x-\frac{i}{k}c\right)^2$ with $\frac{c}{k}\sim 1$.  
Then we can obtain $Z=Z_k$ by calculating $\frac{Z_1}{Z_0}$, $\frac{Z_2}{Z_1}$, $\cdots$, $\frac{Z_k}{Z_{k-1}}$.  
The same method can be applied to any complicated $S(x)$, as long as $e^{-S(x)}$ is real and positive. 

This rather primitive method is actually powerful; for example the partition function of ABJM theory \cite{Aharony:2008ug} 
at finite coupling and finite $N$ has been calculated accurately by using this method \cite{Hanada:2012si}. 
\subsection{Common mistakes}
\hspace{0.51cm}
Let us see some common mistakes below. 
\subsubsection{Don't change step size during the run}\label{sec:don't-change-parameters}
\begin{figure}[htbp]
  \begin{center}
   \includegraphics[width=60mm]{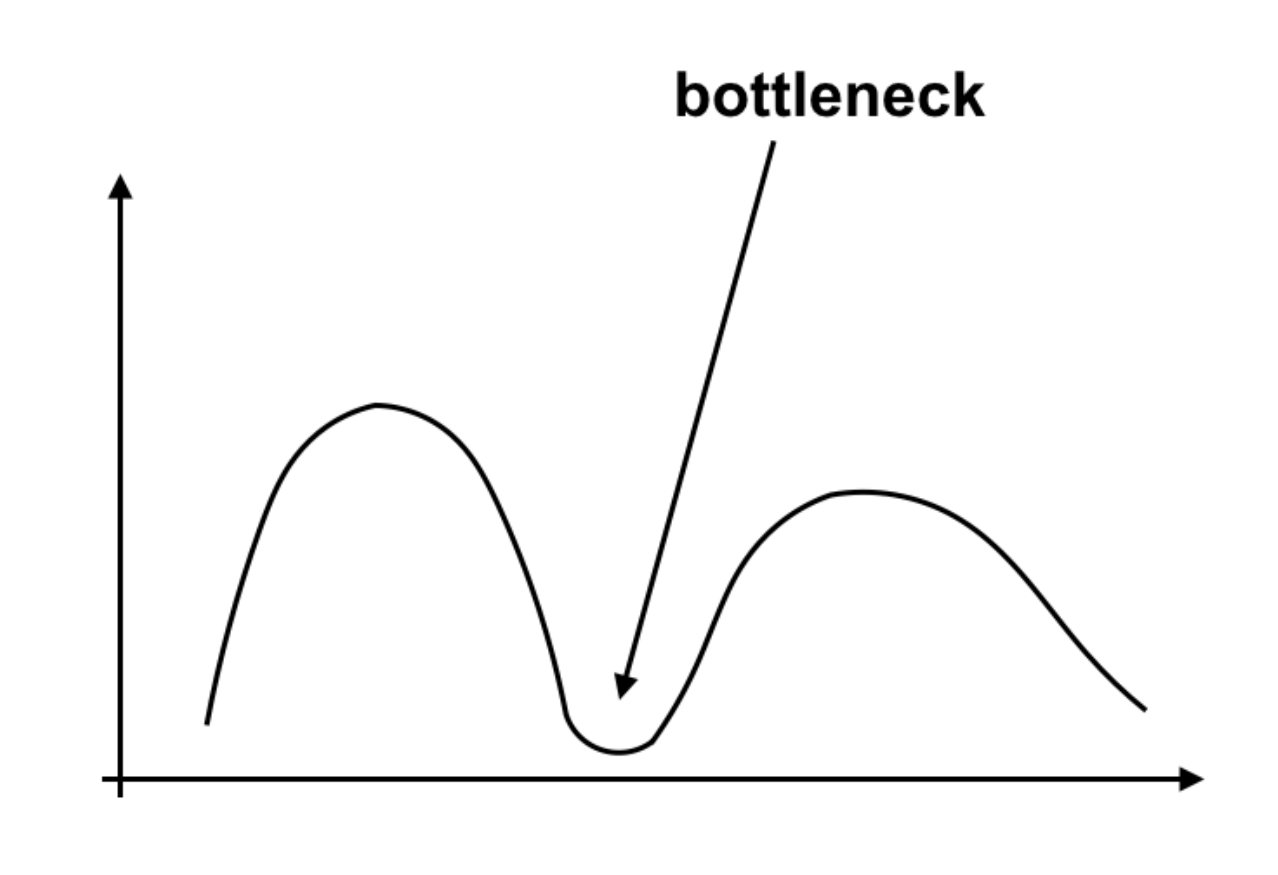}
  \end{center}
  \caption{If the probability distribution has a bottleneck, the acceptance rate goes down there. 
  }\label{fig:bottle-neck}
\end{figure}

\hspace{0.51cm}
Imagine the probability distribution you want to study has a bottleneck 
like in Fig.~\ref{fig:bottle-neck}. For example if $S(x)=-\log\left(e^{-\frac{x^2}{2}}+e^{-\frac{(x-100)^2}{2}}\right)$ then $e^{-S(x)}$ is strongly suppressed between two peaks at $x=0$ and $x=100$. 
By using a small step size $c\sim 1$ you can sample one of the peaks efficiently, but then the other peak cannot be sampled. 
Then in order to go across the bottleneck you would be tempted to change the step size $c$
when you come close to the bottleneck. You would want to make the step size larger so that 
you can jump over the bottle neck,  
or you would want to make the step size smaller so that you can slowly penetrate into the bottle neck.
But if you do so, you obtain a wrong result, because the transition probability 
can depend on the past history.  
{\it You must not change the step size during the simulation}. 

\begin{figure}[htbp]
  \begin{center}
  \rotatebox{0}{
   \includegraphics[width=70mm]{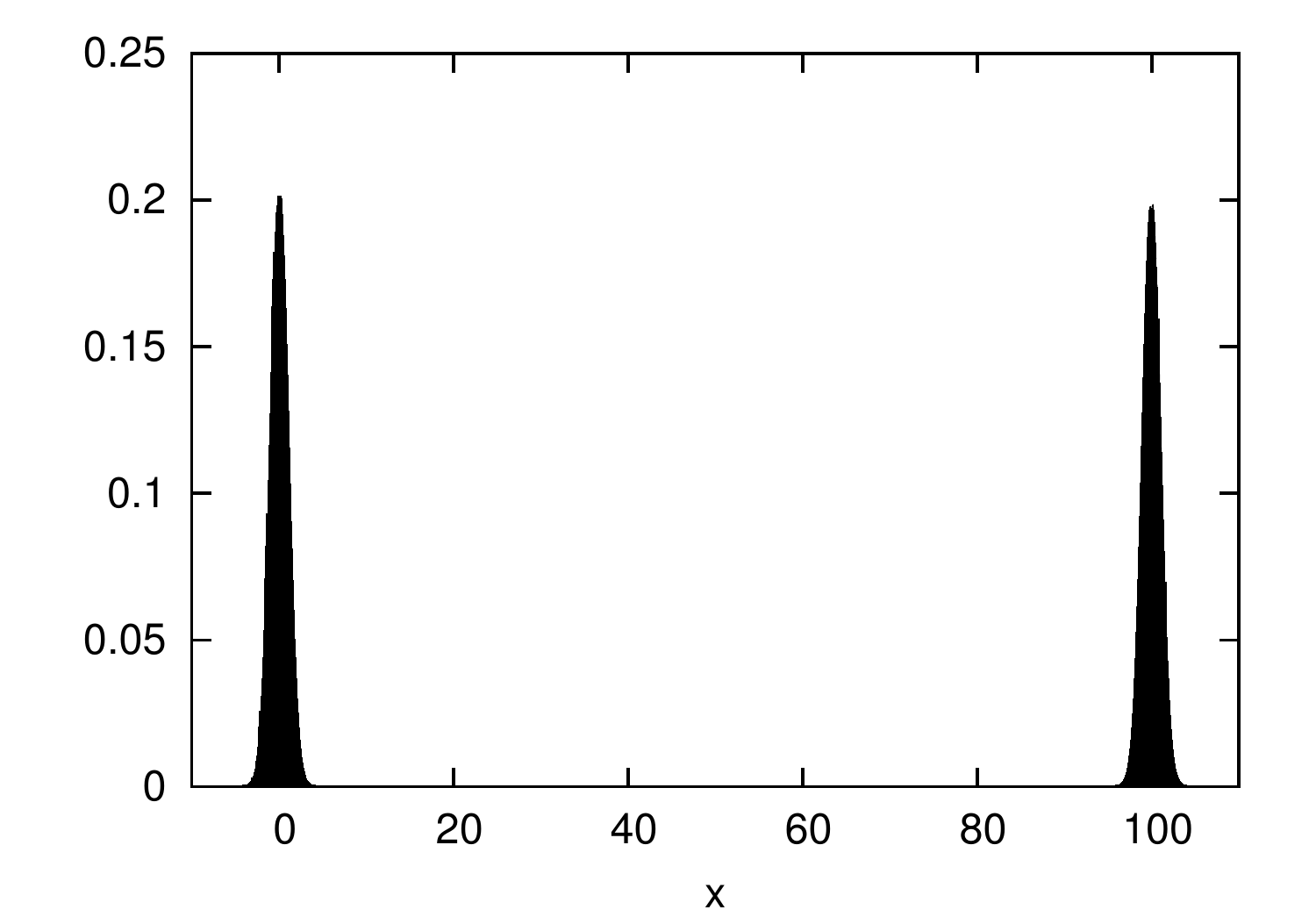}}
  \end{center}
  \caption{A histogram for $S(x)=-\log\left(e^{-\frac{x^2}{2}}+e^{-\frac{(x-100)^2}{2}}\right)$
  with Metropolis, step size $c=1$ for odd steps and $c=100$ for even steps, $10^7$ samples.
  The solid curve (which is actually invisible because it agrees with the histogram too precisely...) is the exact answer, $\frac{e^{-\frac{x^2}{2}}+e^{-\frac{(x-100)^2}{2}}}{2\sqrt{2}}$. 
  }\label{fig:double-peak}
\end{figure}

But it does not mean that you cannot use multiple fixed step sizes; 
it is allowed to change the step size if the conditions listed in Sec.~\ref{sec:MCMC} are not violated. 
For example we can take $c=1$ for odd steps and $c=100$ for even steps; 
see Fig.~\ref{fig:double-peak}. 

Or we can throw a dice, namely randomly choose step size $c=1,2,3,4,5,6$ with probability $1/6$. 
As long as the conditions listed in Sec.~\ref{sec:MCMC}, in particular the detailed balance, are not violated, you can do whatever you want. 

Similar temptation of evil is common in muilti-variable case. For example in the lattice gauge theory simulation 
it often happens that the acceptance is extremely low until the system thermalizes. 
Then we can use smaller step size just to make the system thermalize, 
and then start actual data-taking with a larger step size.\footnote{
Another common strategy to reach the thermalization is to turn off the Metropolis test.
} 
Or it occasionally happens that the simulation is trapped at a rare configuration so that the acceptance rate becomes almost zero. 
In such case, it would be useful to use multiple step sizes (the ordinary and very small).

\subsubsection{Don't mix independent simulations with different step sizes}\label{caution:step-size}
\hspace{0.51cm}
This is similar to Sec.~\ref{sec:don't-change-parameters}: 
{\it when you have several independent runs with different step sizes, 
you must not mix them to evaluate the expectation value}, 
unless you pay extra cares for the error analysis.   
Although each stream are guaranteed to converge to the same distribution, 
if you truncate them at finite number of configurations each stream contains different uncontrollable systematic error.
%

Note however that, if you can estimate the autocorrelation time of each stream reliably
(for that each stream has to be sufficiently long), 
you can mix different streams with proper weights, with a careful error analysis.

\subsubsection{Make sure that random numbers are really random}
\begin{figure}[htbp]
  \begin{center}
  \rotatebox{0}{
   \includegraphics[width=70mm]{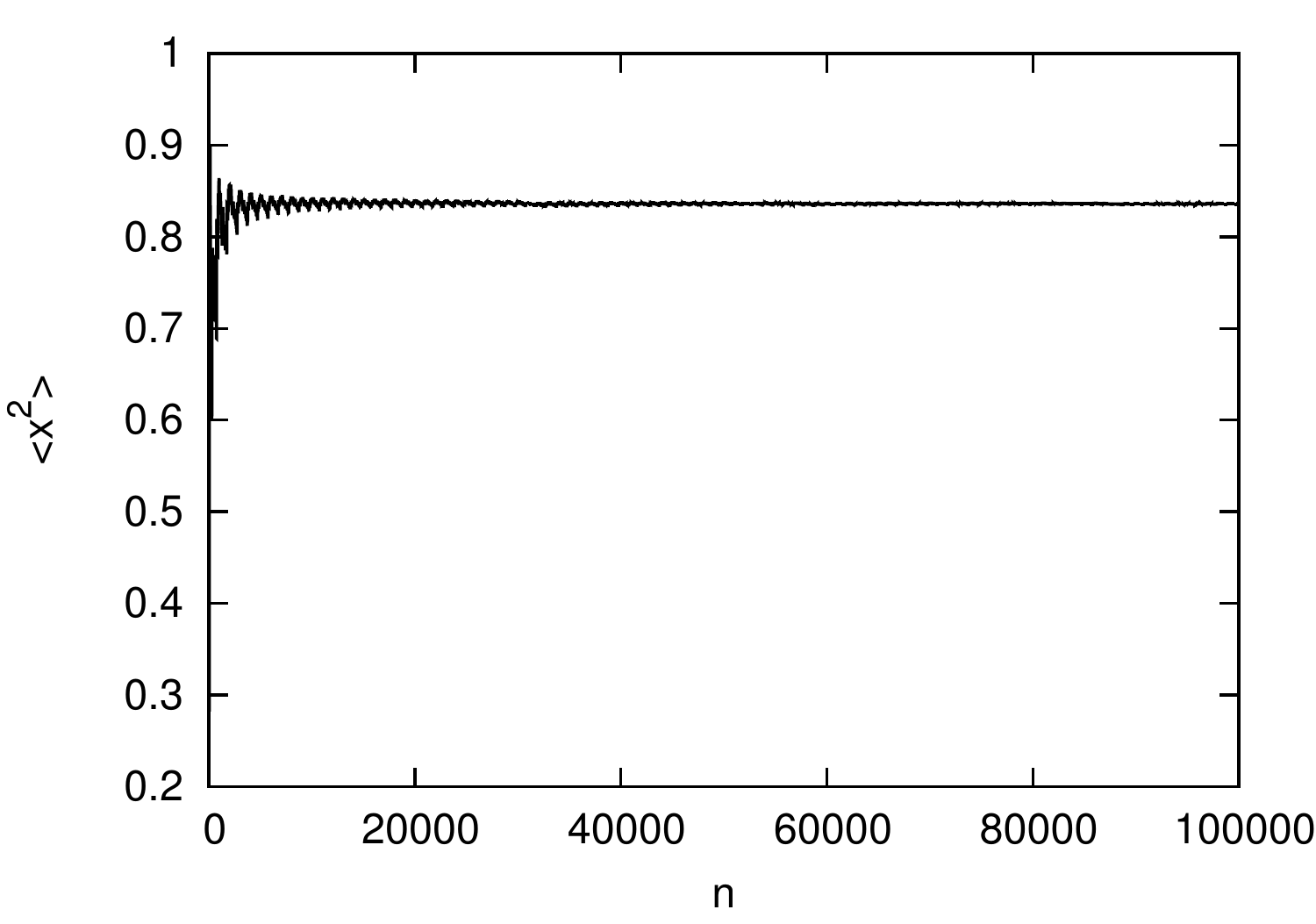}}
  \end{center}
  \caption{Gaussian integral with Metropolis, $c=1$, with {\it non-random} numbers; 
  we chose the same random number sequence every 1000 steps. 
  $\langle x^2\rangle = \frac{1}{n}\sum_{k=1}^n \left(x^{(k)}\right)^2$ converges to a wrong number, 
  which is different from 1. 
  }\label{fig:reset-1000traj}
\end{figure}

\hspace{0.51cm}
In actual simulations, random numbers are not really random, they are just pseudo-random. 
But we have to make sure that they are sufficiently random. 
In Fig.\ref{fig:reset-1000traj}, we used the same sequence of random numbers repeatedly 
every 1000 steps. The answer is clearly wrong. 

This mistake is very common;\footnote{
Yes, I did.
} when one simulates a large system, it will take days or months, 
so one has to split the simulation to small number of steps. Then when one submits a new job 
by mistake one would use the same `random numbers' again, for example by reseting the seed 
of random numbers to the same number.  

\subsubsection{Remark on the use of Mathematica for larger scale simulations}
\hspace{0.51cm}
I saw several people tried to use Mathematica for MCMC of systems of moderate size
(gauge theories consisting of $O(10^3)$ --- $O(10^4)$ variables) and failed to run the code with acceptable speed.
Probably the problem was that unless one understands Mathematica well
one can unintentionally use nice features such as symbolic calculations which are not needed in MCMC.
The same remark could apply to other advanced softwares as well.

The simplest solution to this problem is to keep advanced softwares for advanced tasks, 
and avoid using them for such simple things like MCMC.  
Unless you use a rare special function or something, it is unlikely that you need anything more than C or Fortran. 

But by using those softwares you may be able to save the time for coding. 
As long as the system size you want to study is small, you don't have to to worry; 
10 seconds and 10 minutes are not that different and you will spend more time for the coding anyways. 
When you want to do heavier calculations, make sure to understand the software well and avoid using unnecessary features.\footnote{
Wolfram research can easily solve this issue, I suppose. 
Or perhaps it can easily be avoided by using existing features. If anybody knows how to solve this problem, please let me know. 
Given that Mathematica is extremely popular among physicists, if Mathematica can handle MCMC in physics
it will certainly lower the entrance threshold. 
} 

Mathematica can be a useful tool to generate a C/Fortran code, especially with the HMC algorithm 
Sec.~\ref{sec:HMC}, by utilizing the symbolic calculations. 
I personally think such direction is the right use of Mathematica in the context of MCMC. 

\subsection{Sign problem}\label{sec:sign-problem-one-variable}
\hspace{0.51cm}
So far we have assumed $e^{-S(x)}\ge 0$.
This was necessary because we interpreted $e^{-S(x)}$ as a `probability'.
But in physics we often encounter $e^{-S(x)}< 0$, or sometimes $e^{-S(x)}$ can be complex. 
Then a naive MCMC approach does not work. This is infamous {\it sign problem} (or {\it phase problem}, when $e^{-S(x)}$ is complex). 
Although no generic solution of the sign problem is known, 
there are various theory-specific solutions. 
We will come back to this point in Sec.~\ref{sec:phase_quench_SYM}.

\subsection{What else do we need for lattice gauge theory simulations?}
\hspace{0.51cm}

The advantage of the Metropolis algorithm is clear: it is simple. 
It is extremely simple and applicable to any theory, 
as long as the `sign problem' does not exist. 
When it works, just use it.
For example, for simple matrix model calculations like the one in \cite{Azeyanagi:2007su},\footnote{
In that paper, we studied the symmetry breaking in Twisted Eguchi-Kawai model at large $N$. 
In 1980's people used the best computers available and studied $N\lesssim 16$.   
They did not observe a symmetry breaking. 
In 2006, I wrote a Metropolis code spending an hour or so, 
and studied $25\lesssim N\lesssim 100$ with my laptop.  
Within a few hours I could see a clear signature of the symmetry breaking. 
In order to understand the detail of the symmetry breaking pattern we had to study many parameters, 
so we ran the same code on a cluster machine. 
One of my collaborators was serious enough to write a more sophisticated code to go to much larger $N$, with which we could study $N\gtrsim 100$. 
Note that it is a story from 2006 to 2007; now you can do much better job with Metropolis and your laptop. 
} 
you don't need anything more than Metropolis and your laptop. 

But our budget is limited and we cannot live forever. So sometimes we have to reduce the cost 
and make simulations faster. 
We should use better algorithms, better lattice actions and better observables,  
which are `better' in the following sense:
\begin{itemize}
\item
The autocorrelation length is shorter. 

\item
Easier to parallelize. In lattice gauge theory,  it typically means 
that we should utilize the sparseness of the Dirac operator. 

\item
Find good observables and good measurement methods which are easier to calculate, 
have less statistical fluctuations, and/or show faster convergence to the continuum limit.

\end{itemize}

For quantum field theories, 
especially when the fermions are involved, HMC is effective. 
RHMC is a variant of HMC which is applicable to SYM.

\section{Integration of multiple variables and bosonic QFT}\label{sec:multi-variables}
\hspace{0.51cm}
Once a regularization is given, the path-integral is merely an integral with multiple variables. 
Hence let us start with a simple case of a matrix integral, then proceed to QFT. 

\subsection{Metropolis for multiple variables}\label{sec:Metropolis-multi-variable}
\hspace{0.51cm}
Generalization of the Metropolis algorithm (Sec.~\ref{sec:Metropolis-Gaussian})
to multiple variables $(x_1,x_2,\cdots,x_p)$ is straightforward. 
For example, we can do as follows:
\begin{enumerate}
\item
For all $i=1,2,\cdots,p$, randomly choose $\Delta x_i\in [-c_i,+c_i]$, and shift $x_i^{(k)}$ as $x_i^{(k)}\to x'_i\equiv x_i^{(k)}+\Delta x_i$. Note that the step size $c_i$ can be different for different $x_i$. 

\item
Metropolis test: Generate a uniform random number $r$ between 0 and 1. 
If $r<e^{S[x^{(k)}]-S[x']}$, 
$\{x^{(k+1)}\}=\{x'\}$, i.e. the new configuration is `accepted.' 
Otherwise 
$\{x^{(k+1)}\}=\{x^{(k)}\}$,  i.e. the new configuration is `rejected.' 

\item
Repeat the same for $k+1,k+2,\cdots$. 
\end{enumerate}

One can also do as follows:
\begin{enumerate}
\item
Randomly choose $\Delta x_1\in [-c_1,+c_1]$, and shift $x_1^{(k)}$ as $x_1^{(k)}\to x'_1\equiv x_1^{(k)}+\Delta x_1$. 

\item
Metropolis test: Generate a uniform random number $r$ between 0 and 1. 
If $r<e^{S[x^{(k)}]-S[x']}$, 
$x_1^{(k+1)}=x'_1$, i.e. the new value is `accepted.' 
Otherwise 
$x_1^{(k+1)}=x_1^{(k)}$,  i.e. the new configuration is `rejected.' 
For other values of $i$ we don't do anything, namely $x_{i}^{(k+1)}=x_{i}^{(k)}$ for $i=2,3,\cdots,p$. 

\item
Repeat the same for $i=2,3,\cdots,p$. 

\item
Repeat the same for $k+1,k+2,\cdots$.

\end{enumerate}

\subsubsection{How it works}
\hspace{0.51cm}

Let us consider a one matrix model, 
\begin{eqnarray}
S[\phi]
=
N{\rm Tr}\left(
\frac{1}{2}\phi^2
+
V(\phi)
\right),  \label{action-one-matrix}
\end{eqnarray}
where $\phi$ is an $N\times N$ Hermitian matrix, $\phi_{ji}=\phi_{ij}^\ast$. 
The potential $V(\phi)$ can be anything as long as the partition function is convergent; say $V(\phi)=\phi^4$. 

The code has exactly the same structure as the sample code in Sec.~\ref{sec:Metropolis-Gaussian}; 
we should calculate $S[\phi]$ instead of the Gaussian weight, 
and instead of $x$ we can shift $\phi$ by using $N^2$ real random numbers.

As $N$ gets larger, more and more portion of the integral region becomes unimportant. 
Therefore, if we vary all the components simultaneously, $\Delta S$ is typically large and 
the acceptance rate is very small, unless we take the step size to be small. 
To avoid it, we can vary one component at each time. 
Note that, when only $\phi_{ij}$ and $\phi_{ji}=\phi_{ij}^\ast$ are varied, 
one should save the computational cost by calculating $\phi_{ij}$-dependent part  
instead of $S[\phi]$ itself; the latter costs $O(N^3)$, though the former costs only $O(N^2)$. 
%
%
%

\subsection{Hybrid Monte Carlo (HMC) Algorithm }\label{sec:HMC}
\hspace{0.51cm}
The important configurations are like bottom of a valley; the altitude is the value of the action. 
This is a valley in the phase space, whose dimension is very large. 
So if the configuration is literally randomly varied, like in the Metropolis algorithm, 
the action almost always increases a lot. 
Hence with the Metropolis algorithm the acceptance rate is small unless the step size is extremely small, 
and it causes rather long autocorrelation length. 
The Hybrid Monte Carlo (HMC) algorithm \cite{Duane:1987de} avoids the problem of a long autocorrelation
by effectively crawling along the bottom of the valley;    
this is a `hybrid' of molecular dynamical method and Metropolis algorithm. 

In HMC algorithm, sets of configurations 
$\{x^{(k)}\}$ $(k=0,1,2,\cdots)$ are generated in the following manner. 
Firstly, $\{x^{(0)}\}$ can be arbitrary. 
Once $\{x^{(k)}\}$ is obtained, $\{x^{(k+1)}\}$ is obtained as follows. 
\begin{enumerate}
\item
Randomly generate auxiliary momenta $P_i^{(k)}$, which are `conjugate' to $x_i^{(k)}$, 
with probabilities $\frac{1}{\sqrt{2\pi}}e^{-(P_i^{(k)})^2/2}$. 
To generate Gaussian random numbers, the Box-Muller algorithm is convenient; see Appendix~\ref{sec:Box-Muller}. 

\item
Calculate the `Hamiltonian' $H_i=S[x^{(k)}]+\frac{1}{2}\sum_i (P_i^{(k)})^2$. 

\item
Then we consider `time evolution' along an auxiliary time $\tau$ (which is not the Euclidean time!). 
We set the initial condition to be 
$x^{(k)}(\tau=0)=x^{(k)}$ and  
$P^{(k)}(\tau=0)=P^{(k)}$, 
and use the leap frog method (see below) to calculate 
$x^{(k)}(\tau_{f})$ and 
$P^{(k)}(\tau_{f})$, 
where $\tau_{f}$ is related to the input parameters $\Delta\tau$ and $N_\tau$ by 
$\tau_{f}=N_\tau\Delta\tau$. 

This process is called `molecular evolution.'

\item
Calculate $H_f=S[x^{(k)}(\tau_{f})]+\frac{1}{2}\sum_i (P_i^{(k)}(\tau_{f}))^2$. 

\item
Metropolis test: Generate a uniform random number $r$ between 0 and 1. 
If $r<e^{H_i-H_f}$, 
$x^{(k+1)}=x^{(k)}(\tau_{f})$, i.e. the new configuration is `accepted.' 
Otherwise 
$x^{(k+1)}=x^{(k)}$, i.e. the new configuration is `rejected.' 

\end{enumerate}

At first sight it is a rather complicated algorithm. Why do we introduce such auxiliary dynamical system? 
The key is the `energy conservation'. 

Just like Metropolis, the change of the configuration is random due to the randomness of the choice of auxiliary momenta $P_i$. 
If we just compared the initial and final values of the action, the change would be equally large. 
However in HMC the change of the auxiliary Hamiltonian matters in the Metropolis test. 
(Please accept this fact for the moment, in the next paragraph we will explain the reason.)
If we keep $N_\tau\Delta\tau$ fixed and send $N_\tau$ to infinity, then the Hamiltonian is exactly conserved 
and new configurations are always accepted. By taking $N_\tau\Delta\tau$ to be large, new configurations can be substantially 
different from the old ones. (Of course, calculation cost increase with $N_\tau$. 
So we have to find a sweet spot, 
with moderately large $N_\tau$ and moderately small $\Delta\tau$.)

To check the detailed balance condition $e^{-S[x]}T[\{x\}\to \{x'\}]=e^{-S[x']}T[\{x'\}\to \{x\}]$, note that the leap-frog method is designed so that 
the molecular evolution is reversible; namely, if we start with $x^{(k)}(\tau_{f})$ and 
$-P^{(k)}(\tau_{f})$, the final configuration is $x^{(k)}(\tau=0)$ and $-P^{(k)}(\tau=0)$. 
Hence, if $\{x,p\}$ evolves to $\{x',p'\}$, then (by assuming $H-H'<0$ without loss of generality)
$e^{-S[x]}T[\{x\}\to \{x'\}]\propto e^{-S[x]}e^{-p^2/2}$, 
while $e^{-S[x']}T[\{x'\}\to \{x\}]\propto e^{-S[x']}e^{-p'^2/2}e^{-S[x]-p^2/2+S[x']+p'^2/2}=e^{-S[x]}e^{-p^2/2}$, with the same proportionality factor. 

In fact the HMC algorithm works even when we take different $\tau$ for each $x_i$; 
see Sec.~\ref{sec:multi-step-size}.  
We can use different $\tau$ on depending the field, momentum of the mode, etc.  
The HMC algorithm is powerful especially when we have to deal with fermions, as we will explain later.

\subsubsection{Leap frog method}

The leap frog method is a clever way to discretize the continuum Hamiltonian equation  
keeping the reversibility, which is crucial for assuring the detailed balance condition. 

\begin{figure}[htbp]
  \begin{center}
   \includegraphics[width=80mm]{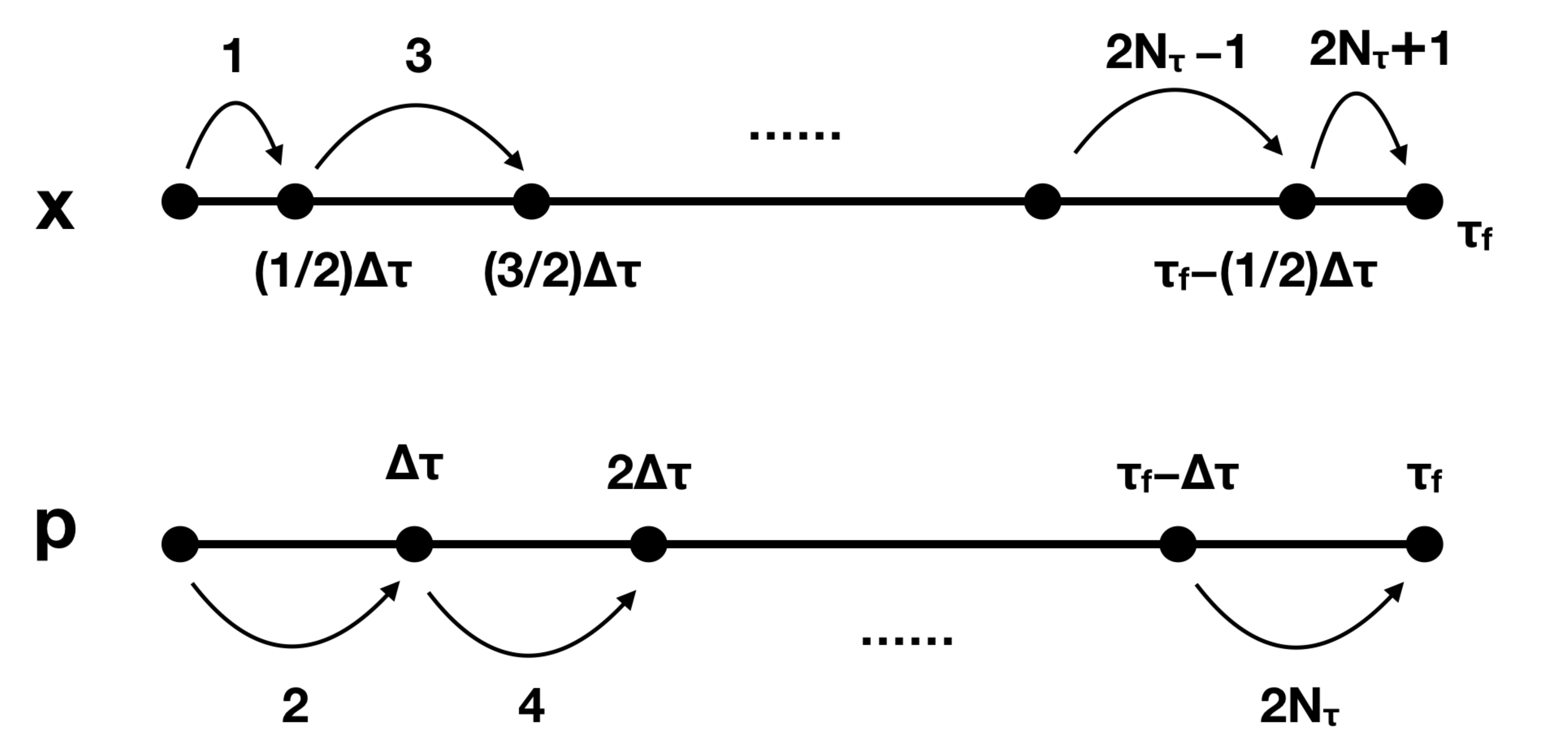}
  \end{center}
  \caption{Leap-frog method. 
  }\label{fig:leap-frog}
\end{figure}

The continuum Hamiltonian equation is given by 
\begin{eqnarray}
\frac{dp_i}{d\tau}
=
-\frac{\partial H}{\partial x_i} 
=
-\frac{\partial S}{\partial x_i}, 
\qquad
\frac{d x_i}{d\tau}
=
\frac{\partial H}{\partial p_i} 
=
p_i.  
\end{eqnarray}
The leap frog method goes as follows\footnote{
Be careful about a factor of $1/2$ and ordering of the operations. 
These are crucial for the reversibility of the molecular evolution
and the detailed balance condition. 
}:
\begin{enumerate}
\item
$x_i(\Delta\tau/2)
=x_i(0)+p_i(0)\cdot\frac{\Delta\tau}{2}$ 
(Step 1 in Fig.~\ref{fig:leap-frog}). 
\item
For $n=1,2,N_\tau-1$, repeat it: 

$p_i(n\Delta\tau)
=p_i((n-1)\Delta\tau)-\frac{\partial S}{\partial x_i}((n-1/2)\Delta\tau)\cdot\Delta\tau$ (Step $2,4,\cdots,2N_\tau-2$ in Fig.~\ref{fig:leap-frog}), 

then

$x_i((n+1/2)\Delta\tau)
=x_i((n-1/2)\Delta\tau)+p_i(n\Delta\tau)\cdot\Delta\tau$  (Step $3,\cdots,2N_\tau-1$ in Fig.~\ref{fig:leap-frog}).

\item
Finally, 

$p_i(N_\tau\Delta\tau)
=p_i((N_\tau-1)\Delta\tau)-\frac{\partial S}{\partial x_i}((N_\tau-1/2)\Delta\tau)\cdot\Delta\tau$ (Step $2N_\tau$ in Fig.~\ref{fig:leap-frog}), 
 
then

$x_i(N_\tau\Delta\tau)
=x_i((N_\tau-1/2)\Delta\tau)+p_i(N_\tau\Delta\tau)\cdot\frac{\Delta\tau}{2}$ (Step $2N_\tau+1$ in Fig.~\ref{fig:leap-frog}). 

\end{enumerate}

\subsubsection{How it works, 1 --- Gaussian Integral}
\hspace{0.51cm}
As the simplest example, let us go back to the one-variable case and start with the Gaussian integral again.\footnote{
This is an extremely stupid example, given that we need the Gaussian random number 
for the HMC algorithm! But I believe it is still instructive. 
} Here is a sample code:

\begin{verbatim}
#include <iostream>
#include <cmath>
#include<fstream>
const int niter=10000;
const int ntau=40;
const double dtau=1e0;
/******************************************************************/
/*** Gaussian Random Number Generator with Box Muller Algorithm ***/
/******************************************************************/
int BoxMuller(double& p, double& q){
  
  double pi;
  double r,s;
  pi=2e0*asin(1e0);
  //uniform random numbers between 0 and 1
  r = (double)rand()/RAND_MAX;
  s = (double)rand()/RAND_MAX;
  //Gaussian random numbers, 
  //with weights proportional to e^{-p^2/2} and e^{-q^2/2}
  p=sqrt(-2e0*log(r))*sin(2e0*pi*s);
  q=sqrt(-2e0*log(r))*cos(2e0*pi*s);
  
  return 0;
}
/*********************************/
/*** Calculation of the action ***/
/*********************************/
// When you change the action, you should also change dH/dx,
// specified in "calc_delh". 
double calc_action(const double x){
  
  double action=0.5e0*x*x;
  
  return action;
}
/**************************************/
/*** Calculation of the Hamiltonian ***/
/**************************************/
double calc_hamiltonian(const double x,const double p){
  
  double ham;
    
  ham=calc_action(x);
  
  ham=ham+0.5e0*p*p;
  
  return ham;
}
/****************************/
/*** Calculation of dH/Dx ***/
/****************************/
// Derivative of the Hamiltonian with respect to x, 
// which is equivalent to the derivative of the action.
// When you change "calc_action", you have to change this part as well. 
double calc_delh(const double x){
  
  double delh=x;
  
  return delh;
}
/***************************/
/*** Molecular evolution ***/
/***************************/
int Molecular_Dynamics(double& x,double& ham_init,double& ham_fin){

  double p;
  double delh;
  double r1,r2;
  
  BoxMuller(r1,r2);
  p=r1;
  
  //*** calculate Hamiltonian ***
  ham_init=calc_hamiltonian(x,p);
  //*** first step of leap frog ***
  x=x+p*0.5e0*dtau;
  //*** 2nd, ..., Ntau-th steps ***
  for(int step=1; step!=ntau; step++){    
    delh=calc_delh(x);
    p=p-delh*dtau;
    x=x+p*dtau;
  }
  //*** last step of leap frog ***
  delh=calc_delh(x);
  p=p-delh*dtau;
  x=x+p*0.5e0*dtau;
  //*** calculate Hamiltonian again ***
  ham_fin=calc_hamiltonian(x,p);
  
  return 0;
}

int main()
{
  double x;
  double backup_x;
  double ham_init,ham_fin,metropolis,sum_xx;
  
  srand((unsigned)time(NULL));
  /*********************************/
  /* Set the initial configuration */
  /*********************************/
  x=0e0;
  /*****************/
  /*** Main part ***/
  /*****************/
  std::ofstream outputfile("output.txt");
  int naccept=0;//counter for the number of acceptance
  sum_xx=0e0;//sum of x^2, useed for <x^2>
  
  for(int iter=0; iter!=niter; iter++){
    
    backup_x=x;
    
    Molecular_Dynamics(x,ham_init,ham_fin);
    metropolis = (double)rand()/RAND_MAX;
    if(exp(ham_init-ham_fin) > metropolis){
      //accept
      naccept=naccept+1;
    }else{
      //reject
      x=backup_x;
	    
    }
    
    /*******************/
    /*** data output ***/
    /*******************/
    sum_xx=sum_xx+x*x;

    // output x, <x^2>, acceptance
    
    std::cout << x << ' ' << sum_xx/((double)(iter+1)) <<  ' ' << 
    ((double)naccept)/((double)iter+1) << std::endl;
     
    outputfile << x << ' ' << sum_xx/((double)(iter+1)) <<  ' ' << 
    ((double)naccept)/((double)iter+1) << std::endl;
    
    
  }
    
  outputfile.close();
  
  return 0;
}
\end{verbatim}

At the beginning of the code, a few parameters are set.
\textbf{niter} is the number of samples we will collect; \textbf{ntau} is $N_\tau$; and \textbf{dtau} is $\Delta\tau$. 

Then several routines/functions are defined: 
\begin{itemize}
\item
\textbf{BoxMuller} generates Gaussian random numbers
by using the Box-Muller algorithm. 
We have to be careful about the normalization of the Gaussian here. It will be a kind of confusing 
when you go to complex variables; see the case of matrix integral in Sec.~\ref{sec:HMC-how-it-works-matrix}. 

\item
\textbf{calc}$\underbar{\ }$\textbf{action} calculates the action $S[x]$. In this case it is just $S[x]=\frac{x^2}{2}$. 
It is called in \textbf{calc}$\underbar{\ }$\textbf{hamiltonian}. 

\item
\textbf{calc}$\underbar{\ }$\textbf{hamiltonian} adds $\frac{p^2}{2}$ to the action and returns the Hamiltonian. 
It is called in \textbf{Molecular}$\underbar{\ }$\textbf{Dynamics}. 

\item
\textbf{calc}$\underbar{\ }$\textbf{delh} returns $\frac{dH}{dx}=\frac{dS}{dx}=x$. 
It is called in \textbf{Molecular}$\underbar{\ }$\textbf{Dynamics}. 

\item
\textbf{Molecular}$\underbar{\ }$\textbf{Dynamics} performs one molecular evolution 
and returns the value of $x$ after the evolution and the values of the Hamiltonian before and after the evolution. 

\end{itemize}

When the action $S[x]$ is changed to more complicated functions, you have to rewrite 
\textbf{calc}$\underbar{\ }$\textbf{action} and \textbf{calc}$\underbar{\ }$\textbf{delh} accordingly. 

In \textbf{main}, the only difference from Metropolis is that \textbf{Molecular}$\underbar{\ }$\textbf{Dynamics} is used 
instead of a naive random change ($x\to x+\Delta x$ with random $\Delta x$), 
and the Metropolis test is performed by using $\Delta H$ instead of $\Delta S$. 

\subsubsection{How it works, 2 --- Matrix Integral}\label{sec:HMC-how-it-works-matrix}
\hspace{0.51cm}
Next let us consider the same example as before, 
\begin{eqnarray}
S[\phi]
=
N{\rm Tr}\left(
\frac{1}{2}\phi^2
+
\frac{1}{4}\phi^4
\right),  
\label{action-matrix-phi4}
\end{eqnarray}
where $\phi$ is $N\times N$ Hermitian, 
and use the convention explained above.  
Then the force terms are
\begin{eqnarray}
\frac{dP_{ij}}{d\tau}
=
-\frac{\partial S}{\partial \phi_{ji}}
=
-\phi_{ij}
-\left(\phi^3\right)_{ij},
\qquad
\frac{d \phi_{ij}}{d\tau}
=
P_{ij}.  
\end{eqnarray} 

The simulation code is very simple. Here is a one in Fortran 90:\footnote{
I realize that people grew up in the 21st century prefer C++. 
Still I personally love Fortran. 
} 

\begin{verbatim}
program phi4

  implicit none
  !---------------------------------
  integer nmat
  parameter(nmat=100)
  integer ninit
  parameter(ninit=0)!ninit=1 -> new config; ninit=0 -> old config
  integer iter,niter
  parameter(niter=10000)
  integer ntau
  parameter(ntau=20)
  double precision dtau
  parameter(dtau=0.005d0)
  integer naccept
  double complex phi(1:NMAT,1:NMAT),backup_phi(1:NMAT,1:NMAT)
  double precision ham_init,ham_fin,action,sum_action
  double precision tr_phi,tr_phi2
  double precision metropolis

  open(unit=10,status='REPLACE',file='matrix-HMC.txt',action='WRITE')
  !*************************************
  !*** Set the initial configuration ***
  !*************************************
  call pre_random
  if(ninit.EQ.1)then
     phi=(0d0,0d0)
  else if(ninit.EQ.0)then
     open(UNIT=22, File ='config.dat', STATUS = "OLD", ACTION = "READ")
     read(22,*) phi
     close(22)
  end if
  sum_action=0d0
  !*****************
  !*** Main part ***
  !*****************  
  naccept=0 !counter for the number of acceptance
  do iter=1,niter

     backup_phi=phi     
     call Molecular_Dynamics(nmat,phi,dtau,ntau,ham_init,ham_fin)
     !***********************
     !*** Metropolis test ***
     !***********************
     call random_number(metropolis)
     if(dexp(ham_init-ham_fin) > metropolis)then
        !accept
        naccept=naccept+1
     else 
        !reject
        phi=backup_phi
     end if
     !*******************
     !*** data output ***
     !*******************
     call calc_action(nmat,phi,action)
     sum_action=sum_action+action

     write(10,*)iter,action/dble(nmat*nmat),sum_action/dble(iter)/dble(nmat*nmat),&
     	&dble(naccept)/dble(iter)

  end do

  close(10)

  open(UNIT = 22, File = 'config.dat', STATUS = "REPLACE", ACTION = "WRITE")
  write(22,*) phi
  close(22)

end program Phi4
\end{verbatim} 

Again, it is very similar to a Metropolis code; randomly change the configuration, perform the Metropolis test, 
randomly change the configuration, perform the Metropolis test,....
In \textbf{Molecular}$\underbar{\ }$\textbf{Dynamics}, random momentum is generated
with the normalization explained below \eqref{auxiliary_momentum_hermitian}, 
the molecular evolution performed, and $H_i$ and $H_f$ are calculated. 
Subroutines \textbf{calc}$\underbar{\ }$\textbf{hamiltonian} and \textbf{calc}$\underbar{\ }$\textbf{force} (which corresponds to 
\textbf{calc}$\underbar{\ }$\textbf{delh} in the previous example) 
return the Hamiltonian and the force term $\frac{\partial H}{\partial\phi_{ji}}=\frac{\partial S}{\partial\phi_{ji}}$;  
it literally calculates products of matrices. 
Another subroutine  \textbf{calc}$\underbar{\ }$\textbf{action} is also simple. 
Let's see them one by one.\footnote{
For routines which are not explained below, please look at the sample code at \url{https://github.com/MCSMC/MCMC_sample_codes}. 
} 

\subsubsection*{\textbf{Molecular}$\underbar{\ }$\textbf{Dynamics}}
\hspace{0.51cm}

\begin{verbatim} 
subroutine Molecular_Dynamics(nmat,phi,dtau,ntau,ham_init,ham_fin)
  
  implicit none
  
  integer nmat
  integer ntau
  double precision dtau
  double precision r1,r2
  double precision ham_init,ham_fin
  double complex phi(1:NMAT,1:NMAT)
  double complex P_phi(1:NMAT,1:NMAT)
  double complex delh(1:NMAT,1:NMAT)
  integer imat,jmat,step
  !*** randomly generate auxiliary momenta ***
  do imat=1,nmat-1
     do jmat=imat+1,nmat
        call BoxMuller(r1,r2)
        P_phi(imat,jmat)=dcmplx(r1/dsqrt(2d0))+dcmplx(r2/dsqrt(2d0))*(0D0,1D0)
        P_phi(jmat,imat)=dcmplx(r1/dsqrt(2d0))-dcmplx(r2/dsqrt(2d0))*(0D0,1D0)
     end do
  end do
  do imat=1,nmat
     call BoxMuller(r1,r2)
     P_phi(imat,imat)=dcmplx(r1)
  end do
  !*** calculate Hamiltonian ***
  call calc_hamiltonian(nmat,phi,P_phi,ham_init)
  !*** first step of leap frog ***
  phi=phi+P_phi*dcmplx(0.5d0*dtau)
  !*** 2nd, ..., Ntau-th steps ***
  step=1
  do while (step.LT.ntau)
     step=step+1
     call calc_force(delh,phi,nmat)
     P_phi=P_phi-delh*dtau  
     phi=phi+P_phi*dcmplx(dtau)
  end do
  !*** last step of leap frog ***
  call calc_force(delh,phi,nmat)
  P_phi=P_phi-delh*dtau
  phi=phi+P_phi*dcmplx(0.5d0*dtau)
  !*** calculate Hamiltonian ***
  call calc_hamiltonian(nmat,phi,P_phi,ham_fin)
  
  return
  
END subroutine Molecular_Dynamics
\end{verbatim}

The inputs are the matrix size \textbf{nmat}$=N$, the matrix \textbf{phi}$=\phi^{(k)}$, 
the step size and number of steps for the molecular evolution, \textbf{dtau}$=\Delta\tau$
and \textbf{ntau}$=N_\tau$. 
The output is \textbf{phi}$=\phi'$ and \textbf{ham}$\underbar{\ }$\textbf{init}$=H_i$, \textbf{ham}$\underbar{\ }$\textbf{fin}$=H_f$. 
Note that the auxiliary momentum is neither input nor output; it is randomly generated every time in this subroutine. 

Firstly random momentum $P_\phi$ is generated. \textbf{BoxMuller($r_1$,$r_2$)} generates random numbers 
$r_1$ and $r_2$ with the Gaussian weight $\frac{e^{-r_1^2/2}}{\sqrt{2\pi}}$, $\frac{e^{-r_2^2/2}}{\sqrt{2\pi}}$. 
Note that $P_\phi$ is Hermitian, $P_\phi=P_\phi^\dagger$. Hence we take $P_{\phi,ii}$ to be real, $P_{\phi,ii}=r_1$, 
and $P_{\phi,ij}=P_{\phi,ji}^\ast=(r_1+ir_2)/\sqrt{2}$ for $i<j$. 
A factor $1/\sqrt{2}$ is necessary in order to adjust the normalization.
    \begin{verbatim}  
  !*** randomly generate auxiliary momenta ***
  do imat=1,nmat-1
     do jmat=imat+1,nmat
        call BoxMuller(r1,r2)
        P_phi(imat,jmat)=dcmplx(r1/dsqrt(2d0))+dcmplx(r2/dsqrt(2d0))*(0D0,1D0)
        P_phi(jmat,imat)=dcmplx(r1/dsqrt(2d0))-dcmplx(r2/dsqrt(2d0))*(0D0,1D0)
     end do
  end do
  do imat=1,nmat
     call BoxMuller(r1,r2)
     P_phi(imat,imat)=dcmplx(r1)    
  end do
      \end{verbatim}  
  
Then we calculate the initial value of the Hamiltonian:
   \begin{verbatim}  
  !*** calculate Hamiltonian ***
  call calc_hamiltonian(nmat,phi,P_phi,ham_init)
    \end{verbatim}  
   
Because we have already taken a backup of $\phi$ before using this subroutine, we do not take a backup here. 

Then we perform the molecular evolution by using the leap frog method.     
  \begin{verbatim}  
  !*** first step of leap frog ***
  phi=phi+P_phi*dcmplx(0.5d0*dtau)
    \end{verbatim}
    Note that we need a factor $1/2$ here! 
    Then we just repeat the leap-frog steps, 
    \begin{verbatim}  
  !*** 2nd, ..., Ntau-th steps ***
  step=1
  do while (step.LT.ntau)
     step=step+1
     call calc_force(delh,phi,nmat)
     P_phi=P_phi-delh*dtau  
     phi=phi+P_phi*dcmplx(dtau)
  end do
    \end{verbatim}
     and we need a factor $1/2$ again at the end:
    \begin{verbatim}  
  !*** last step of leap frog ***
  call calc_force(delh,phi,nmat)
  P_phi=P_phi-delh*dtau
  phi=phi+P_phi*dcmplx(0.5d0*dtau)
    \end{verbatim}

Now the molecular evolution has been done.  
In order to perform the Metropolis test, we need to calculate $H_f$:  
  \begin{verbatim}
  !*** calculate Hamiltonian ***
  call calc_hamiltonian(nmat,phi,P_phi,ham_fin)  
  \end{verbatim}

Next we need to understand other subroutines called in this subroutine. 
We will skip \textbf{BoxMuller} because it is exactly the same as before. 
The other three will be explained below; they are almost trivial as well though. 

\subsubsection*{\textbf{calc}$\underbar{\ }$\textbf{force}}
\hspace{0.51cm}
This subroutine calculates the force term 
\begin{eqnarray}
\frac{\partial H}{\partial\phi_{ji}}
=
\frac{\partial S}{\partial\phi_{ji}}
=
N\left(
\phi+\phi^3
\right)_{ij}. 
\end{eqnarray}
We just do it without using thinking too much, in the following manner:
\begin{verbatim} 
subroutine calc_force(delh,phi,nmat)

  implicit none

  integer nmat
  double complex phi(1:NMAT,1:NMAT),phi2(1:NMAT,1:NMAT),phi3(1:NMAT,1:NMAT)
  double complex delh(1:NMAT,1:NMAT)
  
  integer imat,jmat,kmat
  !*** phi2=phi*phi, phi3=phi*phi*phi ***  
  phi2=(0d0,0d0)
  phi3=(0d0,0d0)
  do imat=1,nmat
     do jmat=1,nmat
        do kmat=1,nmat
           phi2(imat,jmat)=phi2(imat,jmat)+phi(imat,kmat)*phi(kmat,jmat)
        end do
     end do
  end do
  do imat=1,nmat
     do jmat=1,nmat
        do kmat=1,nmat
           phi3(imat,jmat)=phi3(imat,jmat)+phi2(imat,kmat)*phi(kmat,jmat)
        end do
     end do
  end do
  !*** delh=dH/dphi *** 
  delh=phi+phi3
  delh=delh*dcmplx(nmat)

  return

END subroutine Calc_Force
\end{verbatim} 

\subsubsection*{\textbf{calc}$\underbar{\ }$\textbf{hamiltonian}}
\hspace{0.51cm}
This subroutine just returns $H=\frac{1}{2}{\rm Tr}P^2+S[\phi]$. 
Firstly another subroutine \textbf{calc}$\underbar{\ }$\textbf{action}, which calculates the action, is called. 
Then $\frac{1}{2}{\rm Tr}P^2$ is added. 
It is too simple and looks almost stupid, but most things needed for MCMC codes are like this. 
 
\begin{verbatim}   
SUBROUTINE calc_hamiltonian(nmat,phi,P_phi,ham)

  implicit none

  integer nmat
  double precision action,ham
  double complex phi(1:NMAT,1:NMAT)
  double complex P_phi(1:NMAT,1:NMAT)
  integer imat,jmat

  call calc_action(nmat,phi,action)

  ham=action
  do imat=1,nmat
     do jmat=1,nmat
        ham=ham+0.5d0*dble(P_phi(imat,jmat)*P_phi(jmat,imat))
     end do
  end do
  
  return

END SUBROUTINE calc_hamiltonian
\end{verbatim} 

\subsubsection*{\textbf{calc}$\underbar{\ }$\textbf{action}}
\hspace{0.51cm}
This subroutine just returns $S[\phi]$. 
We honestly write down everything. It is tedious but straightforward. You just have to be patient. 
\begin{verbatim} 
SUBROUTINE calc_action(nmat,phi,action)

  implicit none

  integer nmat
  double precision action
  double complex phi(1:NMAT,1:NMAT)
  double complex phi2(1:NMAT,1:NMAT)
  integer imat,jmat,kmat
  !*** phi2=phi*phi ***
  phi2=(0d0,0d0)
  do imat=1,nmat
     do jmat=1,nmat
        do kmat=1,nmat
           phi2(imat,jmat)=phi2(imat,jmat)+phi(imat,kmat)*phi(kmat,jmat)
        end do
     end do
  end do

  action=0d0
  !*** Tr phi^2 term ***
  do imat=1,nmat
     action=action+0.5d0*dble(phi2(imat,imat))
  end do
  !*** Tr phi^4 term ***
  do imat=1,nmat
     do jmat=1,nmat
        action=action+0.25d0*dble(phi2(imat,jmat)*phi2(jmat,imat))
     end do
  end do
  !*** overall normalization ***
  action=action*dble(nmat)

  return
  
END SUBROUTINE calc_action
\end{verbatim}

\subsubsection*{Simulation}
\hspace{0.51cm}

\begin{figure}[htbp]
  \begin{center}
   \includegraphics[width=80mm]{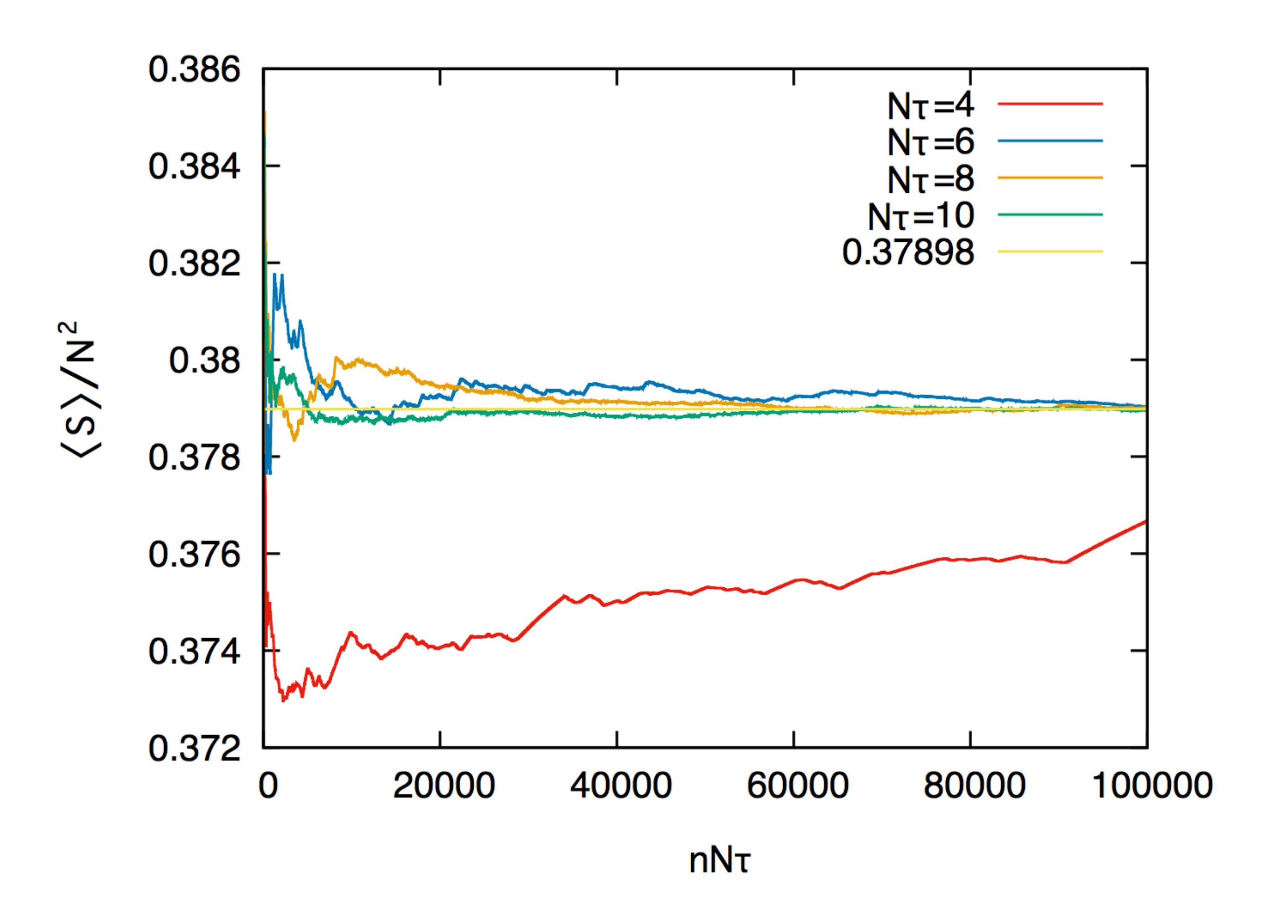}
   \includegraphics[width=80mm]{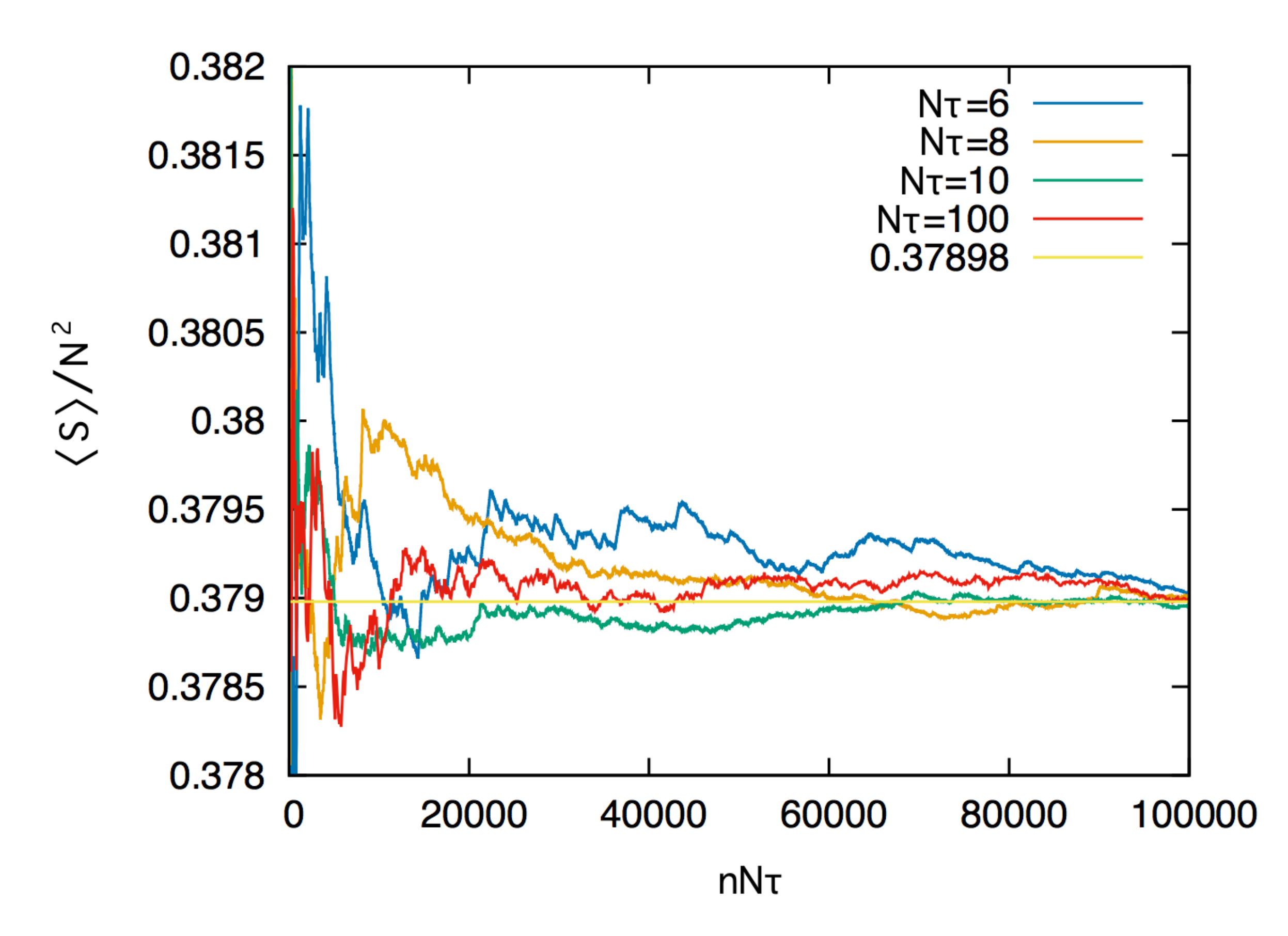}
  \end{center}
  \caption{
  $\langle S\rangle /N^2= \frac{1}{n}\sum_{k=1}^n S[\phi^{(k)}]$ with $N=100$, for several different values of $N_\tau$.
  The horizontal axis is $n\times N_\tau$, which is proportional to the cost (time and electricity needed for the simulation).  
  We can see that $N_\tau=8,10$ are more cost effective; i.e. better convergence with less cost. 
  Note that we started the measurement runs with well thermalized configurations. 
  }\label{fig:HMC-cost}
\end{figure}

In order to see how we can adjust the simulation parameters, 
let us vary $N_\tau$ and $\Delta\tau$ keeping the product $N_\tau\Delta\tau$ to be $0.1$. 
Then the acceptance rate changes as follows shown in Table~\ref{acceptance_HMC}. 
\begin{table}
\begin{center}
\begin{tabular}{|c|c||c|}
\hline
$N_\tau$ & acceptance & acceptance/$N_\tau$\\
\hline
\hline
4 & 0.0633& 0.01583\\
\hline
6 & 0.3418 &  0.05697\\
\hline
8 &  0.6023 & 0.07529 \\
\hline
10 &	0.7393 & 0.07393\\
\hline
20 & 0.9333 & 0.04667 \\
\hline
\end{tabular}
\caption{Acceptance rate for several choices of $N_\tau$, with $N_\tau\Delta\tau=0.1$, matrix size $N=100$.
We started the measurement runs with well thermalized configurations and collected 10000 samples for each parameter choice. 
}\label{acceptance_HMC}
\end{center}
\end{table}
Roughly speaking, the simulation cost is proportional to $N_\tau$. When $N_\tau\Delta\tau$ is fixed, 
the matrix $\phi$ changes more or less the same amount by the molecular evolution, regardless of $N_\tau$. 
Therefore, the rate of change is proportional to the acceptance rate. 
Hence the change per cost is $({\rm acceptance})/N_\tau$. We should maximize it. So we should use $N_\tau=8$ or $10$. 
In Fig.~\ref{fig:HMC-cost} we have plotted $\langle S/N^2\rangle = \frac{1}{n}\sum_{k=1}^n S[\phi^{(k)}]$ for several different values of $N_\tau$
by taking the horizontal axis to be `cost'$=n\times N_\tau$. We can see that $N_\tau=8,10$ are actually cost effective. 

Ideally we should do similar cost analysis varying $N_\tau\Delta\tau$, and estimate the autocorrelation length as well,  
to achieve {\it more independent configurations with less cost}.  

\begin{figure}[htbp]
  \begin{center}
   \includegraphics[width=75mm]{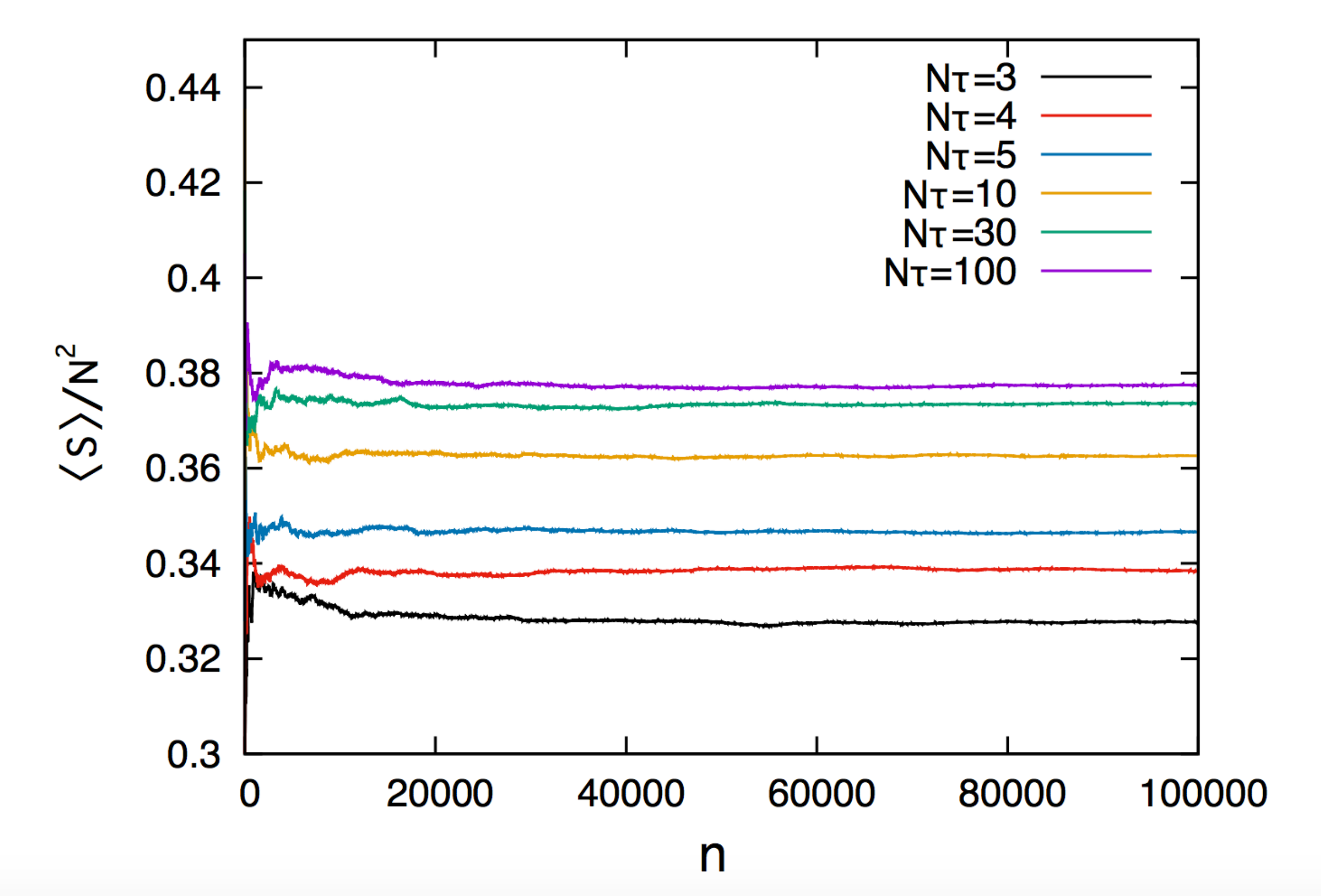}  
   \includegraphics[width=75mm]{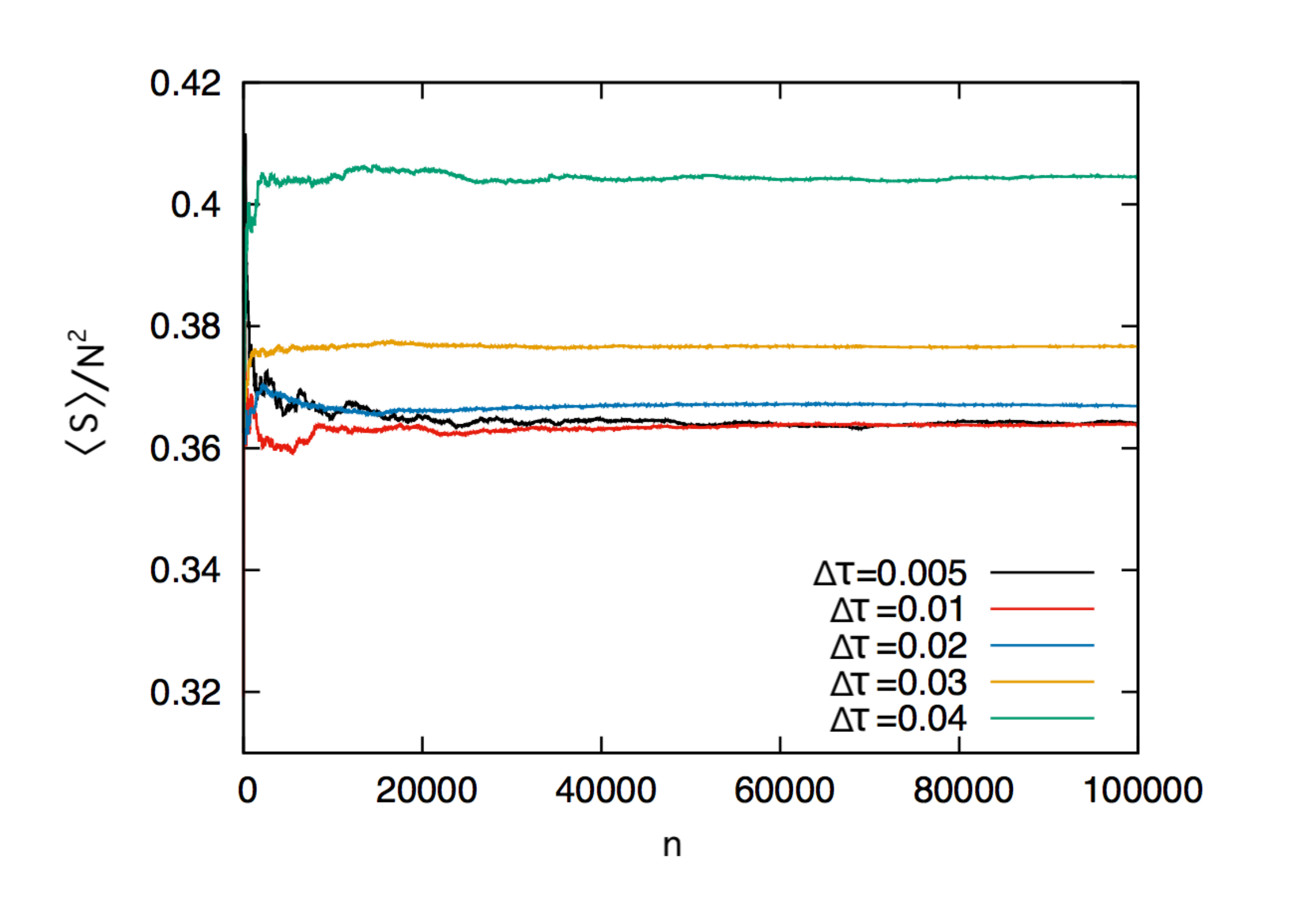}  
   \end{center}
  \caption{
  $\langle S\rangle /N^2= \frac{1}{n}\sum_{k=1}^n S[\phi^{(k)}]$ with $N=10$, for several different values of $N_\tau$ and $\Delta\tau$, 
  {\it without a factor $1/2$ in the final step of the leap frog}.
  (Left) $N_\tau\Delta\tau$ is fixed to $0.1$; (Right) $N_\tau$ is fixed to $10$. 
  Correct expectation value is obtained only at $N_\tau=\infty$ with $N_\tau\Delta\tau$ fixed. 
  }\label{fig:wrong-frog}
\end{figure}

Before closing this section, let us demonstrate the importance of the leap-frog method. 
Let us try a wrong algorithm: 
we omit a factor $1/2$ in the final step in Fig.~\ref{fig:leap-frog}.\footnote{
This was the bug in my first HMC code. It took several days to find it.
} The outcome is a disaster; 
as shown in Fig.~\ref{fig:wrong-frog}, different $N_\tau$ and $\Delta\tau$ give different values. 
Correct expectation value (which agree with the value in Fig.~\ref{fig:HMC-cost} up to a very small $1/N$ correction) is obtained only at $N_\tau=\infty$ with $N_\tau\Delta\tau$ fixed. 
It is also easy to see the importance of the normalization of the auxiliary momentum; you can try it by yourself.  

\subsubsection*{Remarks on the normalization}
\hspace{0.51cm}
Above we have assumed that the variables $x_1,x_2,\cdots$ are real. 
When you have to deal with complex variables, Hermitian matrices etc, 
you can always rewrite everything by using real variables; 
for example one can write a Hermitian matrix $M$ as $M=\sum_a M_a T^a$, 
where $T^a$ are generators and $M_a$ are real-valued coefficients.  
But it is tedious and we have seen so many lattice QCD practitioners, 
who almost always work on SU$(3)$, waste time struggling with SU$(N)$, 
just to fix the normalization. 
So let us summarize the cautions regarding the normalization. 

Let us consider the simplest case again: the Gaussian integral, $S[x]=\frac{x^2}{2}$. 
The Hamiltonian is $H[x,p]=\frac{x^2}{2}+\frac{p^2}{2}$, and the equations of motion are 
\begin{eqnarray}
\frac{dp}{d\tau}
=
-\frac{\partial H}{\partial x}
=
-\frac{\partial S}{\partial x}=-x,
\qquad
\frac{d x}{d\tau}
=
\frac{\partial H}{\partial p}
=
p.  
\end{eqnarray} 
This $p$ should be generated with the weight $\frac{1}{\sqrt{2\pi}}e^{-p^2/2}$. 

Now let $x$ be complex and  $S[x]=|x|^2=\bar{x}x$. 
Let $p$ be the conjugate of $\bar{x}$, then the Hamiltonian is $H[x,p]=\bar{x}x+\bar{p}p$, and  
\begin{eqnarray}
\frac{dp}{d\tau}
=
-\frac{\partial H}{\partial \bar{x}}
=
-\frac{\partial S}{\partial \bar{x}}
=-x,
\qquad
\frac{d x}{d\tau}
=
\frac{\partial H}{\partial \bar{p}}
=
p.  
\end{eqnarray} 
To rewrite it to real variables with the right normalization, we do as follows:
\begin{eqnarray}
x=\frac{x_R+ix_I}{\sqrt{2}}, 
\qquad
p=\frac{p_R+ip_I}{\sqrt{2}}. 
\end{eqnarray}
Then $H[x,p]=\frac{x_R^2+x_I^2+p_R^2+p_I^2}{2}$, and $(x_R, p_R)$ and $(x_I, p_I)$ are conjugate pairs. 
We should generate $p_R$ and $p_I$ with weight $\frac{1}{\sqrt{2\pi}}e^{-(p_R)^2/2}$ and $\frac{1}{\sqrt{2\pi}}e^{-(p_I)^2/2}$. 

Now let $X_{ij}$ be a Hermitian matrix, $X_{ji}=X_{ij}^\ast$. 
The conjugate $P$ is also a Hermitian matrix, and we can take $P_{ji}=P_{ij}^\ast$ to be the conjugate of $X_{ij}$. 
A simple Hamiltonian $H=\frac{1}{2}{\rm Tr}X^2 + \frac{1}{2}{\rm Tr}P^2$ becomes
\begin{eqnarray}
H
=
\frac{1}{2}\sum_i\left(X_{ii}^2+P_{ii}^2\right)
+
\sum_{i<j}\left(X_{ij}X_{ij}^\ast+P_{ij}P_{ij}^\ast\right). 
\label{auxiliary_momentum_hermitian}
\end{eqnarray}
Hence $P_{ii}$ should be generated with the weight $\frac{1}{\sqrt{2\pi}}e^{-(P_{ii})^2/2}$, 
while $P_{ij}$ can be obtained by rewriting it as $P_{ij}=\frac{P_{ij,R}+iP_{ij,I}}{\sqrt{2}}$
and generating $P_{ij,R}$, $P_{ij,I}$ with the weight $\frac{1}{\sqrt{2\pi}}e^{-(P_{ij,R})^2/2}$, $\frac{1}{\sqrt{2\pi}}e^{-(P_{ij,I})^2/2}$. 
The force terms are as follows:
\begin{eqnarray}
\frac{dP_{ij}}{d\tau}
=
-\frac{\partial H}{\partial X_{ji}}
=
-\frac{\partial S}{\partial X_{ji}}
=-X_{ij},
\qquad
\frac{d X_{ij}}{d\tau}
=
\frac{\partial H}{\partial P_{ji}}
=
P_{ij}.  
\end{eqnarray} 

As the final example, let $X_{ij}$ be a complex matrix. 
The conjugate $P$ is also a Hermitian matrix, and we can take $P_{ij}^\ast$ to be the conjugate of $X_{ij}$. 
A simple Hamiltonian $H={\rm Tr}X^\dagger X + {\rm Tr}P^\dagger P$ becomes
\begin{eqnarray}
H
=
\sum_{i,j}\left(X_{ij}X_{ij}^\ast+P_{ij}P_{ij}^\ast\right). 
\end{eqnarray}
Hence $P_{ij}$ can be obtained by rewriting it as $P_{ij}=\frac{P_{ij,R}+iP_{ij,I}}{\sqrt{2}}$
and generating $P_{ij,R}$, $P_{ij,I}$ with the weight $\frac{1}{\sqrt{2\pi}}e^{-(P_{ij,R})^2/2}$, $\frac{1}{\sqrt{2\pi}}e^{-(P_{ij,I})^2/2}$. 
The force terms are as follows:
\begin{eqnarray}
\frac{dP_{ij}}{d\tau}
=
-\frac{\partial S}{\partial X_{ij}^\ast}=-X_{ij},
\qquad
\frac{d X_{ij}}{d\tau}
=
\frac{\partial S}{\partial P_{ij}^\ast}
=
P_{ij}.  
\end{eqnarray} 

The case with generic actions $S$ should be apparent.  
Note that, when you change the normalization of the $p^2$ term in the Hamiltonian, you have to change the width of random Gaussian appropriately. 
Otherwise you will end up in getting wrong answers. 
\subsubsection*{Remark on debugging}
\begin{figure}[htbp]
  \begin{center}
  \scalebox{0.4}{
   \includegraphics{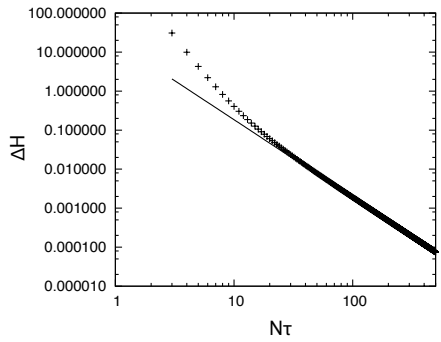}}
   \end{center}
  \caption{$\Delta H = H_f-H_i$ vs $N_\tau$, with $N_\tau\Delta\tau=0.1$ (fixed), 
  in log-log scale. 
  We picked up a thermalized configuration $\{\phi\}$ and  
  a randomly generated auxiliary momentum $\{P_\phi\}$, 
  and used the same $\{\phi,P_\phi\}$ for all $(N_\tau,\Delta\tau)$. 
  The solid line is $18.5N_\tau^{-2}$.
  The action is \eqref{action-matrix-phi4} with $N=100$. 
  When you confirm the conservation of the Hamiltonian, debugging is more or less done. 
  }\label{fig:conservation}
\end{figure}

\hspace{0.51cm}
One extra bonus associated with HMC is that we can use the conservation of the Hamiltonian for debugging. 
It is very rare to make bugs in the calculations of the action and force term consistently; so, practically, 
unless we code both correctly we cannot see the conservation of the Hamiltonian
in the `continuum limit' $N_\tau\to\infty$, $\Delta\tau\to 0$ with $N_\tau\Delta\tau$ fixed.\footnote{
Note that the normalization of the auxiliary momentum and the factor $1/2$ at the first and last steps of 
the leap-frog evolution cannot be tested by the conservation of the Hamiltonian. } 
This is a very good check of the code; see Fig.~\ref{fig:conservation}. 

\subsection{Multiple step sizes}\label{sec:multi-step-size}
\hspace{0.51cm}
Let us conside a two-matrix model
\begin{eqnarray}
S[\phi_1,\phi_2]
=
N{\rm Tr}\left(
V_1(\phi_1)
+
V_2(\phi_2)
+
\phi_1\phi_2
\right),  
\end{eqnarray}
where 
\begin{eqnarray}
V_1(\phi_1)
=
\frac{m_1^2}{2}\phi_1^2+\frac{1}{4}\phi_1^4, 
\qquad
V_2(\phi_2)
=
\frac{m_2^2}{2}\phi_2^2+\frac{1}{4}\phi_2^4. 
\end{eqnarray}

Imagine an extreme situation, $m_1=1$ and $m_2=1000000000$, i.e. 
$\phi_2$ is much heavier. 
Then typical value of $\phi_1$ is much larger than that of $\phi_2$. 
If we use the same step size for both, then in order to raise the acceptance rate for $\phi_2$
we have to take the step size very small, which leads to a very long autocorrelation for $\phi_1$. 

As we have emphasized in Sec.~\ref{caution:step-size}, 
the choice of the step size has to be consistent with the requirements listed in Sec.~\ref{sec:MCMC}, 
but other than that it is completely arbitrary; see Sec.~\ref{sec:Metropolis-multi-variable}.
Hence we can simply use different step sizes for $\phi_1$ and $\phi_2$. 
This simple fact is very important in the simulation of QFT: we should use larger step size for lighter particle. 
Also, in the momentum space, high-frequency modes are `heavier' in that $m^2+p^2$ behaves like a mass. Hence we should use smaller step size for ultraviolet modes, larger step size for infrared modes. This method is called `Fourier acceleration'. 

\subsection{Different algorithms for different fields}\label{sec:hybrid-of-algorithms}
\hspace{0.51cm}
We can even use different update algorithms for different variables, and as we will see, 
this is important when we study systems with fermions.  

Let us consider 
\begin{eqnarray}
S(x,y)=y^2f(x)+g(x) 
\end{eqnarray}
where $f(x)$ and $g(x)$ are complicated functions. 
Then we can repeat the following two steps, 
\begin{itemize}
\item
Update $y$ for fixed $x$, 

\item
Update $x$ for fixed $y$. 

\end{itemize}
In Sec.~\ref{sec:Metropolis-multi-variable} we introduced essentially the same example, 
namely we adopted the Metropolis algorithm and varied $x_1,x_2,\cdots$ one by one. 

If $f(x)>0$, then $z\equiv y\sqrt{f(x)}$ has a Gaussian weight for each fixed $x$. 
In this case, we can use the Box-Muller algorithm (Appendix~\ref{sec:Box-Muller}) 
to generate $z$; then there is no autocorrelation between $y$'s.
Hence we can use the following method:
\begin{itemize}
\item
Update $y$ for fixed $x$, by generating Gaussian random $z$
and setting $y=z/\sqrt{f(x)}$. 

\item
Update $x$ for fixed $y$, by using Metropolis or HMC.  

\end{itemize}

Note that the use of the Box-Muller algorithm does not violate the conditions listed in Sec.~\ref{sec:MCMC}. 
It just gives us a very special Markov chain without autocorrelation. 

\subsection{QFT example 1: 4d scalar theory}\label{sec:4d-scalar}
\hspace{0.51cm}
Let us consider 4d scalar theory, 
\begin{eqnarray}
S[\phi]
=
N\int d^4x
{\rm Tr}\left(
\frac{1}{2}\left(\partial_\mu\phi\right)^2
+
\frac{m^2}{2}\phi^2
+
V(\phi)
\right),  \label{action-4d-scalar}
\end{eqnarray}
where the $N\times N$ Hermitian matrices $\phi(x)$ now depends on the coordinate $x$. 
For simplicity we assume the spacetime is compactified to a square four-torus with circumference $\ell$ 
and volume $V=\ell^4$. It can be regularized by using an $n^4$ lattice with the lattice spacing $a=L/n$ as 
\begin{eqnarray}
S_{\rm lattice}[\phi]
=
Na^4\sum_{\vec{x}}
{\rm Tr}\left(
\frac{1}{2}\sum_\mu\left(
\frac{\phi_{\vec{x}+\hat{\mu}}-\phi_{\vec{x}}}{a}
\right)^2
+
\frac{m^2}{2}\phi^2_{\vec{x}}
+
V(\phi_{\vec{x}})
\right),  \label{action-4d-scalar-lattice}
\end{eqnarray}
where $\hat{\mu}$ stands for a shift of one lattice unit along the $\mu$ direction.
The path-integral is just a multi-variable integral with Hermitian matrices, 
the methods we have already explained can directly be applied.

\subsection{QFT example 2: Wilson's plaquette action (SU(N) pure Yang-Mills)}\label{sec:wilsonnian-action}
\hspace{0.51cm}
Let's move on to 4d SU($N$) Yang-Mills. 
The continuum action we consider is 
\begin{eqnarray}
S_{\rm continuum}
=
\frac{1}{4g_{YM}^2}\int d^4x{\rm Tr}F_{\mu\nu}^2, 
\end{eqnarray}
where the field strength $F_{\mu\nu}$ is defined by $F_{\mu\nu}=\partial_\mu A_\nu-\partial_\nu A_\mu + i[A_\mu,A_\nu]$. 
Typically we take the 't Hooft coupling $\lambda=g_{YM}^2N$ fixed when we take large $N$. 
As a lattice regularization we use Wilson's plaquette action,\footnote{
Often the overall factor $N$ is included in $\beta$ and $\beta'=\beta N$ is used as the lattice coupling. 
It is simply a bad convention when we consider generic values of $N$, 
because the coupling to be fixed as $N$ is varied is not $\beta'$ but $\beta$.  
} 
\begin{eqnarray}
S_{\rm lattice}=-\beta N\sum_{\mu\neq \nu}\sum_{\vec{x}}{\rm Tr}\ U_{\mu,\vec{x}}U_{\nu,\vec{x}+\hat{\mu}}U^\dagger_{\mu,\vec{x}+\hat{\nu}}U^\dagger_{\nu,\vec{x}}. 
\label{eq:plaquette_action}
\end{eqnarray}
Here $U_{\mu,\vec{x}}$ is a unitary variable living on a link connecting to lattice points $\vec{x}$ and $\vec{x}+a\hat{\mu}$, 
where $a$ is the lattice spacing and $\hat{\mu}$ is a unit vector along the $\mu$-th direction. 
It is related to the gauge field $A_\mu(\vec{x})$ by $U_{\mu,\vec{x}}=e^{iaA_\mu(\vec{x})}$.  
The lattice coupling constant $\beta$ is the inverse of the 't Hooft coupling, $\beta=1/\lambda$, 
and it should be scaled appropriately with the lattice spacing $a$ in order to achieve the right continuum limit. 
%
\subsubsection{Metropolis for unitary variables}
\hspace{0.51cm}
The only difference is that, instead of adding random numbers, 
we should multiply random unitary matrices. 
A random unitary matrix can be generated as follows. 
Firstly, we generate random Hermitian matrix $H$ by using random numbers. 
For example we can generate it with Gaussian weight $\sim e^{-{\rm Tr}H^2/2\sigma^2}$.  
Then $V=e^{iH}$ is random unitary centered around $V=1$.  
When $\sigma$ is small, it is more likely to be close to 1. 
Hence $\sigma$ is `step size'. 
Of course, you can use the uniform random number to generate $H$ if you want. 
Regardless, the Metropolis goes as follows:
\begin{enumerate}
\item
Generate $V$ randomly, change $U_{1,\vec{x}}$ to $U_{1,\vec{x}}'=U_{1,\vec{x}}V$, and perform the Metropolis test. 

\item
Do the same for $U_{2,\vec{x}}$, $U_{3,\vec{x}}$ and $U_{4,\vec{x}}$. 

\item
Do the same for other lattice sites. 

\item
Repeat the same procedure many many times. 

\end{enumerate}
\subsubsection{HMC for Wilson's plaquette action}
\hspace{0.51cm}

HMC for unitary variables goes as follows. 
Let us define the momentum $p_\mu^{ij}$ conjugate to the gauge field $A_\mu^{ji}$ by 
\begin{eqnarray}
\frac{dU}{d\tau}=ipU, 
\qquad  
\frac{dU^\dagger}{d\tau}=-iUp. 
\end{eqnarray}
It generates 
\begin{eqnarray}
U\to e^{i\delta A}U, 
\qquad  
U^\dagger\to U^\dagger e^{-i\delta A}. 
\end{eqnarray}
Therefore, 
\begin{eqnarray}
\frac{dp_{ij}}{d\tau}
=
-\frac{\partial S}{d A^{ji}}
=
-i\left(
U
\frac{\partial S}{\partial U}
\right)_{ij}
+
 i\left(
U
\frac{\partial S}{\partial U}
\right)_{ji}^\ast
\end{eqnarray} 
where the second term comes from the derivative w.r.t. $U^\dagger$. 
The discrete molecular evolution can be defined as follows:
\begin{enumerate}
\item
\begin{eqnarray}
& &
U(\Delta\tau/2)
=
\exp\left(
i\frac{\Delta\tau}{2} p(0)
\right)\cdot
U(0), 
\nonumber\\
& &
p(\Delta\tau)
=
p(0)
+
\Delta\tau\cdot\frac{dp}{d\tau}(\Delta\tau/2). 
\nonumber
\end{eqnarray}
%

\item
Repeat the following for $\tau=\Delta\tau,2\Delta\tau,\cdots,(N_\tau-1)\Delta\tau$:
\begin{eqnarray}
& &
U(\tau+\Delta\tau/2)
=
\exp\left(
i\Delta\tau p(\tau)
\right)\cdot
U(\tau-\Delta\tau/2), 
\nonumber\\
& &
p(\tau+\Delta\tau)
=
p(\tau)
+
\Delta\tau\cdot\frac{dp}{d\tau}(\tau+\Delta\tau/2). 
\nonumber
\end{eqnarray}

\item
\begin{eqnarray}
& &
U(N_{\tau}\Delta\tau)
=
\exp\left(
i\frac{\Delta\tau}{2} p(N_{\tau}\Delta\tau)
\right)\cdot
U\Big((N_{\tau}-1/2)\Delta\tau\Big). 
\nonumber
\end{eqnarray}
\end{enumerate}

In order to calculate $e^{i\Delta\tau p}$, it is necessary to diagonalize $p$. 
As long as one considers SU$(N)$ theory with not very large $N$, the diagonalization is not that costly. 
(Note also that, when the fermions are introduced, this part cannot be a bottle-neck, so we do not have to care; 
we should spend our effort for improving other parts.)
In case we need to cut the cost as much as possible, we can approximate it by truncating the Taylor expansion of $e^{i\Delta\tau\cdot p}$ at some finite order.

\section{Including fermions with HMC and RHMC}
\hspace{0.51cm}
Let us consider the simplest example, 
\begin{eqnarray}
S[x,\psi,\bar{\psi}]=\frac{x^2}{2}+\bar{\psi}D(x)\psi, 
\end{eqnarray}
where $\psi$ is a complex Grassmann number and a function $D(x)$ is a `Dirac operator'. 
We can integrate out $\psi$ by hand, so that 
\begin{eqnarray}
Z=\int dx d\psi d\bar{\psi} e^{-S[x,\psi,\bar{\psi}]}
=
\int dx D(x) e^{-x^2/2}
=
\int dx e^{-x^2/2+\log D(x)}. 
\end{eqnarray}
Hence we need to deal with the effective action in terms of $x$, 
$S_{\rm eff}=\frac{x^2}{2}-\log D(x)$. 
This is simple enough so that we don't need anything more sophisticated than the Metropolis algorithm.
However our life becomes a bit more complicated when there are multiple variables, 
$x_{1,2,\cdots, n_x}$ and $\psi_{1,2,\cdots,n_\psi}$. 
Let us consider the action of the following form, 
\begin{eqnarray}
S[x,\psi,\bar{\psi}]=\sum_{i=1}^{n_x}\frac{x_i^2}{2}
+
\sum_{a,b=1}^{n_\psi}\bar{\psi}_aD_{ab}(x)\psi_b.  
\end{eqnarray}
Now the Dirac operator $D_{ab}(x)$ is an $n_\psi\times n_\psi$ matrix, and the partition function becomes 
\begin{eqnarray}
Z=\int [dx] [d\psi] [d\bar{\psi}] e^{-S[x,\psi,\bar{\psi}]}
=
\int [dx] \det D(x)\cdot e^{-\sum_i x_i^2/2}. 
\end{eqnarray}
We can still use the Metropolis algorithm in principle,\footnote{
Here we assumed $\det D(x)>0$ for any $x$, i.e. the sign problem is absent. 
}
but the calculation of the determinant is very costly (cost$\sim n_\psi^3$). Even worse, though $D_{ab}(x)$ is typically sparse, 
it is not easy to utilize the sparseness when one calculates the determinant. 
HMC and RHMC avoid the evaluation of the determinant and enable us to utilize the sparseness of $D_{ab}(x)$. 

\subsection{2-flavor QCD with HMC}\label{sec:2-flavor}
\hspace{0.51cm}
Let us consider 2-flavor QCD, whose action in the continuum is 
\begin{eqnarray}
S[A_\mu,\psi_f,\bar{\psi}_f]
=
\int d^4x\left(
\frac{1}{4}Tr F_{\mu\nu}^2
+
\sum_{f=1}^2
\bar{\psi}^{(f)} D^{(f)}\psi^{(f)}
\right), 
\end{eqnarray}
where $A_\mu (\mu=1,2,3,4)$ is the SU(3) gauge field and 
$D_f=\gamma^\mu D_\mu + m_f$ is the Dirac operator with fermion mass $m_f$. 
For a lattice regularization, we use link variables as in Sec.\ref{sec:wilsonnian-action}, 
and put fermions on sites.
The latice action we consider is  
\begin{eqnarray}
S_{\rm Lattice}[U_\mu,\psi_f,\bar{\psi}_f]
=
S_B[U_\mu]
+
\sum_{\vec{x},\vec{y}}
\sum_{f=1}^2
\bar{\psi}^{(f)}_{\vec{x}\alpha} D^{(f)}_{\vec{x}\alpha,\vec{y}\beta}\psi^{(f)}_{\vec{y}\beta}, 
\end{eqnarray}
where $S_B[U_\mu]$ is the plaquette action \eqref{eq:plaquette_action}. 
There are various choices for a lattice Dirac operator $D^{(f)}_{\vec{x}\alpha,\vec{y}\beta}$; 
we do not specify it here. 

Below we assume $m_1=m_2=m$ (i.e. neglect the difference of the mass of up and down quarks), 
then the partition function is 
\begin{eqnarray}
Z_{\rm Lattice}
=
\int [dU] \left(\det D\right)^2 e^{-S_B}. 
\end{eqnarray}
Usually the determinant is real, and hence 
$\left(\det D\right)^2=\det (D^\dagger D)$. 

Now we introduce {\it pseudo-fermion} $F$, on which the Dirac operator acts just in the same way as on $\psi$.  
But $F$ is a complex bosonic field. Then
\begin{eqnarray}
Z_{\rm Lattice}
=
\int [dU] [dF]  e^{-S_B - F^\dagger (D^\dagger D)^{-1} F}. 
\end{eqnarray} 
At first sight it may look like a stupid way of writing an easy thing in a complicated form. 
However it enables us to avoid the calculation of $\det D$; rather we have to calculate 
a linear equation 
\begin{eqnarray}
(D^\dagger D)\chi=F, 
\end{eqnarray}
for which the sparseness of $D$ can be fully utilized. 
In order to obtain the solution $\chi$, we can use the conjugate gradient method; 
see Appendix \ref{sec:multi-mass_biCG}.

The strategy is simply applying the HMC algorithm to 
\begin{eqnarray}
S[U_\mu,F]
=
S_B[U_\mu]
+
F^\dagger (D^\dagger D)^{-1} F,  
\end{eqnarray}
by using $A_\mu$ and $F$ as dynamical variables. 
The only nontrivial parts are the calculation of the force and Hamiltonian. 
For that, we only need the solution $\chi$ of $(D^\dagger D)\chi=F$. 
Indeed, the force is calculated as 
\begin{eqnarray}
-\frac{\partial S}{\partial A_\mu}
=
-\frac{\partial S_B}{\partial A_\mu}
+
\chi^\dagger \frac{\partial (D^\dagger D)}{\partial A_\mu}\chi
\end{eqnarray}
and 
\begin{eqnarray}
-\frac{\partial S}{\partial F^\dagger}
=
 \chi,  
\end{eqnarray}
and the action is simply 
\begin{eqnarray}
S[U_\mu,F]
=
S_B[U_\mu]
+
F^\dagger\chi.   
\end{eqnarray}

\subsubsection*{A better way of treating $F$}
\hspace{0.51cm}
It is possible to update the pseudo-fermion $F$ more efficiently, 
by using the idea explained in Sec.\ref{sec:hybrid-of-algorithms}. 
This is based on a simple observation that $\Phi\equiv \left(D^\dagger\right)^{-1}F$ has the Gaussian weight and hence can easily be generated randomly by using the Gaussian random number generator.

Hence we can do as follows:

\begin{enumerate}
\item
Randomly generate $\Phi^{(k+1)}$ with the Gaussian weight\footnote{
By using the Box-Muller algorithm explained in Appendix \ref{sec:Box-Muller}, 
real Gaussian random numbers with variance 1 is generated, i.e. the weight of $x,y\in {\mathbb R}$ is 
$e^{-x^2/2}$ and $e^{-y^2/2}$. 
By dividing $\tilde{x}=x/\sqrt{2}$ and $\tilde{y}=y/\sqrt{2}$ have the weights 
$e^{-\tilde{x}^2}$ and $e^{-\tilde{y}^2}$. 
To reproduce the weight is $e^{-\Phi^\dagger\Phi}=e^{-({\rm Re}\Phi)^2-({\rm Im}\Phi)^2}$,  
we should take 
$\Phi=(\tilde{x}+\sqrt{-1}\tilde{y})=(x+\sqrt{-1}y)/\sqrt{2}$. 
If you use a wrong normalization, you end up in a wrong result. 
} $e^{-\Phi^\dagger\Phi}$. 

\item
Calculate the pseudo-fermion $F^{(k+1)}=D^\dagger(U_\mu^{(k)})\cdot \Phi^{(k+1)}$.

\item
Randomly generate auxiliary momenta $P_\mu^{(k)}$, which are conjugate to $A_\mu^{(k)}$, 
with probabilities proportional to $e^{-{\rm Tr}(P_\mu^{(k)})^2/2}$. 

\item
Calculate 
\begin{eqnarray}
H_i=S_B[U^{(k)}]+\sum_\mu {\rm Tr}(P_\mu^{(k)})^2/2+
(F^{(k+1)})^\dagger (D^\dagger(U^{(k)})\cdot D(U^{(k)}))^{-1} F^{(k+1)}.
\end{eqnarray} 
Note that
$(F^{(k+1)})^\dagger (D^\dagger(U^{(k)})\cdot D(U^{(k)}))^{-1} F^{(k+1)}
=
(\Phi^{(k+1)})^\dagger\Phi^{(k+1)}$. 

\item
Then perform the Molecular evolution for $U_\mu$, fixing $F$ to be $F^{(k+1)}$.  
The force is 
\begin{eqnarray}
-\frac{\partial S_B}{\partial A_\mu}
+
\chi^\dagger\frac{\partial (D^\dagger D)}{\partial A_\mu}\chi, 
\end{eqnarray}
where $(D^\dagger D)\chi=F$ as before. 
Then we obtain $U_\mu^{(k+1)}$ and $P_\mu^{(k+1)}$.  

\item
Calculate 
\begin{eqnarray}
H_f=S_B[U^{(k+1)}]+\sum_\mu {\rm Tr}(P_\mu^{(k+1)})^2/2+
(F^{(k+1)})^\dagger (D^\dagger(U^{(k+1)})\cdot D(U^{(k+1)}))^{-1} F^{(k+1)}.
\nonumber\\
\end{eqnarray}

\item
Metropolis test: Generate a uniform random number $r$ between 0 and 1. 
If $r<e^{H_i-H_f}$, 
$U_\mu^{(k+1)}=U_\mu^{(k)}(\tau_{f})$, otherwise $U_\mu^{(k+1)}=U_\mu^{(k)}$.

\item
Repeat 1--7. 

\end{enumerate}

\subsection{(2+1)-flavor QCD with RHMC}
\hspace{0.51cm}
Strange quark has much bigger mass than up and down quarks.
So let us take $m_1=m_2\ll m_3=m_s$.
Let us denote the Dirac operators for up and down by $D$, and the one for strange by $D_s$.   
The partition function is 
\begin{eqnarray}
Z_{\rm Lattice}
=
\int [dU] \left(\det D_s\right) \left(\det D\right)^2 e^{-S_B}. 
\end{eqnarray}
Hence we need to introduce two pseudo fermions $F$ and $F_s$ as\footnote{
$D_s^\dagger D_s$ is used so that the action becomes positive definite. 
} 
\begin{eqnarray}
S[U_\mu,F]
=
S_B[U_\mu]
+
F^\dagger (D^\dagger D)^{-1} F 
+
F_s^\dagger (D_s^\dagger D_s)^{-1/2} F_s.  
\end{eqnarray}
The last term in the right hand side is problematic because 
it is difficult (or costly) to solve $(D^\dagger D)^{1/2}\chi_s=F_s$. 

The Rational Hybrid Monte Carlo (RHMC) algorithm \cite{Clark:2003na} evades this problem by utilizing a rational approximation, 
\begin{eqnarray}
(D^\dagger D)^{-1/2}
\simeq
a_0
+
\sum_{i=1}^Q\frac{a_i}{D^\dagger D + b_i},  
\end{eqnarray}
where $a$'s and $b$'s are positive constants, and $Q$ is typically $10$ or $20$.  
These numbers are adjusted so that the rational approximation is good for all samples appearing in the actual simulation. A good code to find such numbers can be found at \cite{Clark-Kennedy}. 

By using this rational approximation, we replace the last term by 
\begin{eqnarray}
S_{\rm strange}
=
a_0 F_s^\dagger  F_s
+
\sum_{i=1}^Q
a_i F_s^\dagger \frac{1}{D_s^\dagger D_s+b_i}F_s.  
\end{eqnarray}

To determine the force, we need to solve $(D_s^\dagger D_s+\beta_i)\chi_{s,i}=F_s$. 
Rather surprisingly, we do not have to solve these equations $Q$ times; 
the multi-mass CG solver \cite{Jegerlehner:1996pm} solves all $Q$ equations simultaneously, 
essentially without any additional cost. 
See Appendix \ref{sec:multi-mass_biCG}. 


\begin{enumerate}
\item
Randomly generate $\Phi^{(k+1)}$ and $\Phi_s^{(k+1)}$ with the Gaussian weights
$e^{-\Phi^\dagger\Phi}$ and $e^{-\Phi_s^\dagger\Phi_s}$. 

\item
Calculate the pseudo-fermion $F^{(k+1)}=D^\dagger(U_\mu^{(k)})\cdot \Phi^{(k+1)}$. 

\item
Calculate the pseudo-fermion $F_s^{(k+1)}=\left(D^\dagger (U_\mu^{(k)})\cdot D(U_\mu^{(k)})\right)^{1/4}\Phi_s^{(k+1)}$. 
See the end of this section for an efficient way to do it.  

\item
Randomly generate auxiliary momenta $P_\mu^{(k)}$, which are conjugate to $A_\mu^{(k)}$, 
with probabilities proportional to $e^{-{\rm Tr}(P_\mu^{(k)})^2/2}$. 

\item
Calculate 
\begin{eqnarray}
H_i=H[U^{(k)},P^{(k)},F^{(k+1)},F_s^{(k+1)}]
\end{eqnarray}
where
\begin{eqnarray}
H[U,P,F,F_s]
&=&
S_B[U]+\sum_\mu {\rm Tr}(P_\mu)^2/2
\nonumber\\
& &
+
F^\dagger (D^\dagger(U)\cdot D(U))^{-1} F
\nonumber\\
& &
+
a_0 F_s^\dagger  F_s
+
\sum_{i=1}^Q
a_i F_s^\dagger \frac{1}{D_s(U)^\dagger D_s(U)+b_i}F_s.  
\end{eqnarray} 
Note that, although the last line formally seems to agree with $(\Phi_s^{(k+1)})^\dagger \Phi_s^{(k+1)}$ at $\tau=0$, it is not the case due to the approximations used for $(D^\dagger D)^{-1/2}$ and $(D^\dagger D)^{+1/4}$
are not exactly the `inverse'. 

\item
Then perform the Molecular evolution for $U_\mu$, fixing $F$ to be $F^{(k+1)}$.  
The force is 
\begin{eqnarray}
-\frac{\partial S_B}{\partial A_\mu}
+
\chi^\dagger\frac{\partial (D^\dagger D)}{\partial A_\mu}\chi
+
\sum_{i=1}^Q a_i\chi_{s,i}^\dagger\frac{\partial (D_s^\dagger D_s)}{\partial A_\mu}\chi_{s,i}, 
\end{eqnarray}
where $(D^\dagger D)\chi=F$ 
and 
$(D_s^\dagger D_s+b_i)\chi_{s,i}=F_s$. 
Then we obtain $U_\mu^{(k+1)}$ and $P_\mu^{(k+1)}$.  

\item
Calculate 
\begin{eqnarray}
H_f=H[U^{(k+1)},P^{(k+1)},F^{(k+1)},F_s^{(k+1)}]
\end{eqnarray}

\item
Metropolis test: Generate a uniform random number $r$ between 0 and 1. 
If $r<e^{H_i-H_f}$, 
$U_\mu^{(k+1)}=U_\mu^{(k)}(\tau_{f})$, i.e. the new configuration is `accepted.' 
Otherwise 
$U_\mu^{(k+1)}=U_\mu^{(k)}$, i.e. the new configuration is `rejected.' 

\item
Repeat 1 -- 8. 

\end{enumerate}

\subsubsection*{How to calculate $F=\left(D^\dagger D\right)^{1/4}\Phi$}
\hspace{0.51cm}
We can use the rational approximation
\begin{eqnarray}
x^{1/4} \simeq
a'_0 + \sum_{k=1}^{Q'}
\frac{a'_k}{x+b'_k}. 
\end{eqnarray}
Then 
\begin{eqnarray}
F\simeq a'_0\Phi
+
a'_k\chi'_k, 
\end{eqnarray}
where 
\begin{eqnarray}
\frac{1}{D_s^\dagger D_s+b'_k}\Phi=\chi'_k
\end{eqnarray}
is obtained by solving 
\begin{eqnarray}
(D_s^\dagger D_s+b'_k)\chi'_k=\Phi. 
\end{eqnarray}
We can use the multi-mass solver; see Appendix~\ref{sec:multi-mass_biCG}. 

\section{Maximal SYM with RHMC}\label{sec:SYM-simulation}
\hspace{0.51cm}

Sec.~\ref{sec:SYM-action} is for lattice people who knows almost nothing about SYM. 
String theorists, formal QFT people and lattice theorists who already know SYM can skip it.  

\subsection{The theories}\label{sec:SYM-action}
\hspace{0.51cm}
Supersymmetry relates bosons and fermions. Because it changes the spin of the fields, 
if there are too many supercharges (i.e. generators of supersymmetry transformation)
higher spin fields have to be involved. 
In order to construct a gauge theory without gravity, we cannot have massless fields 
with spin larger than one.
It restricts the possible number of supercharges: 
the maximal number is sixteen. 
The {\it maximally supersymmetric Yang-Mills theories} (or {\it maximal} SYM) are the supersymmetric generalizations of 
Yang-Mills theory with sixteen supercharges. 
They can be obtained from SYM in $(9+1)$ dimensions as follows. 

Firstly, why do we care about $(9+1)$ dimensions? It is related to sixteen. 
In $(9+1)$-dimensional Minkowski space, there is a sixteen-dimensional spinor representation, 
which is Majorana and Weyl. Hence the minimal supersymmery transformation has sixteen components; 
and as we have seen above, it is maximal as well. 
The field content is simple: gauge field $A_M (M=0,1,\cdots,9)$ and its superpartner ({\it gaugino}) $\psi_\alpha (\alpha=1,2,\cdots,16)$.   
Both of them are $N\times N$ Hermitian matrices. 
The action is\footnote{We use $(-,+,\cdots,+)$ signature.} 
\begin{eqnarray}
S_{(9+1)\mathchar`-{\rm d}}
=
\frac{1}{g_{YM}^2}\int d^{10}x
{\rm Tr}\left(
-\frac{1}{4}F_{MM'}^2
+
\frac{i}{2}\bar{\psi}\gamma^MD_M\psi
\right), 
\end{eqnarray}
where
$F_{MM'}=\partial_MA_{M'}-\partial_{M'}A_M+i[A_M,A_{M'}]$ and 
$D_M\psi=\partial_M\psi+i[A_M,\psi]$. 
$\gamma_M$ is the left-handed part of the 10d gamma matrices,  
which can be chosen as 
\begin{eqnarray}
\gamma^0
&=&
\textbf{1}\otimes\textbf{1}\otimes\textbf{1}\otimes\textbf{1},
\nonumber\\
\gamma^1
&=&
\sigma_3\otimes\textbf{1}\otimes\textbf{1}\otimes\textbf{1},
\nonumber\\
\gamma^2
&=&
\sigma_2\otimes\sigma_2\otimes\sigma_2\otimes\sigma_2,
\nonumber\\
\gamma^3
&=&
\sigma_2\otimes\sigma_2\otimes\textbf{1}\otimes\sigma_1,
\nonumber\\
\gamma^4
&=&
\sigma_2\otimes\sigma_2\otimes\textbf{1}\otimes\sigma_3,
\nonumber\\
\gamma^5
&=&
\sigma_2\otimes\sigma_1\otimes\sigma_2\otimes\textbf{1},
\nonumber\\
\gamma^6
&=&
\sigma_2\otimes\sigma_3\otimes\sigma_2\otimes\textbf{1},
\nonumber\\
\gamma^7
&=&
\sigma_2\otimes\textbf{1}\otimes\sigma_1\otimes\sigma_2,
\nonumber\\
\gamma^8
&=&
\sigma_2\otimes\textbf{1}\otimes\sigma_3\otimes\sigma_2,
\nonumber\\
\gamma^9
&=&
\sigma_1\otimes\textbf{1}\otimes\textbf{1}\otimes\textbf{1}.
\end{eqnarray}
They are all real and symmetric. 
The fermion $\psi$ is Majorana-Weyl, namely $\psi_\alpha^\dagger =\psi_\alpha$. 

Theories in lower spacetime dimensions can be obtained by the dimensional reduction. 
Let us restrict spacetime to be $(p+1)$-dimensional. 
Namely, we do not allow the fields to depend on $x_{p+1},\cdots,x_9$:
\begin{eqnarray}
A_M(x_0,x_1,\cdots,x_{p},\cdots,x_9)
&=&
A_M(x_0,x_1,\cdots,x_{p}),  
\nonumber\\
\psi_\alpha(x_0,x_1,\cdots,x_{p},\cdots,x_9)
&=&
\psi_\alpha(x_0,x_1,\cdots,x_{p}). 
\end{eqnarray}
Let us use $\mu,\nu,\cdots$ and $I,J\cdots$ to denote $0,1,\cdots,p$ and $p+1,\cdots,9$. 
Also let us use $X_I$ to denote $A_I$. Then
\begin{eqnarray}
F_{\mu\nu}
&=&
\partial_\mu A_\nu
-
\partial_\nu A_\mu
+
i[A_\mu,A_\nu], 
\nonumber\\
F_{\mu I}
&=&
\partial_\mu X_I
-
\partial_I A_\mu
+
i[A_\mu,X_I]
=
\partial_\mu X_I
+
i[A_\mu,X_I]
=
D_\mu X_I, 
\nonumber\\
F_{IJ}
&=&
\partial_I X_J
-
\partial_J X_I
+
i[X_I,X_J]
=
i[X_I,X_J]. 
\end{eqnarray}
Here $D_\mu X_I$ is the covariant derivative; $X_I$ behaves as an adjoint scalar after the dimensional reduction. 
Hence the dimensionally reduced theory has the following action:
\begin{eqnarray}
S_{(p+1)\mathchar`-{\rm d}}
=
S_B + S_F, 
\end{eqnarray}
where
\begin{eqnarray}
S_B
=
\frac{1}{g_{YM}^2}\int d^{p+1}x
{\rm Tr}\left(
-\frac{1}{4}F_{\mu\nu}^2
-
\frac{1}{2}(D_\mu X_I)^2
+
\frac{1}{4}[X_I,X_J]^2
\right)
\end{eqnarray}
and
\begin{eqnarray}
S_F
=
\frac{1}{g_{YM}^2}\int d^{p+1}x
{\rm Tr}\left(
\frac{i}{2}\bar{\psi}\Gamma^\mu D_\mu\psi
-
\frac{1}{2}\bar{\psi}\Gamma^I [X_I,\psi]
\right). 
\end{eqnarray}

The Euclidean theory is obtained by performing the Wick rotation:
\begin{eqnarray}
S_{\rm Euclidean}
&=&
\frac{1}{g_{YM}^2}\int d^{p+1}x
{\rm Tr}\left(
\frac{1}{4}F_{\mu\nu}^2
+
\frac{1}{2}(D_\mu X_I)^2
-
\frac{1}{4}[X_I,X_J]^2
\right.
\nonumber\\
& &
\hspace{3.5cm}
\left.
+
\frac{1}{2}\bar{\psi}D_t\psi
+
\frac{i}{2}\sum_{\mu=1}^3\bar{\psi}\gamma^\mu D_\mu\psi
-
\frac{1}{2}\bar{\psi}\Gamma^I [X_I,\psi]
\right). 
\end{eqnarray}
\subsection{RHMC for SYM}
\hspace{0.51cm}
By integrating out the fermions by hand, we obtain the Pfaffian\footnote{
Pfaffian is defined for $2n\times 2n$ antisymmetric matrices $M$, where $n=1,2,\cdots$. 
Roughly speaking, the Pfaffian is the square root of the determinant: $({\rm Pf} M)^2 = \det M$. 
More precisely, 
${\rm Pf}M=\frac{1}{2^nn!}\sum_{\sigma\in S_{2n}}{\rm sgn}(\sigma)\prod_{i=1}^n M_{\sigma(2i-1)\sigma(2i)}$. 
} of the Dirac operator:
\begin{eqnarray}
Z=\int [dA][dX] ({\rm Pf} D[A,X])\cdot e^{-S_B[A,X]}. 
\label{Z-SYM-full}
\end{eqnarray}
Unfortunately, ${\rm Pf}D$ is not positive definite. 
Hence we use the absolute value of the pfaffian instead:
\begin{eqnarray}
Z_{\rm phase\ quench}=\int [dA][dX] |{\rm Pf} D[A,X]|\cdot e^{-S_B[A,X]}. 
\label{Z-SYM-PQ}
\end{eqnarray}
For the justification of this {\it phase quenching}, 
see Sec.~\ref{sec:sign-problem}. 

In order to evaluate this integral, we use the RHMC algorithm.\footnote{
RHMC has been widely used in lattice QCD. The first application to SYM can be found in \cite{Catterall:2007fp}. 
} 
The starting point is to rewrite 
$|{\rm Pf}{\cal D}|=\left(\det (D^\dagger D)\right)^{1/4}$ as 
\begin{eqnarray}
|{\rm Pf}D|
= \int dF dF^* \exp\left(-((D^\dagger D)^{-1/8} F)^\dagger ((D^\dagger D)^{-1/8} F)\right).   
\end{eqnarray}
If we define $\Phi$ by $\Phi=(D^\dagger D)^{-1/8} F$, then $\Phi$ can be generated by the Gaussian weight $e^{-\Phi^\dagger\Phi}$, 
and $F$ can be obtained by $F={\cal D}^{1/8} \Phi$. 

We further rewrite this expression by using the rational approximation
\begin{eqnarray}
x^{-1/4} \simeq
a_0 + \sum_{k=1}^{Q}
\frac{a_k}{x+b_k}. 
\end{eqnarray}
Then we can replace the Pfaffian with
\begin{eqnarray}
|{\rm Pf}D|
= \int dF dF^* \exp\left(-S_{\rm PF}\right), 
\end{eqnarray} 
where
\begin{eqnarray}
S_{\rm PF} =
a_0 F^\dagger F + \sum_{k=1}^{Q}
a_k F^\dagger (D^\dagger D+b_k)^{-1} F. 
\label{PF-pf}
\end{eqnarray} 

The parameters $a_k$, $b_k$, $a'_k$ and $b'_k$ are real and positive, 
and can be chosen so that the approximation is sufficiently good within the range of 
$D^\dagger D$ during the simulation. 

Now it is clear that everything is the same as RHMC for strange quark except that the powers are different. 
Therefore, the molecular evolution goes as follows:
\begin{enumerate}
\item
Generate $\Phi$ by the Gaussian weight. 

\item
Calculate $F=(D^\dagger D)^{1/8}\Phi$. 

\item
Calculate $H_i$.  

\item
Fix $F$ and update $A_\mu$ and $X_I$.

\item
Calculate $H_f$. 

\item
Do the Metropois test. 

\end{enumerate}

Computationally most demanding part is solving the linear equations 
\begin{eqnarray}
({\cal D}+b_k) \chi_k = 
F \quad
(k=1, \cdots , Q ), 
\end{eqnarray}
which appears in the derivative of $S_{PF}$, 
\begin{eqnarray}
\frac{\partial S_{PF}}{\partial A}
=
-\sum_{k=1}^{Q}
a_k \chi_k^\dagger\frac{\partial \cal{D}}{\partial A} \chi_k, 
\qquad
\frac{\partial S_{PF}}{\partial X}
=
-\sum_{k=1}^{Q}
a_k \chi_k^\dagger\frac{\partial \cal{D}}{\partial X} \chi_k. 
\end{eqnarray}
This is much easier than evaluating the Pfaffian. 

\section{Difference between SYM and QCD}\label{sec:SYM-QCD-difference}
\hspace{0.51cm}
Let us explain important differences between SYM and QCD simulations.\footnote{ 
This section has large overlap with other review articles I have written in the past, 
but I include it here in order to make this article self-contained. }
\subsection{Parameter fine tuning problem and its cure}
\hspace{0.51cm}
Symmetry plays important roles in physics. For example, why is the pion so light? 
Because QCD has approximate chiral symmetry and pion is the Nambu-Goldstone boson associated with this symmetry. 
Finite quark mass breaks chiral symmetry softly, so that a light mass of pion is generated. 
The Wilson fermion breaks chiral symmetry explicitly. If we use it as a lattice regularization, 
the radiative corrections generate pion mass, and hence we have to fine-tune the bare quark mass to keep the pion light. 
This is a well known example of {\it the parameter fine tuning problem}. 
This problem does not exist if we use overlap fermion \cite{Neuberger:1997fp} or domain-wall fermion \cite{Kaplan:1992bt}. 

As another simple example, let us consider the plaquette action \eqref{eq:plaquette_action}. 
It has many {\it exact symmetries} at {\it regularized level}:
\begin{itemize}
\item 
Gauge symmetry $U_{\mu,\vec{x}}\to\Omega_{\vec{x}}U_{\mu,\vec{x}}\Omega_{\vec{x}+\hat{\mu}}^\dagger$, 
where $\Omega_{\vec{x}}$ are unitary matrices defined on sites. 

\item
Discrete translation $\vec{x}\to \vec{x}+\hat{\mu}$

\item
90-degree rotations, e.~g.~$\hat{x}\to\hat{y}, \hat{y}\to -\hat{x}$

\item
Parity $\hat{x},\hat{y},\hat{z}\to -\hat{x},-\hat{y},-\hat{z}$

\item
Charge conjugation $U\to U^\dagger$

\end{itemize}  
The integral measure is taken to be the Haar measure, which respects these symmetries. 
Therefore the radiative corrections cannot break them: these are the symmetries of the quantum theory.\footnote{
In this case, discrete translation and rotation symmetries guarantee the invariance under continuous transformations in the continuum limit. 
} 
What happens if we use a regulator which breaks these symmetries, say a momentum cutoff which spoils gauge symmetry? 
Then we need counter terms, whose coefficients are fine tuned, to restore the gauge invariance in the continuum. 

So now we know the basic strategy for the lattice simulation of supersymmetric theories --- let's keep supersymmetry exactly on a lattice!
But there is a one-sentence proof of a `no-go theorem': because supersymmetry algebra contains infinitesimal translation 
$\{Q_\alpha,\bar{Q}_\beta\}\sim\gamma^\mu P_\mu$, which is explicitly broken on any lattice by construction, it is impossible to keep 
entire algebra exactly. 

Still, it is not the end of the game. Exact symmetry at the regularized level is sufficient, but it is not always necessary. 
Sometimes the radiative corrections can be controlled by using some other symmetries or a part of supersymmetry algebra.

The first breakthrough emerged for 4d ${\cal N}=1$ pure SYM. 
For this theory, the supersymmetric continuum limit is realized 
if the gaugino mass is set to zero. 
This can be achieved if the chiral symmetry is realized on lattice \cite{Kaplan:1983sk,Curci:1986sm}.
Of course it was not really a `solution' at that time, due to the Nielsen-Ninomiya no-go theorem
\cite{Nielsen:1980rz}. 
Obviously, that the author of \cite{Kaplan:1983sk} proposed a way to circumvent the no-go theorem \cite{Kaplan:1992bt} was not a coincidence. 
(Note that the one-parameter fine tuning can be tractable in this case. See \cite{Ali:2018dnd} for recent developments.)

The second breakthrough was directly motivated by the gauge/gravity duality \cite{Maldacena:1997re}.\footnote{
D.~B.~Kaplan and M.~\"{U}nsal told me that their biggest motivation was in the gauge/gravity duality. 
For several years this motivation has not been widely shared among the followers, 
probably because they did not mention it explicitly in their first paper \cite{Kaplan:2002wv}.  
} 
The duality relates supersymmetric theories in various spacetime dimensions, including the ones with less than four dimensions, to string/M-theory \cite{Itzhaki:1998dd}. 
Some prior proposals \cite{Banks:1996vh,deWit:1988wri,Motl:1997th,Dijkgraaf:1997vv} 
related lower dimensional theories to quantum gravity as well. 
Because Yang-Mills in less than four dimensions are super-renormalizable, 
possible radiative corrections are rather constrained. 
In particular, in two dimensions, supersymmetric continuum limit can be guaranteed 
to all order in perturbation theory by keeping some supersymmetries 
in a clever manner \cite{Kaplan:2002wv}. 
 
Yet another breakthrough came from superstring theory. 
Maximally supersymmetric Yang-Mills in $(p+1)$ dimensions describe the low-energy dynamics of 
D$p$-branes \cite{Witten:1995im}. Furthermore higher dimensional D-branes can be made by collecting lower dimensional D-branes \cite{Myers:1999ps}. In terms of gauge theory, higher dimensional theories can be obtained as specific vacua of lower dimensional large-$N$ theories.
Or in other words, spacetime emerges from matrix degrees of freedom \cite{Eguchi:1982nm}.  

Based on these ideas, various regularization methods have been invented. 
A list of regularization schemes for 
4-, 8- and 16-SUSY theories without additional matter (i.e. dimensional reductions of 4d, 6d and 10d ${\cal N}=1$ pure super Yang-Mills) is shown below. 
For more details see \cite{Hanada:2016jok} and references therein. 
\begin{center}
\begin{tabular}{|c|c|c|c|c|}
\hline
Dimensions & \# of SUSY & Regularization scheme & Fine tuning \\
\hline
0+1 & 4,8,16  & lattice & none \\
\hline
0+1 & 4,8,16 & momentum cutoff & none \\
\hline
1+1 & 4,8,16 & lattice & none \\
\hline
2+1 & 4,8,16 & lattice & necessary  \\
\hline
2+1 & 4,8 (large-$N$) & fuzzy sphere & none  \\
\hline
2+1 & 16 & fuzzy sphere & none  \\
\hline
3+1 & 4  & lattice & none  \\
\hline
3+1 & 8,16 & lattice & necessary  \\
\hline
3+1 & 8 (large-$N$) & lattice + fuzzy sphere & none  \\
\hline
3+1 & 16 & lattice + fuzzy sphere & none  \\
\hline
3+1 & 4,8,16 (large-$N$) & Eguchi-Kawai reduction& none  \\
\hline
\end{tabular}

\end{center}

\subsection{Sign problem and its cure}\label{sec:sign-problem}
\hspace{0.51cm}
A crucial assumption of MCMC is that the path-integral weight $e^{-S}$ is real and positive; 
otherwise it is not `probability'.  
This condition is often broken. The simplest example is the path-integral in Minkowski space ---  
the weight $e^{iS}$, where $S$ is real, is complex.  
One may think this problem is gone in Euclidean space; the life is not that easy, 
the action $S$ can be complex after the Wick rotation; it only has to satisfy the reflection positivity, 
which is the counterpart of the reality in Minkowski space.  
Well-known evils which cause the sign problem include the $\theta$-term and 
the Chern-Simons term in Yang-Mills theory, and QCD with finite baryon chemical potential. 
Note that, although it is often called `fermion sign problem', the source of the sign is not necessary the fermionic part. 

Maximal super Yang-Mills have the sign problem, because the pfaffian can take complex values.

There is no known generic solution of the sign problem. 
It has been argued that the sign problem is NP-hard \cite{Troyer:2004ge}, 
and hence, unless you have a reason to think P$=$NP, 
it is a waste of time to invest your time in searching for a generic solution. 
However, theory-specific solutions are not excluded.  
Probably the best case scenario is that somebody comes up with a clever change of variables 
so that the sign problem disappears. 
I actually had such a lucky experience when he studied 
the ABJM matrix model in Ref.~\cite{Hanada:2012si}, 
thanks to smart students who knew important analytic formulas developed in \cite{Kapustin:2010xq}. 
Certain supersymmetric theories seem to allow yet another beautiful solution, as we will explain below.

\subsubsection{Phase reweighting}\label{sec:phase_reweight}
Let us consider SYM as a concrete example. 
The expectation value of an observable ${\cal O}$, which is written in terms of $A_\mu$ and $X_I$, 
with the `full' partition function \eqref{Z-SYM-full} is 
\begin{eqnarray}
\langle {\cal O}(A,X)\rangle_{\rm full}
=
\frac{
\int [dA][dX]({\rm Pf} D[A,X])\cdot e^{-S_B[A,X]}\cdot {\cal O}(A,X)
}{
\int [dA][dX]({\rm Pf}D[A,X])\cdot e^{-S_B[A,X]}}. 
\end{eqnarray} 
By writing ${\rm Pf} D=|{\rm Pf} D|e^{i\theta}$, we can easily see
\begin{eqnarray}
\langle {\cal O}(A,X)\rangle_{\rm full}
=
\frac{\langle {\cal O}(A,X)\cdot e^{i\theta[A,X]}\rangle_{\rm phase\ quench}}{\langle e^{i\theta[A,X]}\rangle_{\rm phase\ quench}},  
\label{phase_reweight} 
\end{eqnarray} 
where $\langle\ \cdot\ \rangle_{\rm phase\ quench}$ is the expectation value with the phase quenched path-integral \eqref{Z-SYM-PQ}. 
By using this trivial identity, in principle we can calculate the expectation value in the full theory. 
This method is called {\it phase reweighting}. 

However there are several reasons we do not want to use the phase reweighting method. 
Firstly, calculation of the pfaffian is numerically very demanding, and although we can avoid explicit evaluation of the Pfaffian 
in the configuration generation with the phase quenched ensemble, in order to calculate \eqref{phase_reweight} 
we must calculate the Pfaffian explicitly. 
Secondly, if the phase factor $e^{i\theta}$ fluctuates rapidly, both the denominator and numerator becomes 
very small, and numerically it is difficult to distinguish them from zero. Then the expression is practically $0/0$, whose error bar is 
infinitely large. The fluctuation is particularly violent when the full theory and phase quenched theory does not have 
substantial overlap (see Sec.~\ref{sec:overlapping-problem}), and it actually happens in finite density QCD. 
As we will argue shortly, the full theory and phase quenched theory are rather close in the case of SYM. 
Still, the phase fluctuation is very large and it is impossible to calculate the average phase at large $N$ and/or large volume.
Roughly speaking, $\theta$ is the imaginary part of the action, which scales as $N^2$ times volume!
  
Although the phase reweighting is an unrealistic approach to SYM, 
it can be a practical tool when the simulation cost is cheaper. 
For example it is a common strategy in certain area of condensed matter physics.

\subsubsection{Phase quench}\label{sec:phase_quench_SYM}
Let us consider the phase quench approximation. Namely, we calculate $\langle {\cal O}(A,X)\rangle_{\rm phase\ quench}$, 
which does not take into account the phase at all, instead of $\langle {\cal O}(A,X)\rangle_{\rm full}$. 
Such an `approximation' can make sense when ${\cal O}$ and $e^{i\theta}$ factorize, 
$\langle {\cal O}e^{i\theta}\rangle_{\rm phase\ quench}\simeq \langle {\cal O}\rangle_{\rm phase\ quench}\times\langle e^{i\theta}\rangle_{\rm phase\ quench}$.

The phase quench approximation almost always fails. Very fortunately, rare exceptions include SYM. 
Firstly, at high temperature and not so large $N$, $\theta$ is close to zero and hence the phase can safely be ignored. 
Rather surprisingly, $\theta$ remains small down to rather low temperature and moderately large $N$, which is relevant for testing the gauge/gravity duality; 
see e.g. \cite{Anagnostopoulos:2007fw,Hanada:2008ez,Filev:2015hia,BHM2012,Catterall:2017lub} for explicit checks for $(0+1)$- and $(1+1)$-dimensional theories. 
At low temperature or with supersymmetric boundary condition, the phase factor does oscillate. Still, as long as the fluctuation is not very rapid, 
we can defeat the sign by brute force. 
Then we can confirm $\langle {\cal O}\rangle_{\rm phase\ quench}= \langle {\cal O}\rangle_{\rm full}$ within numerical error, 
for some observables \cite{BHM2012,Hanada:2011fq}. 
When the phase factor fluctuates rapidly, or the dimension of the Dirac operator is too big, we cannot evaluate the effect of the phase.
Still we can make an indirect argument based on numerical evidence that the phase quench approximation is good 
for certain observables \cite{Berkowitz:2016jlq}. 

Note that the phase quench leads to a wrong result for the type IIB matrix model \cite{Ishibashi:1996xs}, 
which is the dimensional reduction of maximal SYM to zero dimension.

\subsection{Flat direction and its cure}
\hspace{0.51cm}
Many supersymmetric Yang-Mills theories (practically all theories relevant in the context of the gauge/gravity duality) have adjoint scalar fields $X_I$. 
There are flat directions along which scalar matrices commute, $[X_I,X_J]=0$, and the eigenvalues roll to infinity. 
With 2 or 3 noncompact spatial dimensions, we can fix the eigenvalues by hand; it is a choice of moduli, due to the superselection. 
However with compact space\footnote{
More precisely, the torus compactification. For the compactification to a curved manifold, the flat directions are lifted by a mass term associated with the curvature. 
} or lower dimensions the eigenvalue distribution should be determined dynamically. 
There is nothing wrong with this: the flat direction is a feature of the theory, it is not a bug.  
However it causes headache when we perform MCMC for two reasons: firstly, the partition function is not convergent, and hence 
the simulation never thermalizes \cite{Anagnostopoulos:2007fw}. 
Secondly, when the eigenvalues roll too far, the lattice simulation runs into an unphysical lattice artifact \cite{Hanada:2010qg}.  

In order to tame the flat directions, we have to understand its physical meaning. 
$(p+1)$-d U($N$) SYM describe a system of $N$ D$p$-branes sitting parallel to each other, and  
the eigenvalues $(X_1^{ii},X_2^{ii},\cdots,X_{9-p}^{ii})$ describe the location of the $i$-th D$p$-brane.  
When eigenvalues form a bound state, it can be regarded as a black $p$-brane. 
When one of the eigenvalues is separated far from others, the interaction is very weak due to supersymmetry; 
the potential is proportional to $-f(T)/r^{8-p}$ \cite{Danielsson:1996uw,Kabat:1996cu}, 
where $r$ is the distance, $T$ is temperature of the black brane and $f(T)$ is 
a monotonically increasing function which vanishes at $T=0$.    
This potential is not strong enough to trap a D$p$-brane once it is emitted from the black brane.   
In the same manner, phases with multiple bunches of eigenvalues can exist; they describe multiple black branes.  
This is the reason that the flat directions exist;  
the partition function is not convergent because there are too many different classes of configurations. 

Now our task is apparent: we should cut out configurations describing certain physical situation we are interested. 
The first thing we should study is a single black hole or black brane, namely a bound state of all eigenvalues. 
This bound state is only metastable, but as $N$ increases it becomes stable enough so that 
we can collect sufficiently many configurations without seeing the emission of the eigenvalues \cite{Anagnostopoulos:2007fw}.
Large-$N$ behavior of the $(0+1)$-dimensional theory has been studied in this manner in \cite{Anagnostopoulos:2007fw,Hanada:2008ez,Berkowitz:2016jlq}. 
When $N$ is small, we need to introduce a cutoff for the eigenvalues. In \cite{Hanada:2013rga} a cutoff is introduced for $(1/N)\sum_M {\rm Tr}X_M^2$, 
and by carefully changing the cutoff the property of the single bunch phase has been extracted. 
It is also possible to introduce a mass term $Nm^2\int d^{p+1}x {\rm Tr}X_M^2$ and take $m\to 0$ \cite{Hanada:2009hq,Hanada:2010qg}.
Note that one has to make sure that the flat direction is under control when $m^2$ is small, as demonstrated in \cite{Hanada:2009hq,Hanada:2010qg},  
because simulations often pick up the flat direction and end up in the U$(1)^N$ vacuum.
See also \cite{Catterall:2010fx,Giguere:2015cga}\cite{Catterall:2012yq,Catterall:2014vka,Catterall:2015ira,Catterall:2017lub}\cite{August:2018esp}
for the details on how to control the flat directions in actual simulations. 
It is also possible to add slightly more complicated deformation which preserves (at least a part of) supersymmetry, 
which is so-called `plane wave deformation' \cite{Berenstein:2002jq}. 
See \cite{Berenstein:2002jq,Kim:2006wg} for various $(0+1)$-d theories
and \cite{Hanada:2010kt,Hanada:2010gs,Hanada:2011qx} for $(1+1)$-d lattice construction. 
See also \cite{Kato:2011yh} for other cases. 

\section{Conclusion}
\hspace{0.51cm}
In this article I have tried to explain the essence of Markov Chain Monte Carlo. 
Simple algorithms like the Metropolis algorithm is good enough for solving nontrivial problems 
in hep-th literature on a laptop. 

Serious simulations of super Yang-Mills require more efforts, but the basic idea is the same. 
We just have to update the configurations more efficiently. 
For that, we can use efficient techniques developed in lattice QCD community: HMC, RHMC and many more which I haven't explained in this article. 
They can be implemented to our simulation code by using  $+,-,\times,\div,\sin,\cos,\exp,\log,\sqrt{\ \ }$, ``if" and loop. 
There are some challenging issues specific to supersymmetric theories, most notably the problem associated with flat directions, 
but we already know basic strategies to handle them. 
 
In this article I did not explain the complete details of lattice SYM simulation. 
The detail of the simulations of BFSS matrix model can be found in the article available at 
\cite{MCSMC-code}. 

\section*{Acknowledgement}
\hspace{0.51cm}
I thank Tatsuo Azeyanagi, Valentina Forini, Anosh Joseph, Michael Kroyter, So Matsuura, R.~Loganayagam, Joao Penedones, Enrico Rinaldi, David Schaich, Masaki Tezuka and Toby Wiseman. 
I was in part supported by JSPS  KAKENHI  Grants  17K14285. 
Many materials in this article were prepared for 
``Nonperturbative and Numerical Approaches to Quantum Gravity, String Theory and Holography", 
which was held from 27 January 2018 to 03 February 2018
at International Center for Theoretical Science. 
I thank Anosh Joseph and R.~Loganayagam for giving me that wonderful opportunity. 
This work was also partially supported by the Department of Energy, award number DE-SC0017905.

\appendix

\section{Multi-mass CG method }\label{sec:multi-mass_biCG}

\subsection{(Single-mass) CG method }
Let us first introduce the ordinary (or `single-mass') conjugate gradient method. 
The notation is that of \cite{NumericalPecipes}. 
We want to solve the linear equation 
\begin{eqnarray}
A\vec{x}=\vec{b},  
\end{eqnarray}
where $A=D^\dagger D$. 
We construct a sequence of approximate solutions 
$\vec{x}_1,\vec{x}_2,\cdots$ which (almost always) 
converges to the solution. 

We start with an initial trial solution $\vec{x}_1$, which is arbitrary.  
From this we define $\vec{r}_1,\vec{\bar{r}}_1,\vec{p}_1,\vec{\bar{p}}_1$ as 
\begin{eqnarray}
\vec{r}_1=
\vec{\bar{r}}_1=
\vec{p}_1=
\vec{\bar{p}}_1
=
\vec{b}-A\vec{x}_1. 
\end{eqnarray}

Then, we construct $\vec{x}_k$ as follows: 
\begin{enumerate}
\item
$
\alpha_k
=
\frac{\vec{r}^\dagger_k\cdot\vec{r}_k}{\vec{p}^\dagger_k\cdot A\vec{p}_k}
$. 
\item
$
\vec{x}_{k+1}=\vec{x}_k+\alpha_k\vec{p}_k. 
$
\item
$
\vec{r}_{k+1}
=
\vec{r}_k
-
\alpha_k A\vec{p}_k
$. 

\item
$
\beta_k
=
\frac{\vec{r}^\dagger_{k+1}\cdot\vec{r}_{k+1}}{\vec{r}^\dagger_k\cdot\vec{r}_k}
$. 
\item
$
\vec{p}_{k+1}
=
\vec{r}_{k+1}
+
\beta_k\vec{p}_{k}
$. 

\end{enumerate}

Note that $\vec{r}_{k}=\vec{b}-A\vec{x}_{k}$.  
The norm of $\vec{r}_k$ converges to zero, or equivalently, $\vec{x}_{k}$ converges to the solution. 
\subsection{Multi-mass CG method }
Let $A_\sigma$ be `shifted' version of $A$: 
\begin{eqnarray}
A_\sigma=A+\sigma\cdot\textbf{1}. 
\end{eqnarray}
If $A$ were the Laplacian, $\sigma$ would be interpreted as a mass term. 
The multi-mass CG solver \cite{Jegerlehner:1996pm} enables us to solve $A_\sigma\vec{x}=\vec{b}$ for 
many different values of $\sigma$ simultaneously, with only negligibly small additional cost. 

The key idea is that, 
from the iterative series for $A$, 
\begin{eqnarray}
\vec{r}_{k+1}
&=&
\vec{r}_k
-
\alpha_k A\vec{p}_k,
\nonumber\\
\vec{p}_{k+1}
&=&
\vec{r}_{k+1}
+
\beta_k\vec{p}_{k}, 
\label{nonshifted}
\end{eqnarray}
it is possible to construct a similar series for the shifted operator, 
\begin{eqnarray}
\vec{r}_{k+1}^\sigma
&=&
\vec{r}_k^\sigma
-
\alpha_k^\sigma A_\sigma\vec{p}_k^\sigma,
\nonumber\\
\vec{p}_{k+1}^\sigma
&=&
\vec{r}_{k+1}^\sigma
+
\beta_k^\sigma\vec{p}_{k}^\sigma,
\label{shifted} 
\end{eqnarray}
where 
\begin{eqnarray}
\vec{r}_k^\sigma
=
\zeta_k^\sigma\vec{r}_k. 
\end{eqnarray}
Indeed, (\ref{shifted}) can be satisfied by taking\footnote{
From 
(\ref{nonshifted}) we have 
\begin{eqnarray}
\vec{r}_{k+1}
=
\left(1+\frac{\alpha_k\beta_{k-1}}{\alpha_{k-1}}\right)
\vec{r}_k
-
\alpha_k A\vec{r}_k
-
\frac{\alpha_k\beta_{k-1}}{\alpha_{k-1}}
\vec{r}_{k-1}. 
\end{eqnarray}
Comparing the coefficients with those in a similar equation obtained from 
(\ref{shifted}), we obtain (\ref{shifted2}). 
} 
\begin{eqnarray}
\alpha_k^\sigma 
&=&
\alpha_k\cdot\frac{\zeta_{k+1}^\sigma}{\zeta_{k}^\sigma}, 
\nonumber\\
\beta_k^\sigma 
&=&
\beta_k\cdot\left(\frac{\zeta_{k+1}^\sigma}{\zeta_{k}^\sigma}\right)^2, 
\nonumber\\
\zeta_{k+1}^\sigma
&=&
\frac{
\zeta_{k}^\sigma\zeta_{k-1}^\sigma\alpha_{k-1}
}{
\alpha_{k-1}\zeta_{k-1}^\sigma(1+\alpha_k\sigma)
+
\alpha_k\beta_{k-1}(\zeta_{k-1}^\sigma-\zeta_{k}^\sigma)
}. 
\label{shifted2}
\end{eqnarray}
Therefore, we can generalize the usual BiCG solver in the following manner: 
\begin{enumerate}
\item
$
\alpha_k
=
\frac{\vec{r}^\dagger_k\cdot\vec{r}_k}{\vec{p}^\dagger_k\cdot A\vec{p}_k}
$. 
\item
Calculate $\zeta_{k+1}^\sigma$ and $\alpha_k^\sigma$ using (\ref{shifted2}). 
\item
$
\vec{x}_{k+1}^\sigma=\vec{x}_k^\sigma+\alpha_k^\sigma\vec{p}_k^\sigma. 
$
\item
$
\vec{r}_{k+1}
=
\vec{r}_k
-
\alpha_k A\vec{p}_k
$. 

\item
$
\beta_k
=
\frac{\vec{r}^\dagger_{k+1}\cdot\vec{r}_{k+1}}{\vec{r}^\dagger_k\cdot\vec{r}_k}
$. 
\item
Calculate $\beta_k^\sigma$ using (\ref{shifted2}). 
\item
$
\vec{p}_{k+1}
=
\vec{r}_{k+1}
+
\beta_k\vec{p}_{k}
$, 

$
\vec{p}_{k+1}^\sigma
=
\vec{r}_{k+1}^\sigma
+
\beta_k^\sigma\vec{p}_{k}^\sigma
$. 
\end{enumerate}
Then, $\vec{x}^\sigma_k$ is an approximate solution with the residual vector $\vec{r}_k^\sigma$, 
\begin{eqnarray}
\vec{r}_k^\sigma
=
\vec{b}-A_\sigma\vec{x}_k^\sigma. 
\end{eqnarray}

Note that we cannot start with an arbitrary initial condition, because (\ref{shifted}) is not satisfied then. 
We choose the following special initial condition in order to satisfy (\ref{shifted}): 
\begin{eqnarray}
& &
\vec{x}_1
=
\vec{x}_1^\sigma
=
\vec{p}_0
=
\vec{p}_0^\sigma
=
\vec{0},
\nonumber\\
& &
\vec{r}_1
=
\vec{r}_1^\sigma
=
\vec{r}_0
=
\vec{r}_0^\sigma
=
\vec{p}_1
=
\vec{p}_1^\sigma
=
\vec{b},
\nonumber\\
& &
\zeta_0^\sigma
=
\zeta_1^\sigma
=
\alpha_0
=
\alpha_0^\sigma
=
\beta_0
=
\beta_0^\sigma
=
1. 
\end{eqnarray}

As long as we stick to this initial condition, it is hard to implement any preconditioning. 
If you know any preconditioning which can be used with multi-mass CG solver, please let us know. 
\section{Box-Muller method}\label{sec:Box-Muller}
Let $p$ and $q$ be uniform random numbers in $[0,1]$. 
There are many random number generators which give you such $p$ and $q$. 
Then, $x = \sqrt{-2\log p}\sin(2\pi q)$ and $y = \sqrt{-2\log p}\cos(2\pi q)$
are random numbers with weights $\frac{e^{-x^2/2}}{\sqrt{2\pi}}$ and $\frac{e^{-y^2/2}}{\sqrt{2\pi}}$. 

\section{Jackknife method: generic case}\label{sec:Jackknife-generic}
In Sec.~\ref{sec:Jackknife}, we assumed that a quantity of interest can be calculated for each sample.
Let us consider more generic cases; for example in order to determine the mass of particle excitation we need to calculate two-point function 
by using many samples and then extract the mass from that, 
and hence the mass cannot be calculated sample by sample.

In the Jackknife method, we first divide the configurations to bins with width $w$; 
the first bin is $\{x^{(1)}\}, \{x^{(2)}\},\cdots,\{x^{(w)}\}$, the second bin is $\{x^{(w+1)}\}, \{x^{(w+2)}\},\cdots,\{x^{(2w)}\}$, etc. 
Suppose we have $n$ bins. 
Then we define the average of an observable $f(x)$ with $k$-th bin removed, 
\begin{eqnarray}
\overline{f}^{(k,w)}
\equiv
\left({\rm the\ value\ calculated\ after\ removing} k\mathchar`-{\rm th\ bin}\right). 
\end{eqnarray}
The average value is defined by 
\begin{eqnarray}
\overline{f}
\equiv
\frac{1}{n}\sum_{k}\overline{f}^{(k,w)}. 
\end{eqnarray}
The Jackknife error is defined by 
\begin{eqnarray}
\Delta_w
\equiv
\sqrt{\frac{n-1}{n}\sum_{k}\left(\overline{f}^{(k,w)}-\overline{f}\right)^2}. 
\end{eqnarray}


\end{document}